\documentclass[12pt, a4paper]{article}
\pdfoutput=1
\usepackage[dvipdfmx]{graphicx}
\usepackage{amssymb}
\usepackage{amsmath}
\usepackage{bm}
\usepackage{color}
\usepackage{cite}
\usepackage{slashed}
\usepackage{subfigure}
\usepackage{epstopdf}            
\usepackage{epsfig}
\usepackage{here}

\newcommand{\beq}{\begin{equation}}
\newcommand{\eeq}{\end{equation}}

\newcommand{\pa}{\partial}

\usepackage[colorlinks=true, linkcolor=black, citecolor=black, urlcolor=black]{hyperref}

\begin{document}

\begin{titlepage}

\begin{flushright}
KEK-TH-2044
\end{flushright}

\vskip 1.5cm

\begin{center}

{\Large
{\bf 
Flavor physics in the multi-Higgs doublet models induced by the left-right symmetry}
}

\vskip 2cm

Syuhei Iguro$^{1}$,
Yu Muramatsu$^{2}$,
Yuji Omura$^{3}$
and
Yoshihiro Shigekami$^{1,4}$

\vskip 0.5cm

{\it $^1$Department of Physics,
Nagoya University, Nagoya 464-8602, Japan}\\[3pt]

{\it $^2$ Institute of Particle Physics and Key Laboratory of Quark and Lepton Physics (MOE), Central China Normal University, Wuhan, Hubei 430079, People$^\prime$s Republic of China }\\[3pt]

{\it $^3$
Kobayashi-Maskawa Institute for the Origin of Particles and the
Universe, \\ Nagoya University, Nagoya 464-8602, Japan}\\[3pt]

{\it $^4$
Theory Center, IPNS, KEK, Tsukuba, Ibaraki 305-0801, Japan}\\[3pt]

\vskip 0.5cm

\begin{abstract}
In this paper, we discuss the multi-Higgs doublet models, that could be effectively induced by the extended Standard Model (SM). In particular, we focus on the phenomenology in the supersymmetric model with left-right (LR) symmetry, where the down-type and the up-type Yukawa couplings are unified and the Yukawa coupling matrices are expected to be hermitian. In this model, several Higgs doublets are introduced to realize the realistic fermion mass matrices, and
the heavy Higgs doublets have flavor changing couplings with quarks and leptons.
The LR symmetry is assumed to break down at high energy to realize the Type-I seesaw mechanism. 
The supersymmetry breaking scale is expected to be around 100 TeV to achieve the 125 GeV Higgs.
In such a setup, the flavor-dependent interaction of the Higgs fields
becomes sizable, so that we especially discuss the flavor physics induced by the heavy Higgs fields in our work.
Our prediction depends on the structure of neutrinos, e.g., the neutrino mass ordering.
We demonstrate how the flavor structure of the SM affects the flavor violating couplings.
In our analysis, we mainly focus on the four-fermi interaction induced by the scalar exchanging,
and we propose a simple parameterization for the coefficients.
Then, we find the correlations among the flavor observables and, for instance, see that our prediction for the $\mu \to 3 e$ process could be covered by the future experiment, in one case where the neutrino mass hierarchy is normal. 
\end{abstract}

\end{center}
\end{titlepage}

\section{Introduction}
There are a lot of candidates for new physics.
Many possible extensions of the Standard Model (SM) have been considered
to explain the origins of the parameters in the SM. For instance,
the Grand Unified Theory (GUT) reveals the origin of the
SM gauge symmetry; the left-right (LR) symmetry can resolve the
strong CP problem \cite{Babu:1989rb}.
Such new physics is often assumed to reside at the very high scale,
so that we need to find out the fragments of the hypotheses at the low scale
to verify them experimentally. 
In our paper, we focus on the extended SM with the LR symmetry and supersymmetry (SUSY), and investigate the
possibility that the LR symmetry is tested in flavor physics.

In the supersymmetric models with the LR symmetry (supersymmetric LR model), the Yukawa couplings are unified,
and the LR symmetry, that exchanges the left-handed and the right-handed fields, is conserved.
In order to realize the realistic Yukawa coupling,
extra Higgs fields are often introduced.
We write down several Yukawa couplings between the Higgs fields and the matter fields, i.e. quarks and leptons.
(See, for instance, Ref. \cite{Babu:2008ep}.)
After the LR symmetry breaking, extra Higgs $SU(2)_L$ doublets
are generated, and the light mode of the Higgs doublets contributes to the electroweak (EW) symmetry breaking.
Then, the realistic mass matrices consist of the vacuum expectation values (VEVs) and the several Yukawa couplings.
In this scenario, there is no reason that only one Higgs doublet 
exists around the EW scale. If anything, we can expect that there are additional Higgs doublets at the low scale.
For instance, in the minimal supersymmetric SM (MSSM), the MSSM could effectively lead the extended SM with one extra Higgs doublet, namely the type-II two Higgs doublet model (2HDM) below the supersymmetry (SUSY) breaking scale.\footnote{We note that the extra Higgs mass is governed by the soft SUSY breaking terms. 
If the SUSY particles are very heavy, 
it would be a non-trivial issue to make the mass hierarchy between the extra Higgs doublet and the
SUSY particles. One scenario to make the hierarchy is shown in Ref. \cite{Kawamura:2017qey}. } 
In the supersymmetric LR model, there are four Higgs doublets, as discussed in Sec. \ref{sec2}.
The LR symmetry may break down at very high energy to generate the heavy neutrino Majorana mass.
The four Higgs doublets remain the small mass scales even after the LR symmetry breaking, because they decouple with the LR symmetry breaking sector.
Then, the induced effective models would be namely multi-Higgs doublet models, where the extra Higgs doublets
couple to quarks and leptons. For instance, the Higgs fields induced by the LR symmetric model are summarized in Refs. \cite{Huitu:1994zm,Babu:2014vba,Dev:2016dja,Frank:2014kma,Frank:2011jia}.
Note that the couplings do not respect the condition for the minimal flavor violation,
so that large tree-level flavor changing neutral currents (FCNCs) involving the Higgs doublets are predicted \cite{Glashow:1976nt}. Such FCNCs can be a good probe to test the LR symmetric model.
The extra Higgs mass scale is governed by the soft SUSY breaking terms, so that
it is related to the SUSY breaking scale. 
Based on the recent LHC results and the discovery of the Higgs boson around
125 GeV, the extra Higgs masses are expected to be ${\cal O}$(100) TeV \cite{ArkaniHamed:2004fb,Giudice:2004tc,ArkaniHamed:2004yi,Giudice:2011cg,Wells,Hall,Ibe,Ibe2,ArkaniHamed}.
Then, the size of the flavor violation induced by the tree-level scalar exchanging may be enough large for the flavor physics experiments to reach. 


The extended SM with additional Higgs doublets have been discussed in the bottom-up approach, as well.
There is a rich phenomenology even in the 2HDM,
so that a lot of aspects of 2HDM have been widely investigated.  
In the bottom-up approach, we can simply classify 2HDMs according to
the type of the Yukawa couplings between the two Higgs doublet fields and the SM fermions.
For instance, in the type-II 2HDM, one Higgs doublet couples to the up-type and
the other couples to the down-type quarks. This setup is known as the one that can forbid the FCNC at the tree level. 
On the other hand, we can consider a generic 2HDM, namely type-III 2HDM,
where two Higgs doublet fields couple to both up-type and down-type quarks.
In this case, tree-level FCNCs involving scalars are generally predicted and 
we need to assume that the FCNCs are enough suppressed to evade the strong bounds from flavor physics.
In the bottom-up approach, there are many free parameters in the type-III 2HDM, so that
we are sure that we can discuss interesting physics assuming some specific Yukawa coupling alignments. 
For instance, the flavor-violating Higgs decays \cite{Omura:2015nja,Omura:2015xcg,Sierra:2014nqa,Crivellin:2015mga,deLima:2015pqa}, the magnetic dipole moment of muon \cite{Omura:2015nja,Omura:2015xcg}, the $B$ physics \cite{Ko:2012sv, Ko:2017lzd,Iguro:2017ysu,Iguro:2018qzf,Crivellin2012ye,Celis:2012dk,Tanaka2012nw,Crivellin:2013wna,Crivellin:2015hha,Cline:2015lqp,
Hu:2016gpe,Arnan:2017lxi,Arhrib:2017yby} and 
the top physics \cite{Ko:2011vd,Ko:2011di} have been studied in this framework, motivated by
the experimental results. This approach, however, raises a question what the underlying theory of
the type-III 2HDM with the specific Yukawa couplings is, even if the model is confirmed at some experiments.

Based on this background, in this paper, we discuss flavor physics in the multi-Higgs doublet model with the
{\it unification constraint}, where the realistic mass matrices for quarks and leptons
are given by the linear combination of the VEVs of Higgs doublets and the several unified Yukawa couplings. 
It is interesting that the FCNCs of the scalars are written down by the mixing angles of the scalars, the CKM matrices and
the fermion masses because of the {\it unification constraint}. Then, we can derive the explicit predictions against flavor violating processes. 
In the supersymmetric LR model, the Higgs doublets can be expected to obtain the masses via the soft SUSY breaking terms, since they decouple to the LR breaking sector. In our setup, the LR breaking scale is very large to generate large Majorana right-handed neutrino masses for the Type-I seesaw mechanism. 
Then, the effective model at the low energy is interpreted as a multi-Higgs doublet model with the tree-level FCNCs.
 In addition, we consider the split SUSY scenario in which the sfermion masses are ${\cal O}$(100) TeV, while the gaugino masses are ${\cal O}$(1) TeV. In this setup, the extra Higgs masses are expected to be around the sfermion mass scale to realize the 125 GeV Higgs mass as mentioned above.
Note that the FCNCs are also induced by the sfermions at the loop-level. These contributions are subdominant in our discussion. Therefore, we do not discuss about such contributions and we focus only on the tree-level FCNCs induced by the extra Higgs in this paper.
As we see in Sec. \ref{sec2}, the bigger the hierarchy between 
the LR breaking scale and the EW breaking scale is,  the larger FCNCs are predicted in the lepton sector.
Although the many possibilities of the scalar mass spectrum and the scalar mixing would hinder
the search for the explicit predictions of the FCNCs, our setup makes it simpler to analyze the Wilson coefficients of the
four-fermi interactions induced by the scalar exchanging.
In fact, we find a simple description of the four-fermi interactions. Then we propose that 
we can discuss the correlations among the flavor observables in this model, using the simple parameterization.

Note that there are many studies on Higgs physics \cite{Babu:2014vba,Dev:2016dja,Huitu:1994zm,Frank:2014kma} and flavor physics \cite{Frank:2011jia} in the supersymmetric LR model. The flavor physics in the non-supersymmetric LR model
has been done in Refs.\cite{Ball:1999mb,Kiers:2002cz,Haba:2017jgf,Zhang:2007da,Guadagnoli:2010sd,Blanke:2011ry,Bertolini:2014sua,Bernard:2015boz,ValeSilva:2016czu,FileviezPerez:2017zwm}. In those references, the LR symmetry breaking scale is much lower than in our model. In our case, the LR symmetry breaks down at the very high scale to realize the Type-I seesaw mechanism. What we stress in this paper is that our setup makes it much simpler to study the phenomenology and
we can discuss flavor physics even in such a high-scale scenario. In Ref. \cite{Frank:2011jia}, the $\Delta F=2$ processes
are mainly studied. In our paper, we study the leptonic meson decay and the charged lepton flavor violation (LFV)
as well as the $\Delta F=2$ processes.

In Sec. \ref{sec2}, we explain our setup and discuss how the realistic Yukawa couplings can be derived in the supersymmetric LR model, and then study the induced Yukawa couplings of the extra Higgs doublets in the effective models.
In Sec. \ref{sec3}, we study phenomenology, especially flavor physics, in the multi-Higgs doublet models 
with the {\it unification constraint} of the LR model.
Sec. \ref{summary} is devoted to summary. In Appendix \ref{appendix3},
the alignment of the lepton Yukawa couplings required by the neutrino observations are shown.
In Appendix \ref{appendix2}, the supersymmetric LR model, that can realize the Type-I seesaw scenario,
is introduced. The complementary discussions about the four-fermi interactions and the corrections from the renormalization group (RG) are summarized, in Appendix \ref{appendix1}, Appendix \ref{appendix4}
and Appendix \ref{app_MZto100TeV}.

\section{Multi-Higgs doublet models effectively induced by the LR model}
\label{sec2}

First of all, let us briefly explain how to realize the realistic Yukawa couplings in the supersymmetric LR model.

We begin with the brief introduction of the Yukawa interaction in the Standard Model.
The Yukawa couplings for quark and lepton masses are given by
\beq
\label{eq;Yukawa-h}
Y^u_{ij}\overline{\Hat Q^i_L} \, \widetilde h \, \Hat u^j_R + Y^d_{ij}\overline{\Hat Q^i_L} \, h \, \Hat d^j_R + Y^e_{ij}\overline{\Hat L^i} \, h \, \Hat e^j_R +h.c.,
\eeq
where $\Hat Q^i_L$, $\Hat u^i_R$, $\Hat d^i_R$, $\Hat L^i$ and $\Hat e^i_R$
are the SM quarks and leptons in the interaction base. $h$ denotes the Higgs doublet, and $\widetilde h$ is defined
as $\tau_2 h^*$. The neutral component of $h$ gains the mass about 125 GeV after the EW symmetry breaking.
$Y^u_{ij}$, $Y^d_{ij}$ and $Y^e_{ij}$ are the Yukawa couplings described as
\beq
\label{eq;Yukawa}
Y^u_{ij}= V^\dagger_{ik} \frac{m^u_{k} \sqrt{2}}{v} V_{R \, kj}, ~ Y^d_{ij}=  \delta_{ij} \, \frac{m^d_{j} \sqrt{2}}{v} , ~ Y^e_{ij}=  \delta_{ij}\frac{m^e_{i} \sqrt{2}}{v},
\eeq 
where $v \simeq 246$ GeV is the SM Higgs VEV, $V$ is the CKM matrix and $V_R$ is the unitary matrix to rotate the right-handed quarks. If $Y^u$ is hermitian, $V_R$ is identical to $V$.

Now, we extend the Higgs sector, assuming that the high-energy model is the supersymmetric LR model whose gauge group is $SU(3)_C \times SU(2)_L \times SU(2)_R \times U(1)_{B-L}$.
In this model, $Y^u_{ij}=Y^d_{ij}$ is predicted if there is only one bi-doublet chiral superfield.
This relation conflicts with the experimental results, so that
we need some improvements to realize the realistic Yukawa couplings effectively.
One simple way is to introduce extra Higgs fields and extra Yukawa couplings to the extended SM.
Let us explain the idea in a supersymmetric LR model below.

\subsection{The supersymmetric LR model}
\label{sec;SUSYLRmodel}
In the LR model, the right-handed up-type and down-type fermions are unified into $SU(2)_R$ doublet fields, and the $SU(2)_L$ doublet Higgs is embedded into a bi-doublet scalar field denoted as $\Phi$ in this paper.
If we consider the supersymmetric LR model, at least two bi-doublet chiral superfileds are needed for the realistic Yukawa couplings since the potential is described by the holomorphic function, namely superpotential. Therefore, the superpotential for the visible sector is given by 
\beq
\label{eq:superpotential}
W_{vis}= Y^{a \, *}_{ji}  \Hat Q^i_L \tau_2 \Phi_a \tau_2 \Hat Q^{c \,j}_R +Y^{l \, a \, *}_{ji}  \Hat L^i \tau_2 \Phi_a \tau_2 \Hat L^{c \,j}_R + \lambda^{\nu}_{ij} \,  \Hat  L_R^{c \,i} \tau_2 \Delta_R \Hat L^{c \,j}_R  + \mu^{ab} \, Tr \left( \tau_2 \Phi^T_a  \tau_2 \Phi_b\right),
\eeq
where $\Hat Q^{c \,j}_R$ and $\Hat L_R^{c \,j} $ denote $(\Hat d^{c \,j}_R, \, -\Hat u^{c \,j}_R)^T$ and $(\Hat e^{c \,j}_R, \, -\Hat n^{c \,j}_R)^T$. $\Phi_a$ ($a=1, \, 2$) can be decomposed as $\Phi_a = (\widetilde H^a_u, \,H^a_d)$, where $H^a_u$ and $H^a_d$ are the $SU(2)_L$ doublets in this notation.
As a result, we obtain four Higgs doublet fields after the $SU(2)_R$ symmetry breaking.
The third term in Eq. (\ref{eq:superpotential}) effectively generates the Majorana mass term for the right-handed neutrino, and
the last term corresponds to the $\mu$-term of the Higgs superfields.
In our analysis, the Yukawa couplings, $Y^a_{ij}$ and $Y^{l \, a}_{ij}$, are defined in the base where 
$ \mu^{ab}$ is in the diagonal form: $ \mu^{ab}= \mu^a \delta_{ab}$.

Let us consider the scenario that $\Delta_R$ develops the very large VEV for the very heavy right-handed neutrino.
This can be easily realized by introducing a singlet field, $S$:
\beq
W_{SB}= m(S) \, Tr\left( \Delta_R \overline \Delta_R \right ) + w(S).
\eeq  
We can find the supersymmetric vacuum that breaks down $SU(2)_R \times U(1)_{B-L}$ to $U(1)_Y$.
This type of model has been proposed in Ref. \cite{Babu:2008ep}. The other setup has been discussed 
in Refs. \cite{Aulakh:1997ba,Kuchimanchi:1995vk,Mohapatra:1995xd}.
The matter contents and
the charge assignment are summarized in Table \ref{table;LRmodel}.
$\Delta_{L}$ and $\overline \Delta_{L}$ are also introduced to respect the LR symmetry.
$S$ is a gauge singlet superfield to induce the LR symmetry breaking. The detail of the breaking
is summarized in Appendix \ref{appendix2}.

Note that the terms involving $\Delta_{L}$ and $\overline \Delta_{L}$ can be also written as 
\beq
\label{eq:WDeltaL}
W_{\Delta_L}= (m_L+m(S)) \, Tr\left( \Delta_L \overline \Delta_L \right )+ \lambda^{\nu \, *}_{ij}  L^i \tau_2 \Delta_L  L^j.
\eeq
$m_L$ is the soft LR breaking term, so that $\Delta_L$ and $\overline{\Delta}_L$ are integrated out at some scale in our setup.
Except for the soft LR breaking term, the total superpotential, $W = W_{vis} + W_{SB} + W_{\Delta_L}$, is LR symmetric which means that our lagrangian is invariant under the parity transformation: $Q_L \rightarrow Q^{c \,*}_R$, $L_L \rightarrow L^{c \,*}_R$, $\Delta_L \rightarrow \Delta_R^*$, $\overline \Delta_L \rightarrow \overline \Delta_R^*$, $\Phi_a \rightarrow \Phi_a^{\dagger}$ and $S \rightarrow S^*$.
Because of this LR symmetry, the Yukawa coupling matrices $Y^a_{ij}$ and $Y^{l \, a}_{ij}$ are the hermitian matrices.
In general, the $\Delta_L$ and $\overline \Delta_L$ exchanging causes the LFV processes,
governed by $\lambda^\nu_{ij}$.
Besides, $\Delta_L$ and $\overline \Delta_L$ would enhance the LR breaking effect after the spontaneous LR symmetry breaking, if they are light. 
In our setup, we introduce the soft LR breaking terms, and $\Delta_L$ and $\overline \Delta_L$ are integrated out
at the high scale. We safely ignore the contribution to the LFV, but the more detailed analysis should be done
taking into account the LR breaking effect. Our main motivation is, however, to find the correlation among 
the flavor observables concerned with not only leptons but also quarks.
Then,  we discuss the flavor physics induced by the Higgs doublets
and demonstrate how largely the flavor violating processes are enhanced by the supersymmetric LR extension.
The detailed analysis including the LR breaking effect will be done near future.

\begin{table}[t]
\begin{center}
  \begin{tabular}{|c|c|} \hline
    Field & $SU(3)_C \times SU(2)_L \times SU(2) _R \times U(1)_{B-L}$ \\ \hline \hline
    $Q^i_L$ & $({\bf 3}, \, {\bf 2}, \, {\bf 1}, \, 1/3  )$ \\
    $Q^{c \,i}_R$ & $(\overline{\bf 3}, \, {\bf 1}, \, {\bf 2}, \, -1/3  )$ \\
    $L^i$ & $({\bf 1}, \, {\bf 2}, \, {\bf 1}, \, -1  )$ \\
    $L^{c \,i}_R$ & $({\bf 1}, \, {\bf 1}, \, {\bf 2}, \, 1  )$ \\
    $\Phi_{1,2}$ & $({\bf 1}, \, {\bf 2}, \, {\bf 2}, \, 0  )$ \\  \hline
    $\Delta_{R}$ & $({\bf 1}, \, {\bf 1}, \, {\bf 3}, \, -2  )$ \\
    $\overline \Delta_{R}$ & $({\bf 1}, \, {\bf 1}, \, {\bf 3}, \, 2  )$ \\
    $\Delta_{L}$ & $({\bf 1}, \, {\bf 3}, \, {\bf 1}, \, 2  )$ \\
    $\overline \Delta_{L}$ & $({\bf 1}, \, {\bf 3}, \, {\bf 1}, \, -2  )$ \\
    $S$ & $({\bf 1}, \, {\bf 1}, \, {\bf 1}, \, 0  )$
    \\ \hline
  \end{tabular}
  \caption{Matter contents and the charge assignment of the $SU(2)_R \times U(1)_{B-L}$  model.}
  \label{table;LRmodel}
\end{center}
\end{table}

There are two bi-doublet superfields, $\Phi_a$, in our model.
$\Phi_a$ can not couple to $\Delta_R$ at the renormalizable level because of the $U(1)_{B-L}$ symmetry, so that the Higgs doublets from $\Phi_a$ do not gain the large masses from the VEV of $\Delta_R$.
Then, the masses of the scalars from $\Phi_a$ dominantly come from the soft SUSY breaking terms and the $\mu$-terms
that are expected to be around the sfermion scale in our scenario.
We could expect that the $SU(2)_R$ breaking effect is mediated by the mediators for the SUSY breaking effects.\footnote{Note that we may wonder how SUSY is broken and how the $SU(2)_R$ breaking effect is mediated.
See, for instance, Ref. \cite{Kobayashi:2017fgl}.} In our study, we simply assume that 
the $SU(2)_R$ breaking effect appears in the soft SUSY breaking terms and discuss the mass terms for the Higgs doublets which are associated with the $SU(2)_R$ breaking, below.

In this supersymmetric LR model, there are two up-type Higgs doublets $(H^{a}_u)$ and two down-type Higgs doublets $(H^{a}_d)$ originated from $\Phi_a$ $(a=1,\,2)$.
The masses of the four Higgs doublets are given by not only the supersymmetric masses but also
the soft SUSY breaking terms. Let us define the mass squared as 
\beq
(M^2_H)_{IJ} \Hat H^{\dagger}_I \Hat H_J,
\eeq
where $(M^2_H)_{IJ}$ is a $4 \times 4$ hermitian matrix and $\Hat H_I$ ($I=1,\,2,\,3,\,4$) denotes
($\Hat H_1$, $\Hat H_2$, $\Hat H_3$, $\Hat H_4$)=($H^1_u$, $H^2_u$, $H^1_d$, $H^2_d$), respectively. In this notation, $(M^2_H)_{IJ}$ is given by
\beq
\label{HiggsMassMatrix}
 (M^2_H)_{IJ} = 
\begin{pmatrix} m^2_{H^1_u}+|\mu_1|^2  &  m^2_{12 \, u} & B_{11} & B_{12} \\
m^{2 \, *}_{12 \, u} & m^2_{H^2_u}+|\mu_2|^2 & B_{21} & B_{22}  \\ 
B^*_{11} & B^*_{21}  & m^2_{H^1_d}+|\mu_1|^2  & m^2_{12 \, d}  \\   
B^*_{12} & B^*_{22} & m^{2 \, *}_{12 \, d} & m^2_{H^2_d}+|\mu_2|^2   \end{pmatrix}.
\eeq
Here, $H^a_u$ and $H^a_d$ are the supersymmetric mass eigenstates: $\mu^{ab}=\mu^a \delta_{ab}$.
The other parameters in Eq. (\ref{HiggsMassMatrix}) denote the soft SUSY breaking parameters.
In order to realize the EW symmetry breaking, sizable $B_{ab}$ is required. 
In addition, $m^2_{H^a_u}$ and $m^2_{H^a_d}$ should satisfy some conditions to cause the EW symmetry breaking and to avoid the unbounded-from-below vacua.
In our study, we do not discuss the origin of the soft SUSY breaking terms and simply assume that the conditions are satisfied. In our setup, the soft SUSY breaking terms are ${\cal O}(100)$ TeV, since we consider the split SUSY scenario whose sfermion scale is ${\cal O}(100)$ TeV,
so that the fine-tuning of the supersymmetric mass is required to realize the 125 GeV Higgs mass,
as well known in the MSSM. The other massive scalars from $\Phi_a$ are, on the other hand, expected to be ${\cal O}(100)$ TeV.

The VEVs of $\Hat H_I$ are aligned as
\beq
\langle \Hat H_I \rangle= \frac{v}{\sqrt 2} U_{Ih},
\eeq
where $U_{Ih}$ is the four-dimensional vector that satisfies $\sum_I U_{Ih} U^*_{Ih}=1$.
Finding the directions orthogonal to $U_{Ih}$, we define another base for the Higgs doublets:
\beq
\Hat H_I =  U_{Ih} \, h +  U_{IA} \, H_A ~ (A=1,\,2,\,3),
\eeq
where $ U_{IA}$ satisfies 
\beq
U^*_{IA} U_{IB} = \delta_{AB}, ~ U^*_{Ih}U_{IA}=0.
\eeq
In this base, only $h$ develops a non-vanishing VEV, $\langle h \rangle = (0, \, v/\sqrt{2})^T$.

$h$ would correspond to the mass eigenstate given by $(M^2_H)_{IJ}$: $(M^2_H)_{IJ}U_{Jh}= -\mu^2_h U_{Ih}$.
The other states, $H_A$, could be also interpreted as the mass eigenstates of $(M^2_H)_{IJ}$, so that
the mass squared for the Higgs fields is described as
\beq
\Hat U^\dagger_H M^2_H \Hat U_H= diag(-\mu^2_h, \, M^2_{H_1}, \, M^2_{H_2}, \, M^2_{H_3}),
\eeq
where the unitary matrix, $\Hat U_H$, is defined as $\Hat U_H = (U_{Ih} \, U_{I1}\, U_{I2}\, U_{I3})$.
Note that the exact masses of the heavy scalars would be deviated from $M^2_{H_A}$,
because of the contributions of 4-point couplings, e.g. $|h|^2 |H_A|^2$, to the masses squared.
These contributions are, however, suppressed, compared to $(M^2_H)_{IJ}$, if $M_{H_A}$
is much larger than the EW scale as expected in our model.
Then, we discuss the phenomenology, assuming $H_A$ are the mass eigenstates
with $M_{H_A}$. The mass differences among the scalars in each $H_A$ are negligible, in this assumption.

Now, we write down the Yukawa couplings involving $h$ and $H_A$.
The Yukawa couplings of $h$ correspond to the SM Yukawa couplings, e.g., $Y^u_{ ij}$ and $Y^d_{ ij}$.
The relation between $Y^a$ and the realistic Yukawa couplings can be obtained by
\beq
\begin{pmatrix} Y^u_{ ij}  \\ Y^d_{ ij}  \end{pmatrix}  = \begin{pmatrix} U^*_{1h} & U^*_{2h} \\  U_{3h} & U_{4h} \end{pmatrix}\begin{pmatrix} Y^1_{ ij}  \\ Y^2_{ ij} \end{pmatrix}.
\eeq
Note that the strong CP problem would arise if $U_{Ih}$ is complex.
When the Yukawa couplings of $H_A$ with quarks are defined as
\beq
Y^u_{A \, ij}\overline{\Hat Q^i_L} \, \widetilde H_A \, \Hat u^j_R + Y^d_{A \, ij}\overline{\Hat Q^i_L} \, H_A \, \Hat d^j_R +h.c.,
\eeq
$Y^u_{A \, ij}$ and $Y^d_{A \, ij}$ are related to $Y^u_{ij}$ and $Y^d_{ij}$ as
\beq
 \label{eq;heavy-Yukawa-LR-quark-SUSY}
 \begin{pmatrix} Y^u_{A\, ij}  \\ Y^d_{A\, ij}   \end{pmatrix}  = \frac{1}{\Delta_h} \begin{pmatrix} U^*_{1A} &  U^*_{2A} \\  U_{3A} & U_{4A} \end{pmatrix} \begin{pmatrix}  U_{4h} & - U^*_{2h} \\  -U_{3h} & U^*_{1h} \end{pmatrix}\begin{pmatrix}  Y^u_{ ij}  \\  Y^d_{ ij}\end{pmatrix},
\eeq
where $\Delta_h= U^*_{1h} U_{4h}-U_{3h}U^*_{2h}$ is defined.
Note that $\Delta_h$ is vanishing in the $SU(2)_R$ symmetric limit.
Similarly, the Yukawa couplings of $H_A$ for the leptons,
\beq
Y^\nu_{A \, ij}\overline{\Hat L^i} \, \widetilde H_A \, \Hat \nu^j_R + Y^e_{A \, ij}\overline{\Hat L^i} \, H_A \, \Hat e^j_R +h.c.,
\eeq
are related to $Y^\nu_{ij}$ and $Y^e_{ij}$:
\beq
 \label{eq;heavy-Yukawa-LR-lepton-SUSY}
 \begin{pmatrix} Y^\nu_{A\, ij}  \\ Y^e_{A\, ij}   \end{pmatrix}  = \frac{1}{\Delta_h} \begin{pmatrix} U^*_{1A} &  U^*_{2A} \\  U_{3A} & U_{4A} \end{pmatrix} \begin{pmatrix}  U_{4h} &-U^*_{2h} \\  -U_{3h} & U^*_{1h} \end{pmatrix}\begin{pmatrix}  Y^\nu_{ ij}  \\  Y^e_{ ij}\end{pmatrix}.
\eeq
There are many parameters in the Yukawa couplings of the heavy scalars: $U_{IA}$ and $U_{Ih}$. In addition, there are three mass parameters, $M_{H_A}$.
The mass parameters could be expected to be around the sfermion scale $(\sim {\cal O}(100) \, {\rm TeV})$, since they correspond to
the soft SUSY breaking terms. The mass spectrum, however, depends on the mediation mechanism. 

After the $SU(2)_R$ breaking, the Majorana mass terms,  $(M_\nu)_{ij}$, would be effectively generated as
\beq
 (M_\nu)_{ij} \,  \overline{\Hat \nu^{c \, i}_R} \Hat \nu^j_R  + h.c..
\eeq 
Assuming that the magnitude of $(M_\nu)_{ij}$ is very large compared to the EW scale,
the tiny neutrino masses of the active neutrinos are given by
\beq
\label{neutrino}
\Hat m^{\nu}_{ij}   = v^2 Y^\nu_{  ik} \left (M_\nu^{-1} \right)_{kl} Y^\nu_{  jl}.
\eeq
In our base, the Yukawa coupling matrix for the charged lepton, $Y^e_{ij}$, is in the diagonal form, so that
$\Hat m^{\nu}_{ij}$ is described as
\beq
\Hat m^{\nu}_{ij} = (V_{PMNS})_{ik} m^\nu_k (V^T_{PMNS})_{kj}= v^2 (U_\nu)_{ik} y^{\nu}_k \left ( U^{\nu \, \dagger}_R  M_\nu^{-1} U^{\nu \, *}_R  \right)_{kl} y^{\nu}_l (U^T_{\nu})_{lj},
\eeq
where $V_{PMNS}$ is the PMNS matrix and $Y^\nu_{ij}$ is defined as
\beq
Y^\nu_{ij}=  \left ( U_{\nu} \right )_{ik} y^{\nu}_k  \left ( U^{\nu \, \dagger}_R \right )_{kj}.
\eeq
Thus, $U^{\nu}$ is not identical  to $V_{PMNS}$, unless $U^{\nu \, \dagger}_R  M_\nu^{-1} U^{\nu \, *}_R$ is 
in the diagonal form.

 Another important point is the relation between $U_{\nu}$ and $U^{\nu}_R$.
If the LR symmetry is assumed to be conserved at high energy, $Y^\nu_{ij}$ would be the hermitian matrix if the radiative corrections can be safely ignored.\footnote{There are also other contributions to the LR breaking effects: e.g., the one from the $SU(2)_L$ triplet which is introduced to respect the LR symmetry.}
In such a case, we can simply estimate the sizes of $Y^\nu$ and $M_\nu$.
In Appendix \ref{appendix3}, we quantitatively estimate the Dirac neutrino Yukawa couplings, assuming that the Majorana mass terms
are also hierarchical.
$Y^\nu$ and $M_\nu$ are shown assuming the mass hierarchy in $M_\nu$. We denote $(M_\nu)_{ii}$ as $M_{\nu i}$ in the following.


\begin{table}[t]
\begin{center}
  \begin{tabular}{|c|c||c|c|} \hline
 $m_e$& $0.511$ MeV \cite{PDG}  &  $\sin^2 \theta_{12}$ & 0.321$^{+0.018}_{-0.016}$  \cite{neutrino1}     \\ 
    $m_\mu$ & $105.658$ MeV \cite{PDG} & $\sin^2 \theta_{23}$ (NO) & 0.430$^{+0.020}_{-0.018}$   \cite{neutrino1}  \\ 
      $m_{\tau}$&1776.86$\pm 0.12$ MeV  \cite{PDG}    &  $\sin^2 \theta_{23} $ (IO)& 0.596$^{+0.017}_{-0.018}$  \cite{neutrino1}  \\
 $\tau_\mu$& $3.33781\times10^{15}$ MeV$^{-1}$  \cite{PDG} &    $\sin^2 \theta_{13}$ (NO)& 0.02155$^{+0.00090}_{-0.00075}$  \cite{neutrino1} \\ 
 $\tau_\tau$& $(441.0\pm0.8)\times10^6$ MeV$^{-1}$  \cite{PDG} &$\sin^2 \theta_{13}$ (IO)& 0.02140$^{+0.00082}_{-0.00085}$  \cite{neutrino1} \\ 
  $\Delta m^2_{12}$& $(7.56 \pm 0.19)\times 10^{-5}$ eV$^2$ \cite{neutrino1}   &   $\delta/\pi$ (NO) &   1.40$^{+0.31}_{-0.20}$ \cite{neutrino1} \\ 
 $|\Delta m^2_{13}|$ (NO)& $(2.55 \pm 0.04) \times 10^{-3}$ eV$^2$   \cite{neutrino1}   &  $\delta/\pi$ (IO) & 1.44$^{+0.26}_{-0.23}$ \cite{neutrino1} \\
 $|\Delta m^2_{13}|$ (IO)& $(2.49 \pm 0.04) \times 10^{-3}$ eV$^2$  \cite{neutrino1}    &  & \\ \hline
  \end{tabular}
 \caption{The input parameters for leptons in our analysis. The notation
 for the lepton mixing is following Ref. \cite{PDG}. NO (IO) is short for the normal (inverted) ordering neutrino mass hierarchy. The central values are used in our study.}
  \label{table;input2}
  \end{center}
\end{table}


$Y^\nu$ would be large if the Majorana mass is very heavy.
The Majorana mass term is originated from the $SU(2)_R$ breaking, so that
the high $SU(2)_R$ breaking scale, that is assumed in our study, leads sizable Yukawa couplings involving heavy scalars, according to Eq. (\ref{eq;heavy-Yukawa-LR-quark-SUSY}). 
For example, if $M_{\nu 1} \gg M_{\nu 2,3}$ and $M_{\nu 2,3} \sim \mathcal{O}(10^{13 \mathchar`- 14})$ GeV are assumed, we can obtain $Y^{\nu}$ as
\begin{align}
Y^{\nu} &\simeq \begin{pmatrix}
1 & -0.05 + 0.04 i & 0.05 + 0.03 i \\
-0.05 - 0.04 i & 0.1 & 0.2 + 0.008 i \\
0.05 - 0.03 i & 0.2 - 0.008 i & 0.1 \\
\end{pmatrix} \qquad \qquad \text{(normal ordering),} \\
Y^{\nu} &\simeq \begin{pmatrix}
1 & -0.003 - 0.03 i & -0.09 + 0.5 i \\
-0.003 + 0.03 i & 0.1 & -0.1 + 0.002 i \\
-0.09 - 0.5 i & -0.1 - 0.002 i & 0.02 \\
\end{pmatrix} \qquad \text{(inverted ordering)}.
\end{align}

As we discuss in Sec. \ref{sec;4-fermi}, if we focus on the four-fermi couplings,
we find that those parameter dependences on the Yukawa couplings in Eq. (\ref{eq;heavy-Yukawa-LR-lepton-SUSY})
lead simple forms to the Wilson coefficients, that contribute to the flavor physics.
We derive the coefficients in Sec. \ref{sec3} and discuss the flavor physics, using the simplified parametrization of the Wilson coefficients.

\subsection{The induced four-fermi couplings}
\label{sec;4-fermi}
Before the phenomenology, we derive the effective couplings induced by the heavy scalars with the Yukawa couplings in Eqs. (\ref{eq;heavy-Yukawa-LR-quark-SUSY}) and (\ref{eq;heavy-Yukawa-LR-lepton-SUSY}).
Integrating out the heavy scalars, we obtain the four-fermi couplings.
In our study, we assume that the components of $H_A$ in the supersymmetric LR models
are degenerate. Then, the couplings by the heavy neutral scalar exchanging are given as follows:
\beq
\label{4-fermi}
{\cal L}^n_{eff}=\frac{1}{m^2_{H_A}} Y^{f \, \dagger }_{A\, ij}  Y^{F}_{A\, kl} \, \left (\overline{\Hat f^{ i}_R}  \Hat f^{j}_L \right ) \,  \left ( \overline{\Hat F^{k}_L} \Hat F^{ l}_R \right ),
\eeq
where $f$ and $F$ denote $u$, $d$, $e$ or $\nu$, respectively.
Using the relation between $(M^2_{H})_{IJ}$ and $\Hat U_H$, the coefficients in front of the four-fermi operators can be simplified. The explicit expressions are summarized in the Appendix \ref{appendix1}.

Defining the dimensional parameters, $\Lambda_{\alpha \beta}$, we write down the down-type quark couplings in Eq. (\ref{4-fermi}) as
\beq
\label{4-fermi-down}
(C^d_4)^{ij}_{kl} =  \begin{pmatrix} Y^{u \, \dagger}_{ij} &  y^{d }_i \,  \delta_{ij} \end{pmatrix}
\begin{pmatrix}  \Lambda^{-2}_{uu} & \Lambda^{-2}_{ud} \\  \Lambda^{-2}_{du} &  \Lambda^{-2}_{dd} \end{pmatrix} \begin{pmatrix} Y^u_{kl} \\ y^{d }_{k}  \,  \delta_{kl} \end{pmatrix},
\eeq
where $Y_u=V^\dagger y^{u } V_R$ is in Eq. (\ref{eq;Yukawa}) and $y^f_k$ satisfies $y^f_k=\sqrt{2} m^f_k/v$.
We change the base of the down-type quark into the mass base denoted by $d^i$, so that
$(C^d_4)^{ij}_{kl}$ corresponds to the coefficient of $(\overline{ d^{ i}_R}  d^{j}_L  ) \,  (\overline{d^{k}_L} d^{ l}_R  )$.
Note that $\Lambda_{\alpha \beta}$ $(\alpha,\,\beta=u,\,d)$ are related to $(M^{-2}_H)_{IJ}$ and $U_{Ih}$ and summarized in Appendix \ref{appendix1}.

When we discuss the flavor violating processes, such as the $\Delta F=2$ processes,
we find that $\Lambda^{-2}_{uu}$ is only relevant in our model according to Eq. (\ref{4-fermi-down}).
As shown in Eq. (\ref{Lambda-uu}), $\Lambda^{-2}_{uu}$ is described as
\beq
 \Lambda^{-2}_{uu} = \frac{1}{|\Delta_h|^2} (U^1_{\bot})^*_{I} (M^{-2}_H)_{IJ} (U^1_{\bot})_J,
\eeq
where $(U^1_{\bot})_I$ denotes the vector orthogonal to $U_{Ih}$: $( (U^1_{\bot})_1$, $(U^1_{\bot})_2$, $(U^1_{\bot})_3$, $(U^1_{\bot})_4$)$=(0$, $0$, $-U^*_{4h}$, $U^*_{3h}$) satisfying $(U^1_{\bot})^*_I U_{Ih}=0 $.
$U_{Ih}$ would denote one mass eigenstate of $M^{2}_H$ whose mass squared is $-\mu^2_h$, so that $(U^1_{\bot})_I$ would be described by one linear combination of the other three mass eigenstates of  $M^{2}_H$ with the masses, $m^2_{H_A}$. Then, $\Lambda^{-2}_{uu} $ is positive and could not be vanishing, as far as all $M^{2}_{H_A}$ are not extremely large. As discussed in Sec. \ref{DeltaF2}, large deviations from the SM predictions are actually derived 
in the $\Delta F=2$ processes.

The four-fermi coupling in the charged lepton sector has the structure similar to the one in the down-type quark sector.
Replacing $Y^u$ and $Y^d$ with $Y^\nu$ and $Y^e$ respectively,
$(C^e_4)^{ij}_{kl}$, that is the coefficient of $(\overline{ e^{ i}_R}  e^{j}_L  ) \,  (\overline{e^{k}_L} e^{ l}_R  )$,
is given by
\beq
(C^e_4)^{ij}_{kl}=\begin{pmatrix} (Y^{\nu \, \dagger})_{ij} &  y^e_i \, \delta_{ij} \end{pmatrix}
\begin{pmatrix}  \Lambda^{-2}_{uu} & \Lambda^{-2}_{ud} \\  \Lambda^{-2}_{du} &  \Lambda^{-2}_{dd} \end{pmatrix} \begin{pmatrix} (Y^{\nu })_{kl} \\ y^{e }_{k}  \,  \delta_{kl} \end{pmatrix}.
\eeq
Note that $Y^{\nu}$ is the source of the flavor violation in the charged lepton sector.
This means that there is a possibility that the observable, the PMNS matrix, 
in the neutrino physics connects with the charged LFV processes.
The detail is shown in Sec. \ref{LFV}.


The coefficients of the other four-fermi interactions, that induce the LFV decays of mesons, 
are given by
\beq
\label{eq;summary-deltaF1}
{\cal L}^{l}_{eff}=(C^{ue}_4)^{kl}_{ij} \left (\overline{u^{ i}_L}  u^{j}_R \right ) \,  \left ( \overline{e^{k}_L} e^{ l}_R \right )+(C^{de}_4)^{kl}_{ij} \left (\overline{d^{ i}_R}  d^{j}_L \right ) \,  \left ( \overline{e^{k}_L} e^{ l}_R \right )+h.c.,
\eeq
where $(C^{de}_4)_{ij}^{kl}$ and $(C^{ue}_4)_{ij}^{kl}$ are described as
\beq
(C^{de}_4)_{ij}^{kl} =\begin{pmatrix} Y^{u \, \dagger }_{ij} &  y^{d }_i \delta_{ij} \end{pmatrix}
\begin{pmatrix}  \Lambda^{-2}_{uu} & \Lambda^{-2}_{ud} \\  \Lambda^{-2}_{du} &  \Lambda^{-2}_{dd} \end{pmatrix} \begin{pmatrix} (Y^\nu )_{kl} \\ y^{e }_{k} \delta_{kl} \end{pmatrix}
\eeq
and 
\beq
(C^{ue}_4)_{ij}^{kl} = \begin{pmatrix} y^{u }_{i} \delta_{ij} &  ( V y^{d} V^\dagger_R )_{ij} \end{pmatrix}
\begin{pmatrix}  \left  (\Lambda^{(u e) }_{uu} \right )^{-2} & \left (\Lambda^{(u e) }_{ud} \right)^{-2}  \\   \left (\Lambda^{(u e) }_{du} \right )^{-2}  &   \left (\Lambda^{(u e)}_{dd} \right )^{-2}  \end{pmatrix} \begin{pmatrix} (Y^{\nu })_{kl} \\ y^{e }_{k} \delta_{kl} \end{pmatrix},
\eeq
where $\Lambda^{(u e) }_{\alpha \beta}$ $(\alpha,\,\beta=u,\,d)$ are also related to $(M^{-2}_H)_{IJ}$ and $U_{Ih}$ as shown in Appendix \ref{appendix1}.
We emphasize that these parameters $\Lambda_{\alpha \beta}$ and $\Lambda_{\alpha \beta}^{(ue)}$ ($\alpha$, $\beta=u$, $d$) 
can be calculated once the mass matrix for Higgs doublets in Eq. (\ref{HiggsMassMatrix}) is given.
The other operators and corresponding couplings that are not shown here is summarized in Appendix \ref{appendix1}.

Before the concrete study on flavor physics, let us comment on the LR breaking contributions from the RG to the Yukawa couplings.
In the supersymmetric case, the RG corrections to the leptonic Yukawa couplings ($Y^{l \, a}_{ij}$) are only given by the wave faction renormalization factors:
\beq
Y^{l \, a}_{ij} (\mu)=(Z^{\dagger }_{L})^{ik} \, Y^{l \, b}_{km} (\Lambda)  \, Z^{mj}_{R_l} \,  Z^{H_d}_{ba}.
\eeq
The each $Z$ factor in the right-handed side corresponds to the wave faction renormalization factor of the each field denoted in the subscript. Even in the one-loop correction, the LR breaking effect appears since right-handed neutrinos are integrated out
and the other fields such as $SU(2)_L$ triplet, $\Delta_L$, may reside at low energy.
$Z_{L}$ and $Z_{R}$ are not identical because of the effect, so that $Y^{l \, a}_{ij} (\mu)$ is not hermitian matrix below the LR breaking scale. \footnote{Once the hermitian condition is violated, the strong CP phase of QCD may arise through the RG correction from the wave faction renormalization factors of Higgs doublets, $Z^{H_d}_{ab}$ and $Z^{H_u}_{ab}$. The strong CP problem will be taken into consideration in the future work.}
When we discuss the phenomenology in our model, we focus on the parameter region that the LR breaking effect in $Y^{l \, a}_{ij} (\mu)$ is approximately parameterized as
\beq
Y^{l \, a}_{ij} (\mu) = \Hat Y^{l \, a}_{ ik} (\mu) Z^l_{kj}, 
\eeq
using the hermitian matrix, $\Hat Y^{l \, a}$, and extra parameters $Z^l_{ij}$ which is generally a $3 \times 3$ matrix.
In our study on phenomenology, $Z^l_{ij}$ is assumed to be in a diagonal form: $Z^l_{ij}=Z^l_i \, \delta_{ij}.$
This situation can be realized, assuming that one element of $Y^{l \, a}_{ij}$ is dominant in the each RG equation at the one-loop level. 
When only one element of $Y^{l \, a}_{ij}$ is close to unit and the RG runs from $10^{12}$ TeV to $10^2$ TeV, the RG correction is about 20 \%. Although the correction highly depends on the setup at the high scale, we simply treat $Z^l_{i}$ as real free parameters satisfying $0.8 \leq Z^l_{i} \leq 1.2$\footnote{$Z^l_i$ is expected to be less than unit, taking into account only the $(Y^l_{ a})_{ik}$ contributions.} in our analysis.

The RG corrections to the quark Yukawa couplings can be evaluated in the same manner:
\beq
Y^{u \, a}_{ij} (\mu)=(Z^{\dagger }_{Q_L})^{ik} \, Y^{ b}_{km} (\Lambda)  \, Z^{mj}_{u_R} \,  Z^{H_u}_{ba}, ~ Y^{d \, a}_{ij} (\mu)=(Z^{\dagger }_{Q_L})^{ik} \, Y^{ b}_{km} (\Lambda)  \, Z^{mj}_{d_R} \,  Z^{H_d}_{ba}.
\eeq
The contribution of the quark Yukawa couplings in $Z_{Q_L}$, $ Z_{u_R}$ and $ Z_{d_R}$ respects the 
LR symmetry at the one-loop level, ignoring the contribution of the $SU(2)_L \times U(1)_Y$ gauge interactions and $Y^{l \, a}$ interaction.
The LR breaking effects induced by the gauge interaction are flavor universal, so that the flavor structure is not modified.
Thus, in quark sector, we could expect that the prediction of the LR model is still valid at the low scale. The detail discussion about the RG correction is summarized in Appendix \ref{appendix4}.

\section{Flavor physics}
\label{sec3}
In this section, we discuss the phenomenology, especially flavor physics, in our models.
We simplify the RG corrections focusing on the cases (i), (ii) and (iii) as in Sec. \ref{sec;SUSYLRmodel} and Appendix \ref{appendix3}, and numerically study our predictions derived from the LR symmetry at high energy.

There are many parameters, e.g. the scalar masses and the mixing. In our study, we discuss the phenomenology 
using the dimensional parameters defined in Sec. \ref{sec;4-fermi}. Using the parameterization, we do not need to touch the detailed setup concerned with the masses and the mixing. Then, the parameters relevant to our study about flavor physics are as follows:
\beq
\label{eq;parameter}
\Lambda_{uu}, \, \Lambda_{ud}, \, \Lambda_{du}, \, \Lambda^{(ue)}_{uu}, \, \Lambda^{(ue)}_{du}, \,  \Lambda^{(ue)}_{dd}.
\eeq
$\Lambda_{ud}= \Lambda^*_{du}$ is predicted, as shown in Eq. (\ref{eq;Lambdadu}).
The other parameters can be, in principle, independent each other, so that we discuss the constraints and the impacts of the each parameter on flavor physics. Note that we also assume that the sfermion scale is ${\cal O}(100)$ TeV, to
avoid the strong constraint on the SUSY particles from the LHC experiments and to obtain the 125 GeV Higgs mass \cite{ArkaniHamed:2004fb,Giudice:2004tc,ArkaniHamed:2004yi,Giudice:2011cg,Wells,Hall,Ibe,Ibe2,ArkaniHamed}.
This means that the extra scalar masses are also expected to be much higher than the EW scale,
and then the flavor-violating processes induced by the one-loop diagrams, such as $b \to s \gamma$,
are safely negligible and we discuss only the tree-level processes induced by the extra Higgs.
Note that the branching ratio of $b \to s \gamma$ limits the new physics scale if the scalars are below 1 TeV:
the lower bound is about 580 GeV in the Type-II 2HDM \cite{Misiak:2017bgg}.  

The first three parameters in Eq. (\ref{eq;parameter}) contribute to the $\Delta F=2$ processes, the LFV and the leptonic decays of the mesons. In particular, $\Lambda_{uu}$ is strongly constrained by the $K$-$\overline{K}$ mixing.
In the LFV and the leptonic meson decay, $ \Lambda_{ud}$ may significantly contribute to the observables.
 
 The other three parameters, $\Lambda^{(ue)}_{uu}$, $\Lambda^{(ue)}_{du}$ and $\Lambda^{(ue)}_{dd}$,
 suppress the couplings between up-type quarks and leptons. Then, they contribute to the
 $\mu-e$ conversion process significantly. 
 We comment on the other observables in flavor physics and the collider experiments.

We note that the RG correction from the LR breaking scale to the sfermion scale around 100 TeV
is approximately evaluated, as explained in the last of Sec. \ref{sec;4-fermi}.
The details are shown in Appendix \ref{appendix4}.
The correction, in fact, depends on the detailed setup, such as the mass spectrum.
In our study, we simply multiply a numerical factor as the correction including not only the RG
but also the threshold corrections. The RG correction from the sfermion scale
to the observed scale is evaluated at the one-loop level. The sfermion scale is expected to be about 100 TeV
to obtain the SM Higgs mass. Note that all gaugino masses are assumed to be 1 TeV, to introduce the dark matter candidate. 

The Yukawa couplings for the quark and lepton masses are run from $M_Z$ to 100 TeV,
and the Wilson coefficients of the four-quark interactions are evaluated at the scale, $\Lambda_{uu}$.
Then, the RG corrections from $\Lambda_{uu}$ to the low energy are taken into account at the one-loop level.
In the four-lepton interactions, the RG corrections are ignored.
In the four-fermi couplings concerned with the leptonic meson decays,
the RG effect could be interpreted as the same as the one for the quark mass.
We calculate the Wilson coefficients at 100 TeV, using the Yukawa couplings 
derived from the realistic quark mass matrices.
\footnote{A procedure for evaluating the Yukawa couplings for the quark and lepton masses at the 100 TeV is summarized in Appendix \ref{app_MZto100TeV}.} Then, the RG corrections are included in our analysis. Below, we explain our results in the each process.

\subsection{$\Delta F=2$ processes}
\label{DeltaF2}
First, we summarize our predictions for the $\Delta F=2$ processes.
The $\Delta F=2$ processes are consistent with the SM predictions, although the
predictions suffer from large uncertainties. In our model, the neutral Higgs exchanging modifies the SM prediction at the tree level:
\begin{eqnarray}
{\cal H}^{\Delta F=2}_{eff}=-( C^d_{4})_{ij} (\overline{d^i_L} d^j_R)  (\overline{d^i_R}  d^j_L) +h.c., \label{eq;deltaF2}
\end{eqnarray}
where the Wilson coefficient $( C^d_{4})_{ij}$ is given by 
\beq
( C^d_{4})_{ij} = ( C^d_{4})^{ij}_{ij}=  Y^{u \,*}_{ji}   Y^u_{ij} \,  \Lambda^{-2}_{uu},
\eeq
where $Y^u$ is given in Eq. (\ref{eq;Yukawa}). Assuming that the LR symmetry is assigned,
$  Y^{u \,*}_{ji}$ is identified with $ Y^u_{ij}$. 
In our analysis, we use the following values at 100 TeV:
\begin{align}
Y^u_{12} &= -(7.8 + 1.0 i) \times 10^{-4}, \\
Y^u_{13} &= (5.6 + 2.4 i) \times 10^{-3}, \\
Y^u_{23} &= -2.9 \times 10^{-2} + 5.4 i \times 10^{-4},
\end{align}
obtained from the RG flow explained in Appendix \ref{app_MZto100TeV}.


We investigate the bound from the $K$-$\overline{K}$ mixing.
In the $K$ physics,  $\epsilon_K$ and $\Delta M_K$ generally give stringent bounds.
They are approximately evaluated as \cite{Buras:2008nn}
\beq
\epsilon_K= \frac{\kappa_\epsilon e^{i \varphi_\epsilon} }{\sqrt{2} (\Delta M_K)_{\rm exp}} \, {\rm Im}(M^K_{12}), ~ \Delta M_K =2  {\rm Re}(M^K_{12}),
\eeq
where $\kappa_\epsilon$ and $\varphi_\epsilon$ are $\kappa_\epsilon=0.94 \pm 0.02$ and $\varphi_\epsilon=0.2417 \times \pi$ \cite{Buras:2010pza,Buras:2008nn}. $(\Delta M_K)_{\rm exp}$ is the experimental value given in Table \ref{table;input} and $M^K_{12}$ includes both the SM contribution and our prediction \cite{Iguro:2017ysu}:
\begin{eqnarray}
M^{K \, *}_{12}&=&\left ( M^K_{12} \right )^*_{\rm SM} - ( C^d_4)_{sd}  \times \frac{1}{4} \left ( \left ( \frac{m_K}{m_s + m_d} \right )^2+\frac{1}{6} \right )m_K F^2_K \Hat{B}_{K}. \nonumber \\
&&
\end{eqnarray}
The first term is the SM prediction described by 
$(M^K_{12})_{\rm SM}$ \cite{Inami:1980fz}, 
\beq
(M^K_{12})_{\rm SM}^*= \frac{G^2_F}{12 \pi^2} F^2_K \Hat{B}_K m_K M^2_W \left \{ \lambda^2_c \eta_1S_0( x_c) +   \lambda^2_t \eta_2 S_0(x_t) + 2   \lambda_c  \lambda_t \eta_3 S(x_c, x_t)  \right \},
\eeq
where $x_i$ and $ \lambda_i$ denote $(m^u_i)^2/M^2_W$ and $V^*_{is} V_{id}$, respectively. $\eta_{1,2,3}$ correspond to the NLO and NNLO QCD corrections. The input parameters for the quark mixing and masses are
summarized in Table \ref{table;input}. The input parameters for the $\Delta F=2$ processes are
shown in Table \ref{table;input1-2}. We use the central values to estimate the SM predictions.
Note that the functions which appear in $K$-$\overline{K}$ mixing are defined as \cite{Inami:1980fz}
\begin{eqnarray}
S_0(x)&=&  \frac{4x -11x^2+x^3}{4(1-x)^2} - \frac{3x^3 \log x}{2(1-x)^3}, \\
S(x,y)&=&\frac{-3xy}{4(y-1)(x-1)} - \frac{xy(4-8y+y^2) \log y}{4(y-1)^2(x-y)}  \nonumber \\
&&+\frac{xy(4-8x+x^2) \log x}{4(x-1)^2(x-y)}.
\end{eqnarray}

\begin{figure}[!t]
\begin{center}
{\epsfig{figure=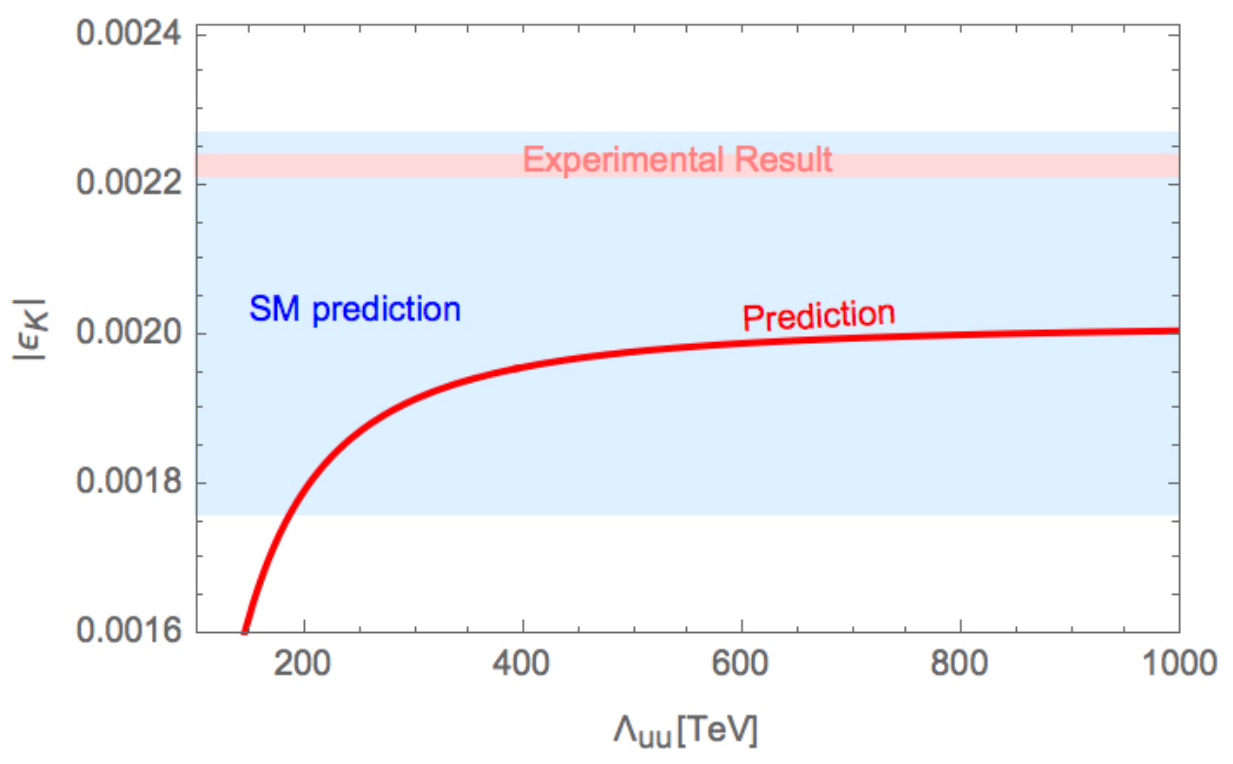,width=0.6\textwidth}}
\caption{$\Lambda_{uu}$ vs $|\epsilon_K|$ in the LR models. The red line corresponds to our prediction.
The light blue band is the SM prediction with 1$\sigma$ errors of $\eta_{1,2,3}.$
The pink band corresponds to the experimental result \cite{PDG} . }
\label{fig;epsilonK}
\end{center}
\end{figure}
In Fig. \ref{fig;epsilonK}, we draw our predictions for $|\epsilon_K|$ 
in the supersymmetric LR models. 
When we assume that the LR symmetry is assigned to our models,
the only parameter is $\Lambda^2_{uu}$. 
Fig. \ref{fig;epsilonK} shows the constraint on the scale, assuming $V_R=V$.
The red line corresponds to the prediction, and the blue band is the SM prediction with 1$\sigma$ errors of $\eta_{1,2,3}.$
The pink band corresponds to the experimental result on $|\epsilon_K|$, given in Table \ref{table;input1-2}.
If we require that our prediction is within the $1 \sigma$-region of the SM prediction,
the lower bound from $\epsilon_K$ is 
\beq
\label{epsilonKbound}
\Lambda_{uu} \gtrsim 180 \, {\rm TeV}.
\eeq

\begin{table}[t]
\begin{center}
  \begin{tabular}{|c|c||c|c|} \hline
 $m_d$(2 GeV) & 4.8$^{+0.5}_{-0.3}$ MeV \cite{PDG}  &  $\lambda$& 0.22509$^{+0.00029}_{-0.00028}$  \cite{CKMfitter}     \\ 
    $m_s$(2 GeV) & 95$\pm 5$ MeV \cite{PDG} & $A$& $0.8250^{+0.0071}_{-0.0111}$  \cite{CKMfitter}  \\ 
      $m_{b}(m_b)$&4.18$\pm 0.03$ GeV  \cite{PDG}    &  $\overline{\rho}$& 0.1598$^{+0.0076}_{-0.0072}$  \cite{CKMfitter} \\ 
  $\frac{2m_{s}}{(m_u + m_d)}$(2 GeV)& 27.5$\pm 1.0$ \cite{PDG}   &  $\overline{\eta}$&   0.3499$^{+0.0063}_{-0.0061}$ \cite{CKMfitter} \\ 
     $m_{c}(m_c)$&1.275$\pm 0.025$ GeV  \cite{PDG}  & $M_Z$ & 91.1876(21) GeV  \cite{PDG}  \\ 
    $m_t(m_t)$& 160$^{+5}_{-4}$ GeV  \cite{PDG}  &  $M_W$ & $80.385(15)$ GeV  \cite{PDG}   \\ 
 $\alpha$ & 1/137.036  \cite{PDG}  &  $G_F$  & 1.1663787(6)$\times 10^{-5}$ GeV$^{-2}$  \cite{PDG}       \\ 
 $\alpha_s(M_Z)$ & $0.1182(12)$ \cite{PDG}   &  & 
   \\ \hline
  \end{tabular}
 \caption{The input parameters in our analysis. The CKM matrix, $V$, is written in terms of $\lambda$, $A$, $\overline{\rho}$ and $\overline{\eta}$ \cite{PDG}.}
  \label{table;input}
  \end{center}
\end{table}

Similarly,  we can discuss the $B_q$-$\overline{B_q}$ mixing, where $q$ is $d$ or $s$. 
The contributions to these observables are also governed by $\Lambda^2_{uu}$.
In the same manner, we can evaluate the $B_d$-$\overline{B_d}$ and $B_s$-$\overline{B_s}$ mixing.
The mass differences of the $B$ mesons in our model are given by \cite{Iguro:2017ysu}
\beq
\Delta M_{B_q}= 2 \left |  M^{B_q}_{12} \right | =2 \left | (M^{B_q}_{12})^*_{\rm SM}  - ( C^d_4)_{bq}  \times \frac{1}{4}  \left (\left ( \frac{m_{B_q}}{m_b + m_q} \right )^2+\frac{1}{6} \right ) m_{B_q} F^2_{B_q} \Hat{B}_{B_q} \right | ~(q=d, \, s).
\eeq
$ (M^{B_q}_{12})_{\rm SM}$ is given by the top-loop contribution \cite{Inami:1980fz}:
\beq
 (M^{B_q}_{12})^*_{\rm SM}=\frac{G^2_F}{12 \pi^2} F^2_{B_q} \Hat{B}_{B_q} m_{B_q} M^2_W  ( V^*_{tb} V_{tq} )^2 \eta_BS_0(x_t) .
\eeq
The time-dependent CP violations, $S_{\psi K}$ and $S_{\psi \phi}$, are evaluated as follows including the new physics contributions:
\beq
S_{\psi K}=- \sin \varphi_{B_d}, ~ S_{\psi \phi}=\sin \varphi_{B_s},
\eeq
where $ \varphi_{B_q}$ is the phase of $M^{B_q}_{12}$: $M^{B_q}_{12}=|M^{B_q}_{12}|e^{i \varphi_{B_q}}$.
The input parameters are summarized in Table \ref{table;input1-2}, and the central values are used in our analyses.
Note that $S_{\psi K}$ and $S_{\psi \phi}$ are experimentally measured well: $S_{\psi K}=0.691 \pm 0.017$ and $S_{\psi \phi}= 0.015 \pm 0.035$ \cite{PDG}.

As far as $\Lambda_{uu}$ is larger than 180 TeV, the deviations of the observables concerned with the
$B_q$-$\overline{B_q}$ mixing are enough suppressed to evade the conflicts with the experimental results.
For instance, the deviations of the mass differences from the SM prediction are at most 0.2 \% and
the deviations of $S_{\psi K}$ and $S_{\psi \phi}$ are much smaller.
We conclude that the strongest bound comes from $\epsilon_K$, that is shown in Eq. (\ref{epsilonKbound}).

\begin{table}
\begin{center}
  \begin{tabular}{|c|c||c|c|} \hline
    $m_K$ & 497.611(13) MeV  \cite{PDG}  & $m_{B_s}$ & 5.3663(6) GeV \cite{PDG}  \\ 
    $\tau_{K_L}$&7.78$\times$10$^{13}$ MeV$^{-1}$ \cite{PDG}&$\tau_{B}$&2.31$\times$10$^{12}$ GeV$^{-1}$ \cite{PDG} \\
    $\tau_{K_S}$&1.36$\times$10$^{11}$ MeV$^{-1}$ \cite{PDG}&$\tau_{B_s}$&2.30$\times$10$^{12}$ GeV$^{-1}$ \cite{PDG} \\
     $F_K$ & 156.1(11) MeV \cite{Aoki:2013ldr} & $m_{B}$ & 5.2795(3) GeV \cite{PDG} \\ 
     $\Hat B_K$ & 0.750(15) \cite{Aoki:2013ldr}  &  $F_{B_s}$ & 228.6 $\pm$ 3.8 MeV \cite{Bazavov:2016nty} \\ 
  $(\Delta M_K)_{\rm exp}$  & 3.484(6)$\times 10^{-12}$ MeV  \cite{PDG} & $F_{B}$ &193.6 $\pm$ 4.2 MeV \cite{Bazavov:2016nty}  \\ 
    $|\epsilon_K|$ & $(2.228(11)) \times 10^{-3}$  \cite{PDG}  &  $\Hat B_{B_s}$ & 1.44(10) \cite{Bazavov:2016nty}\\ 
     $\eta_1$ & 1.87(76) \cite{Brod:2011ty} & $\Hat B_{B}$ &1.38(13)  \cite{Bazavov:2016nty} \\ 
        $\eta_2$& 0.5765(65) \cite{Buras:1990fn}  & $\eta_B$ & 0.55(1) \cite{Buras:1990fn}  \\ 
  $\eta_3$ & 0.496(47) \cite{Brod:2010mj}  &  &  \\

   \hline
  \end{tabular}
 \caption{The input parameters relevant to our analyses on flavor physics.}
  \label{table;input1-2}
  \end{center}
\end{table}

\subsection{Lepton flavor violation}
\label{LFV}
Next, we discuss the charged LFV processes in our model. 
$\Lambda_{uu}$ plays an important role in the LFV processes, as well.
Those processes are induced by the LFV four-lepton couplings:
\begin{eqnarray}
{\cal H}^{LFV}_{eff}=-( C^{e}_{4})_{kl}^{ij}(\overline{e^i_R} e^j_L)  (\overline{e^k_L}  e^l_R) . \label{eq;LFV}
\end{eqnarray}
These operators predict the LFV $\mu$ and $\tau$ decays; e.g., $\mu \to 3 e$ and $\tau \to e \mu \mu$.
The Wilson coefficient $( C^{e}_{4})_{kl}^{ij}$, which contributes to the LFV decays, depends on the Yukawa couplings as 
 \beq
( C^{e}_{4})_{kl}^{ij}=  Y^{\nu \,*}_{ji}  Y^\nu_{kl} \,  \Lambda^{-2}_{uu} + \delta_{ij} \, y^{e}_{i} \,  Y^\nu_{kl} \,  \Lambda^{-2}_{du} +  \delta_{kl} \, y^{e}_{k} \,  Y^{\nu \,*}_{ji} \,  \Lambda^{-2}_{ud}.
\eeq
Note that the charged leptons are mass eigenstates in this description.

In the limit that the light leptons are massless,
the branching ratio of the LFV process is estimated as\cite{Crivellin:2013wna}
\beq
Br(e_i \to  e^+_k e_j e_j)= \frac{m^5_{e_i} \tau_{e_i}}{6144 \pi^3}  \left ( |( C^{e}_{4})^{jk}_{ji} |^2 +|( C^{e}_{4})^{ji}_{jk} |^2 \right ).
\eeq
This description can be applied to $j=k$ case, such as  $\mu \to 3 e$ and  $\tau \to 3 e$.
In the $e_i \to  e^+_k e_k e_j$ ($j \neq k$) processes, the branching ratios are given by
\beq
Br(e_i \to  e^+_k e_k e_j)= \frac{m^5_{e_i} \tau_{e_i}}{6144 \pi^3}  \left ( |( C^{e}_{4})^{kk}_{ji} |^2 +|( C^{e}_{4})^{ji}_{kk} |^2+ |( C^{e}_{4})^{jk}_{ki} |^2 +|( C^{e}_{4})^{ki}_{jk} |^2 \right ).
\eeq
The current experimental bounds are summarized in Table \ref{table;LFVbound}.

$Y^\nu_{ij}$ in the coefficients originates the LFV decays.
As discussed in Sec. \ref{sec;SUSYLRmodel} and Appendix \ref{appendix3},
$Y^\nu_{ij}$ has large off-diagonal elements to reproduce the neutrino mixing.
In order to estimate our predictions explicitly, we focus on the three cases assuming that
the lightest neutrino mass and the Majorana phases are vanishing. 
In particular, $Y^\nu_{11}$ and $Y^\nu_{22}$, that are relevant to
the LFV processes involving light leptons, can be large in the cases (i) and (ii).

\begin{figure}[!t]
\begin{center}
{\epsfig{figure=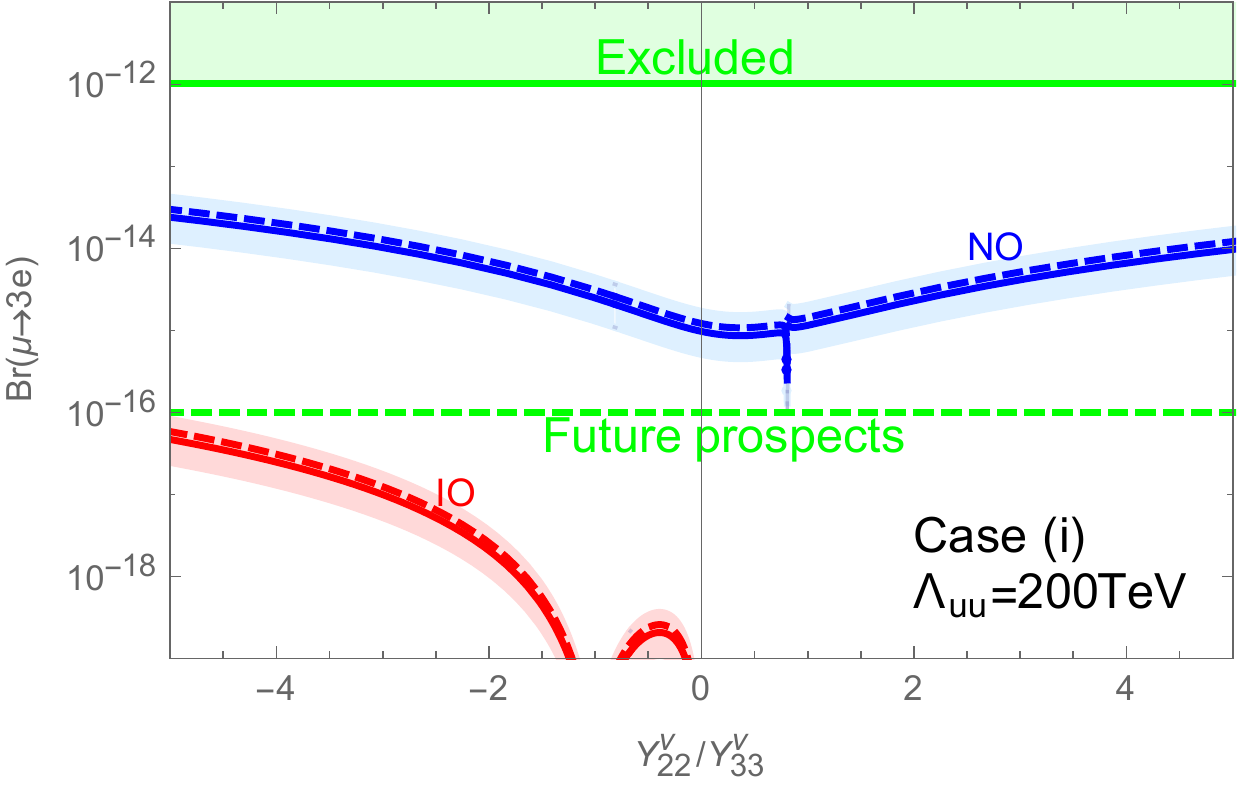,width=0.45\textwidth}}{\epsfig{figure=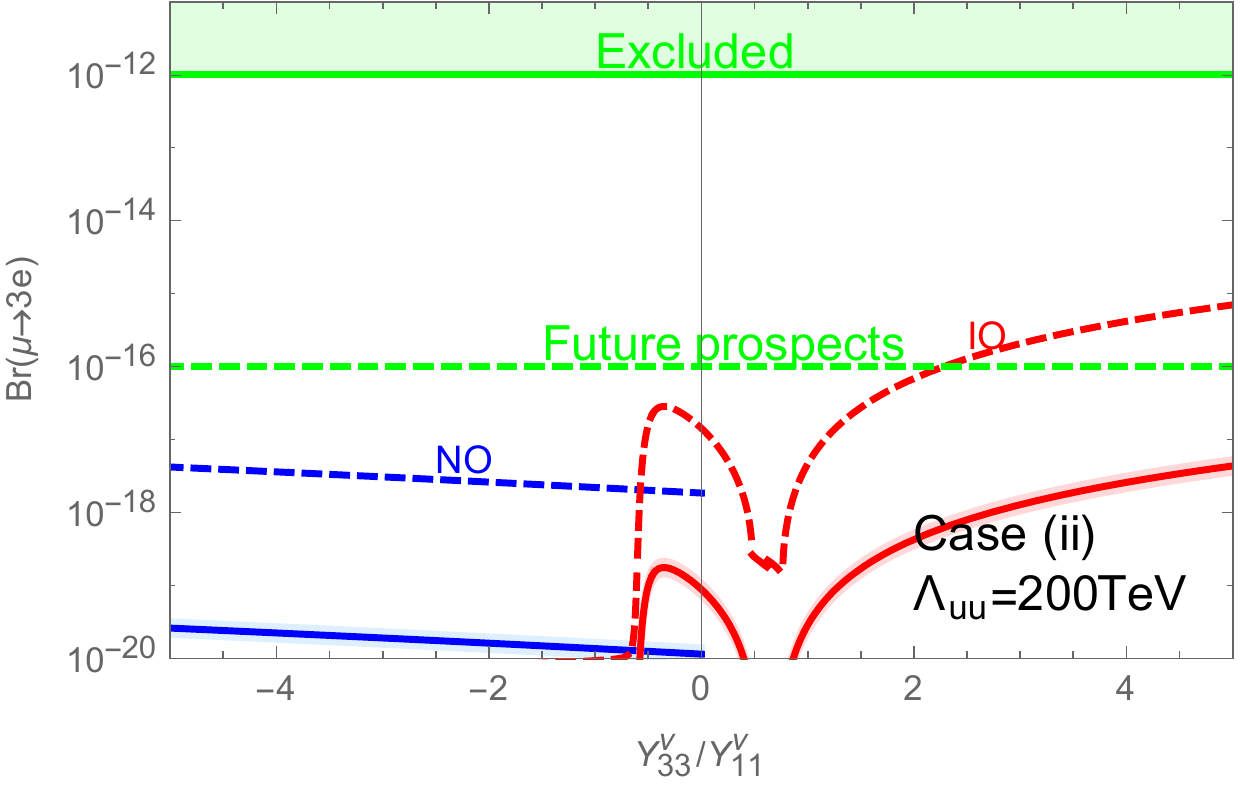,width=0.45\textwidth}}
\caption{Predictions of $\mu \to 3 e$. The thick and dashed blue lines depict our predictions in the NO cases with $(Y^\nu_{11},\, Y^\nu_{33}, \, M^{-1}_{\nu1})$$= (1, \, 0.1, \, 0)$ on the left panel and $(Y^\nu_{11},\, Y^\nu_{22}, \, M^{-1}_{\nu2})$$= (0.01, \, 1, \, 0)$ on the right panel. 
The thick and dashed red line describes our predictions in the IO cases with $(Y^\nu_{11},\, Y^\nu_{33}, \, M^{-1}_{\nu1})= (1, \, 0.02, \, 0)$ on the left panel and $(Y^\nu_{11},\, Y^\nu_{22}, \, M^{-1}_{\nu2})= (0.01, \, 1, \, 0)$ on the right panel, respectively. 
$\Lambda_{uu}$ and $\Lambda_{ud}$ are fixed at $(\Lambda_{uu}, \,\Lambda_{ud})=(200\, {\rm TeV}, \, 100 \,(1)\, {\rm TeV})$ on the thick (dashed) lines. 
The green region is excluded by the experiment \cite{Bellgardt:1987du} and the dashed green line is 
the future prospect \cite{Perrevoort:2016nuv}.  }
\label{fig;mu3e}
\end{center}
\end{figure}

To begin with, we estimate our predictions for $\mu \to 3e$ in the cases (i) and (ii).
As shown in Fig. \ref{fig;neutrinoYukawa}, $Y^\nu_{12}$ can be also relatively large
in those cases, depending on the size of the Majorana neutrino mass.
In order to avoid too large RG corrections for the Yukawa couplings,
we expect all Yukawa couplings to be less than unit. 
In Fig. \ref{fig;mu3e}, our predictions for $\mu \to 3e$ are depicted by 
the thick and dashed blue lines in the NO cases with $(Y^\nu_{11},\, Y^\nu_{33}, \, M^{-1}_{\nu1})= (1, \, 0.1, \, 0)$ on the left panel and 
$(Y^\nu_{11},\, Y^\nu_{22}, \, M^{-1}_{\nu2})= (0.01, \, 1, \, 0)$ on the right panel. 
NO (IO) is short for the normal (inverted) ordering neutrino mass hierarchy. 
Note that the left panel corresponds to the case (i)
and the right panel corresponds to the case (ii).
The thick and dashed red line describes our predictions in the IO cases with $(Y^\nu_{11},\, Y^\nu_{33}, \, M^{-1}_{\nu1})= (1, \, 0.02, \, 0)$ on the left panel and $(Y^\nu_{11},\, Y^\nu_{22}, \, M^{-1}_{\nu2})= (0.01, \, 1, \, 0)$ on the right panel, respectively. 
In those plots, $\Lambda_{uu}$ and $\Lambda_{ud}$ satisfies $\Lambda_{uu}=200$ TeV and $\Lambda_{ud}=100 \,(1)\, {\rm TeV}$ on the thick (dashed) lines. The light blue and red bands depict
the 20 \% corrections from the RG and the threshold.
We note that the green region is excluded by the SINDRUM experiment \cite{Bellgardt:1987du} and the dashed green line is the future prospect proposed by the Mu3e experiment \cite{Perrevoort:2016nuv}. 
As we discuss below, $\Lambda_{ud}$ is constrained strongly by $B_{s} \to \mu \mu$, so that
we can conclude that our model is not excluded as far as $\Lambda_{uu}$ is larger than 200 TeV.
Interestingly, the NO case with $|M_{\nu 2,3}| \ll |M_{\nu 1}|$ predicts the sizable branching ratio of $\mu \to 3e$,
as far as the heavy Majorana neutrinos reside above ${\cal O}(10^{13})$ GeV.
Our predictions can be covered by the future experiment \cite{Perrevoort:2016nuv}.
If the right-handed neutrinos are lighter, $Y^\nu_{33}$ is smaller and the off-diagonal element $Y^\nu_{12}$ is also suppressed.
In the case with vanishing $M^{-1}_{\nu 2}$ (Case (ii)), 
$Y^\nu_{22}$ can be large and the predictions are different from the case (i), reflecting the difference between the
neutrino mass spectrums. If the active neutrino is in the IO,
the future experiment may cover our region, depending on the heavy Majorana mass scale.

\begin{table}[!t]
  \begin{center}
    \begin{tabular}{|c|c|c|}
       \hline
   processes  & upper bounds on $Br$ & future prospects \\
          \hline \hline
       $\tau \to 3e$& $2.7\times10^{-8}$   \cite{PDG} & $5\times10^{-10}$ \cite{Belle2}   \\
       \hline
       $\tau^- \to e^-\mu^+\mu^-$&$2.7\times10^{-8}$   \cite{PDG} & $5\times10^{-10}$ \cite{Belle2}\\
       \hline
       $\tau \to e^+\mu^-\mu^-$&$1.7\times10^{-8}$   \cite{PDG} & $3 \times10^{-10}$ \cite{Belle2}\\
       \hline
       $\tau \to \mu^-e^+e^-$&$1.8\times10^{-8}$   \cite{PDG} & $3 \times10^{-10}$ \cite{Belle2}\\
       \hline
       $\tau^- \to \mu^+e^-e^-$&$1.5\times10^{-8}$  \cite{PDG}& $3 \times10^{-10}$\cite{Belle2} \\
       \hline
       $\tau \to 3\mu$&$2.1\times10^{-8}$   \cite{PDG}& $4 \times10^{-10}$ \cite{Belle2} \\
       \hline
       $\mu \to 3e$&$1.0\times10^{-12}$   \cite{Bellgardt:1987du} & $1 \times10^{-16}$ \cite{Perrevoort:2016nuv} \\
       \hline
      \end{tabular}
    \caption{Summary of the upper bounds on the branching ratios ($Br$) of the LFV decays at 90 $\%$ CL. The future prospects for the LFV processes are also shown on the last column. }
    \label{table;LFVbound}
  \end{center}
\end{table}

\begin{figure}[!t]
\begin{center}
{\epsfig{figure=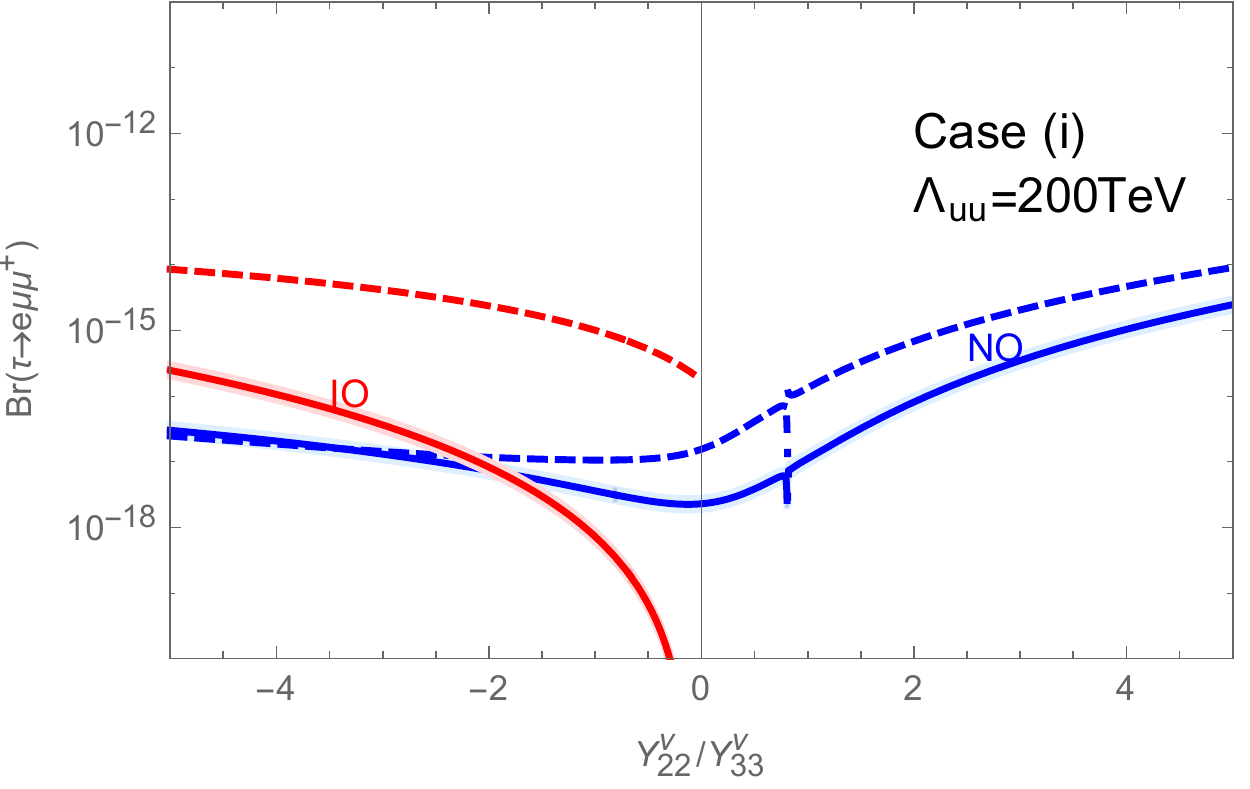,width=0.45\textwidth}}{\epsfig{figure=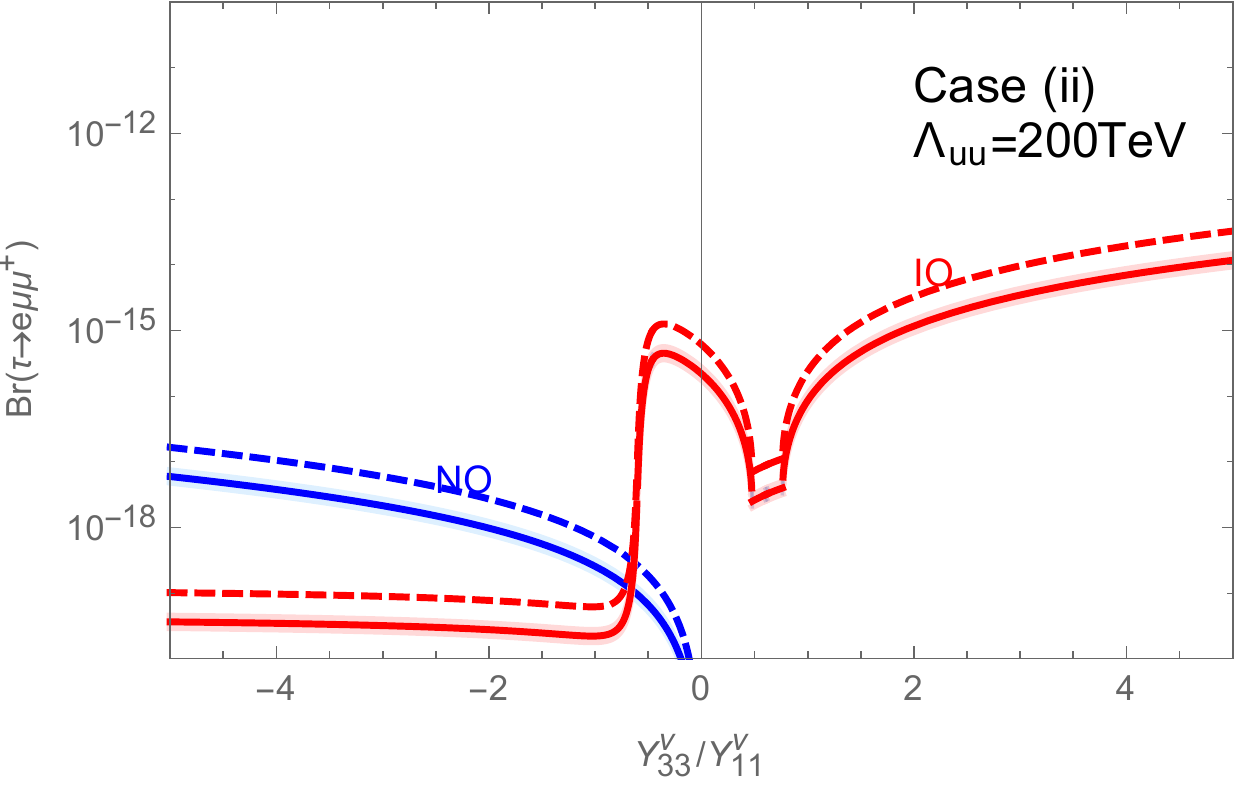,width=0.45\textwidth}}
\caption{ Predictions of $\tau \to e \mu^+ \mu$. The thick and dashed blue lines
depict our predictions in the NO cases with $(Y^\nu_{11},\, Y^\nu_{33}, \, M^{-1}_{\nu1})= (1, \, 0.1, \, 0)$ on the left panel and $(Y^\nu_{11},\, Y^\nu_{22}, \, M^{-1}_{\nu2})= (0.01, \, 1, \, 0)$ on the right panel. 
The thick and dashed red line describes our predictions in the IO cases with $(Y^\nu_{11},\, Y^\nu_{33}, \, M^{-1}_{\nu1})= (1, \, 0.02, \, 0)$ on the left panel and $(Y^\nu_{11},\, Y^\nu_{22}, \, M^{-1}_{\nu2})= (0.01, \, 1, \, 0)$ on the right panel, respectively. 
$\Lambda_{uu}$ and $\Lambda_{ud}$ are fixed at $(\Lambda_{uu}, \,\Lambda_{ud})=(200\, {\rm TeV}, \, 100 \,(6)\, {\rm TeV})$ on the thick (dashed) lines.  }
\label{fig;taue2mu}
\end{center}
\end{figure}

Similarly, the LFV $\tau$ decay, such as $\tau \to e \mu^+ \mu$, is also predicted in our model.
Although the prediction tends to be small compared to the future prospect
of the experiments, it would be worth estimating the size of our prediction.
Fig. \ref{fig;taue2mu} shows our prediction of $\tau \to e \mu^+ \mu$ in the case (i) (left) and (ii) (right).
The parameters are the same as in Fig. \ref{fig;mu3e}, on the each line.
As we see, our prediction is at most ${\cal O}(10^{-14})$, that is much below the current experimental bound
and the future prospect in Table \ref{table;LFVbound}. The other LFV $\tau$ decays
are also suppressed as in this process.




\subsection{Leptonic meson decays}
\label{LeptonicDecay}
In this section, we discuss the leptonic meson decays, based on the results on the $\Delta F=2$ processes and
the LFV processes.
In our model, the leptonic meson decays are given by the following operators,
\begin{eqnarray}
{\cal H}^{\Delta F=1}_{eff}=-( C^{de}_{4})^{kl}_{ij} (\overline{d^i_R} d^j_L)  (\overline{e^k_L}  e^l_R)- C^{ij}_{{\rm SM}}  (\overline{d^i_L} \gamma^\mu  d^j_L)  (\overline{e^k} \gamma_\mu \gamma_5 e^k)+h.c.. \label{eq;deltaF1}
\end{eqnarray}
The Wilson coefficients $( C^{de}_{4})^{kl}_{ij}$ with $i \neq j$ are obtained from the neutral scalars exchanging:
\beq
\label{C4de}
( C^{de}_{4})^{kl}_{ij} = \widetilde  Y^{u \,*}_{ji}  Y^\nu_{kl} \,  \Lambda^{-2}_{uu} + \widetilde  Y^{u \,*}_{ji}  y^e_{k} \,\delta_{kl} \,  \Lambda^{-2}_{ud}.
\eeq
$C^{ij}_{{\rm SM}}$ is the SM prediction. Using the operators and the coefficients, the leptonic $B_q$ ($q=s, \, d$) decays are described as follows: 
\begin{eqnarray}
Br(B_q \to e_k \overline{e_l})&=&\frac{\tau_{B_q}}{128 \pi} (m_{e_k}+m_{e_l})^2 m_{B_q} F^2_{B_q} \sqrt{ \left(1- \frac{( m_{e_k}+m_{e_l})^2}{m^2_{B_q}} \right )\left(1- \frac{ (m_{e_k}-m_{e_l})^2}{m^2_{B_q}} \right ) } \nonumber \\
&&\times \Biggl \{ \left | \frac{R_{B_q}   }{ m_{e_k}+m_{e_l}} \{   ( C^{de}_{4})^{kl}_{bq} +( C^{de}_{4})^{lk \, *}_{qb} \}  - \delta_{kl} \,C^{bq}_{{\rm SM}} \right |^2 \left (1- \frac{(m_{e_k}-m_{e_l})^2}{m^2_{B_q}} \right )  \nonumber \\
&& + \left | \frac{R_{B_q}   }{ m_{e_k}+m_{e_l}}  \{   ( C^{de}_{4})^{kl}_{bq} -( C^{de}_{4})^{lk \, *}_{qb} \} \right |^2  \left (1- \frac{(m_{e_k}+m_{e_l})^2}{m^2_{B_q}} \right )   \Biggr \},
\end{eqnarray}
where $R_{B_q}$ is defined as
\beq
\label{eq;RM}
R_{B_q} =\frac{m^2_{B_q}}{m_b +m_q}.
\eeq
 In the $B_q$ decays, the SM prediction is described as
\beq
C^{bq}_{{\rm SM}} =- \frac{G^2_F M^2_W}{\pi^2} \sum_{n =c,t}V^*_{nb} V_{nq} \, \eta_Y \, \frac{x_n}{4} \left \{ \frac{4-x_n}{1-x_n} +\frac{3x_n}{(1-x_n)^2} \ln x_n\right \},
\eeq
where $x_n$ is defined as $x_n=m^2_n/M^2_W$ and $\eta_Y$ corresponds to the NLO correction: $\eta_Y=1.0113$ \cite{Buras:2012ru}. Note that the lepton flavor violating decay is vanishing in the SM.

In our model, sizable $Y^\nu$ and relatively small $\Lambda_{ud}$ may largely deviate the SM predictions
in the leptonic $B$ decays. The SM predictions are consistent with the experimental results \cite{Buras:2013uqa,Beneke:2017vpq}.

\begin{figure}[!t]
\begin{center}
{\epsfig{figure=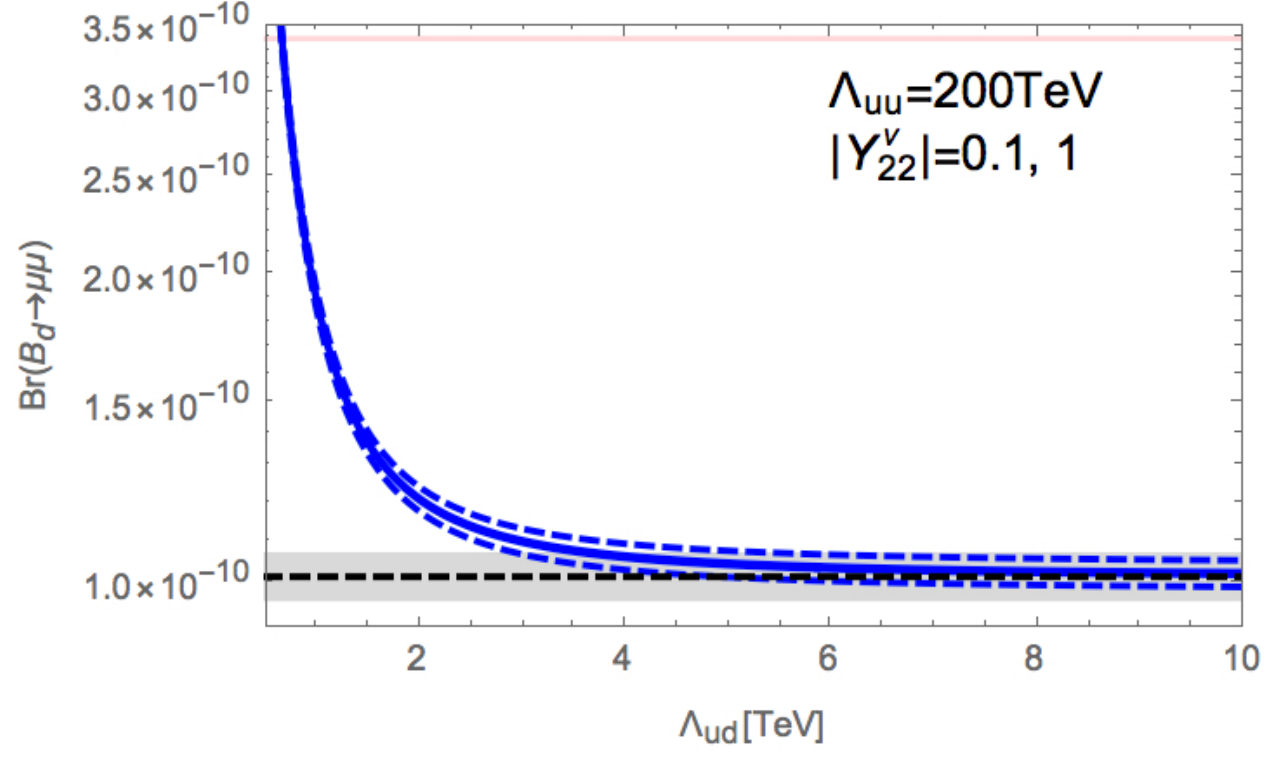,width=0.47\textwidth}} {\epsfig{figure=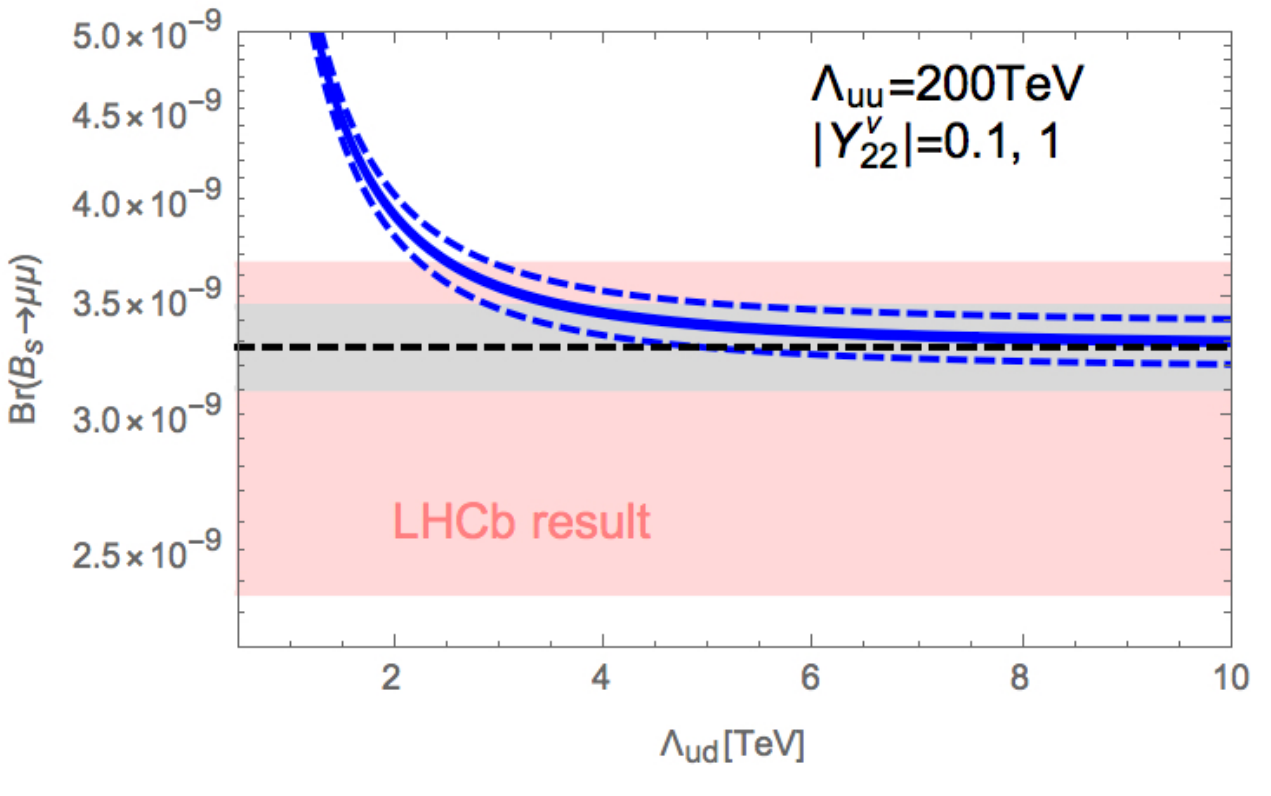,width=0.45\textwidth}} 
\caption{The deviations of $Br(B_d \to \mu \mu$) (left) and $Br(B_s \to \mu \mu$) (right) in our model. 
The thick and dashed blue lines
depict our predictions in the case with $|Y^\nu_{22}|=0.1, \, 1$. 
Note that the SM predictions of $Br(B_q \to \mu \mu)$ using our input parameters are depicted by the dashed black lines
with the gray band, taking into account the errors of the form factors. }
\label{fig;Btoll}
\end{center}
\end{figure}

Note that the latest experimental results given by the LHCb collaboration are
$Br(B_d \to \mu \mu)< 3.4  \times 10^{-10}$ and $Br(B_s \to \mu \mu)=(3.0 \pm 0.6^{+0.3}_{-0.2}) \times 10^{-9}$ 
\cite{Aaij:2017vad}. On the left and right panels in Fig. \ref{fig;Btoll}, the LHCb results are shown as 
the pink line and the pink band, respectively. 

Figs. \ref{fig;Btoll} show the deviations of $Br(B_q \to \mu \mu)$ from the SM predictions given by the input parameters
in Table \ref{table;input} and Table \ref{table;input1-2}. 
The relevant parameter in $Y^\nu$ is only $Y^\nu_{22}$, so that we can discuss the predictions using $Y^\nu_{22}$, $\Lambda_{uu}$
and $\Lambda_{ud}$. Then, Figs. \ref{fig;Btoll} are drawn fixing $\Lambda_{uu}$ at $200$ TeV.
On the two thick (dashed) lines, $Y^\nu_{22}$ satisfies $|Y^\nu_{22}|=  0.1$ $(1)$, respectively.
If we require the deviations are within the $1\sigma$ error of the experimental result, 
we obtain the lower bound on $\Lambda_{ud}$ as
\beq
\left | \Lambda_{ud}  \right |\gtrsim \, 2 \,{\rm TeV}.
\eeq
We stress on that this bound does not depend on the size of $Y^\nu_{22}$.
We note that the LFV decay of $B_q$ meson is also predicted but the prediction is below
the current experimental bound \cite{PDG}, as far as $ \Lambda_{uu} $ is larger than 200 TeV.

By analogy with the $B_q$ decay, we can obtain the leptonic $K$ decays, e.g., $K \to ll^\prime.$
In Ref. \cite{Isidori:2003ts}, the short-distance contribution of $Br(K_L \to \mu \mu)$ has been studied:
$Br(K_L \to \mu \mu) \leq 2.5 \times 10^{-9}$. 
Our prediction of the branching ratio is about $4.6 \times 10^{-10}$, when 
$\Lambda_{ud} =2 $ TeV and $\Lambda_{uu}=200 $ TeV. 
Note that our SM prediction is $3.7 \times 10^{-10}$. 
Thus, our model is safe for this leptonic $K$ decay, as far as the constraints from the leptonic $B_q$ decays
and $\epsilon_K$ are satisfied. The LFV decay of $K_L$ is also experimentally constrained as 
$Br(K_L \to e\mu)  \le4.7\times10^{-12}$ \cite{PDG}.
In our model, the predicted branching ratio is about $1.0 \times 10^{-13}$
when $\Lambda_{uu}=200 $ TeV and $Y^\nu_{12}=1$. Thus, 
we conclude that the leptonic $B$ decay gives the stronger bound on our model.

We can also discuss the constraint from $D \to ll^\prime$, but the current experimental bound is not too strong
to draw crucial bounds on $\Lambda^{(ue)}_{uu}$ and $\Lambda^{(ue)}_{du}$ scale parameters.
The stringent bounds on the scales in the up sector may come from the Drell-Yan process at the LHC \cite{Aaboud:2017buh}. In the case (i) with vanishing $M^{-1}_{\nu1}$, we obtain the lower limit as
\beq
\left| \Lambda^{(ue)}_{du} \right | \gtrsim \sqrt{|Y^\nu_{11}|} \times 115\, {\rm GeV},  ~\left| \Lambda^{(ue)}_{uu} \right | \gtrsim \sqrt{|Y^\nu_{11}|} \times 384\, {\rm GeV}.
\eeq
Note that we can also find the lower bound on $\Lambda_{du}$ from this process at the LHC \cite{Aaboud:2017buh}:
\beq
\left| \Lambda_{du} \right | \gtrsim \sqrt{|Y^\nu_{11}|} \times 533\, {\rm GeV}.
\eeq
Here, we estimate the cross sections, using CALCHEP \cite{Belyaev2012qa}, and
adopt the conservative bound: the lower limit on the contact interaction scale normalized by $\sqrt{4 \pi}$ is 40 TeV \cite{Aaboud:2017buh}. 


\subsection{$\mu-e$ conversion process in nuclei}
\label{mutoe conversion}
Among the operators in Eq. (\ref{eq;summary-deltaF1}), we find the lepton flavor violating coupling that
induces the $\mu-e$ conversion in nuclei:
\begin{eqnarray}
{\cal L}_{eff}^{\mu\rm{-}e}&=&\sum_{q=d,s} ( C^{de}_{4})^{e \mu}_{qq} (\overline{q_R} q_L)  (\overline{e_L}  \mu_R)+( C^{de}_{4})^{ \mu  e \, *}_{qq} (\overline{q_L} q_R)  (\overline{e_R}  \mu_L)  \nonumber \\
&&+( C^{ue}_{4})^{e \mu}_{uu} (\overline{u_L} u_R)  (\overline{e_L}  \mu_R)
+( C^{ue}_{4})^{\mu e \, *}_{uu} (\overline{u_R} u_L)  (\overline{e_R}  \mu_L).  
\end{eqnarray}
As discussed in Sec. \ref{sec;4-fermi}, the LFV processes induced by $(C^{ue}_{4})^{kl}_{ij}$ are governed by
$\Lambda^{ue}_{uu}$, $\Lambda^{ue}_{du}$ and $\Lambda^{ue}_{dd}$, while $(C^{de}_{4})^{kl}_{ij}$
depends on $\Lambda_{uu}$ and $\Lambda_{du}$.
We have found that $\Lambda_{uu}$ and $\Lambda_{du}$ are strongly constrained by 
the $\Delta F=2$ processes and the leptonic $B_s$ decay.
Then, we expect that the $\mu-e$ conversion process dominantly depends on $(C^{ue}_{4})^{kl}_{ij}$.

\begin{figure}[!t]
\begin{center}
{\epsfig{figure=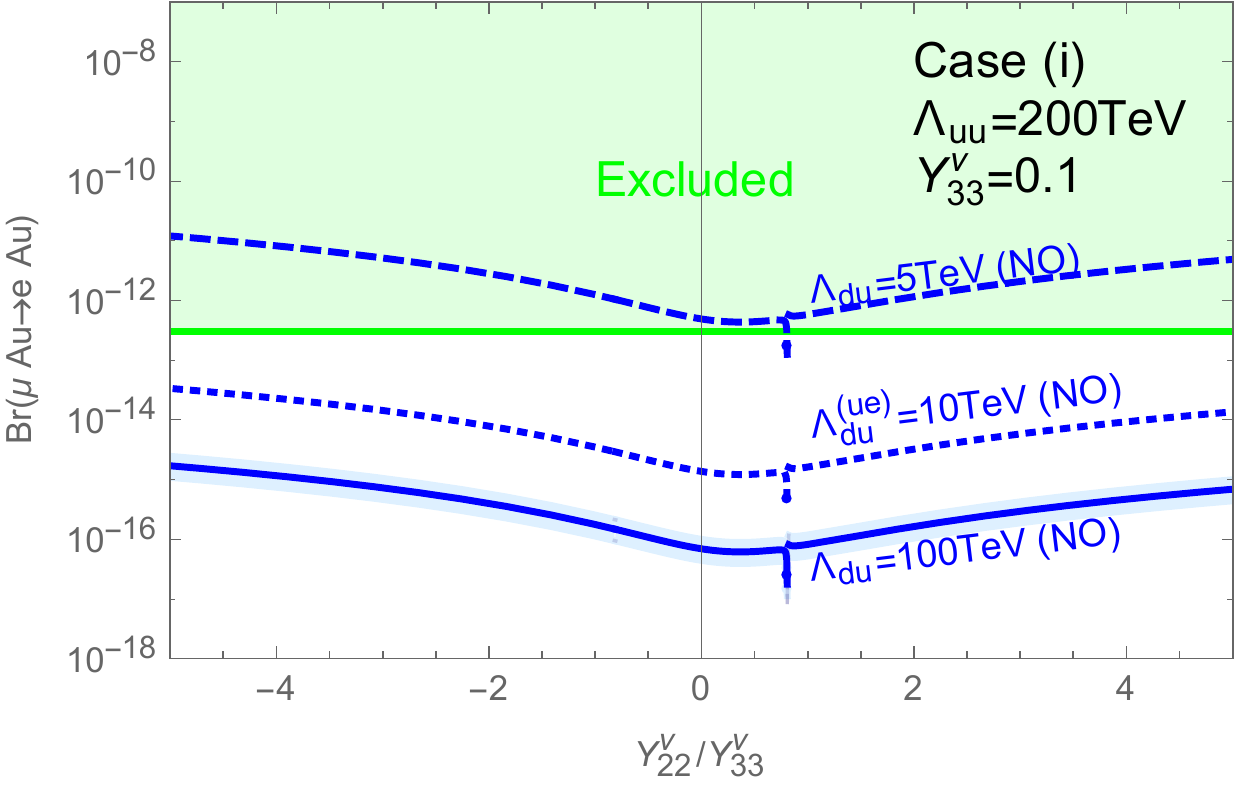,width=0.45\textwidth}} {\epsfig{figure=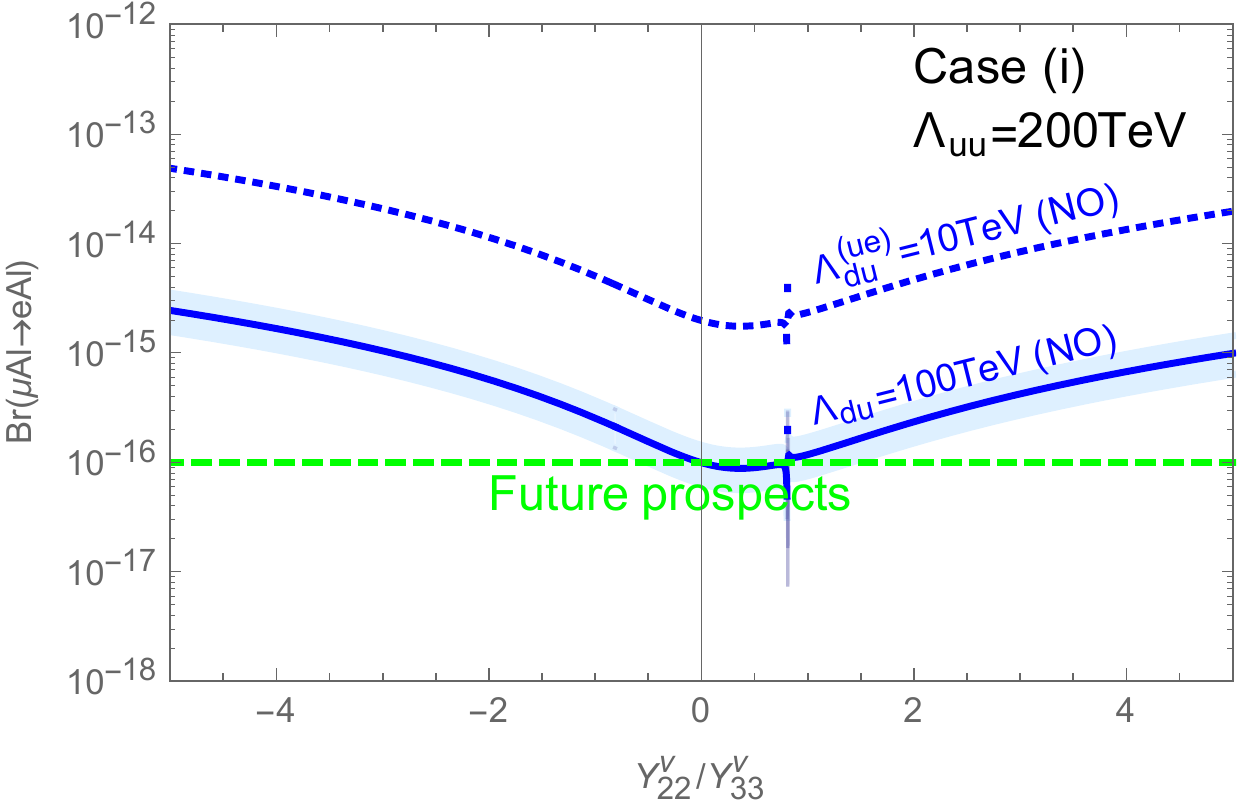,width=0.45\textwidth}} 
\caption{The predictions of $Br(\mu \, N\rightarrow e \, N)$ $(N=$Au (left), Al (right)).  
The thick and dashed blue lines
depicts our predictions in the NO case with $\Lambda_{du}=100$ TeV and $5$ TeV, respectively.
$Y^\nu_{33}$ is fixed at $Y^\nu_{33}=0.1$ and $M^{-1}_{\nu1}=0$ is satisfied. $\Lambda^{(ue)}_{uu}$ and $\Lambda^{(ue)}_{du}$
are fixed at $(\Lambda^{(ue)}_{uu}, \, \Lambda^{(ue)}_{du})=(100 \, {\rm TeV}, \, 100 \,(10) \, {\rm TeV})$ on the
(dotted) lines. The green region is excluded by the experiment \cite{Bertl:2006up} and the dashed green line corresponds to
the future prospect proposed in Ref. \cite{Kuno:2013mha}. 
 }
\label{fig;mueconversion}
\end{center}
\end{figure}

The branching ratio of the $\mu-e$ conversion can be calculated in our model, 
based on the results in Ref. \cite{Kitano:2002mt}:
\begin{align}
Br&(\mu N\to eN)=\frac{\omega_{conv}}{\omega_{capt}},\\
\omega_{conv}&=2G_F^2\biggl(|\tilde g_{L}^{(p)} S^{(p)}+ \tilde g_{L}^{(n)}S^{(n)})|^2
+|\tilde g_{R}^{(p)} S^{(p)}+ \tilde g_{R}^{(n)}S^{(n)})|^2 \biggl),\\
\tilde g_{L,R}^{(p)}&=\sum_q G_S^{(q,p)} g_{L,R(q)},~~\tilde g_{L,R}^{(n)}=\sum_q G_S^{(q,n)} g_{L,R(q)},\\
g_{L(u)}&=g_{R(u)}=\frac{\sqrt 2}{2G_F}(C_4^{ue})^{e\mu}_{uu},~g_{L(d,s)}=g_{R(d,s)}=\frac{\sqrt 2}{2G_F}(C_4^{(de,se)})^{e\mu}_{(dd,ss)},
\end{align}
here, the index $q$ runs for $u$, $d$, $s$ and $G_S^{(d,p)}=G_S^{(u,n)}=4.3$, $G_S^{(u,p)}=G_S^{(d,n)}=5.1$, $G_S^{(s,p)}=G_S^{(s,n)}=2.5$.
We can study the typical values of $Br(\mu  \, {\rm Au} \to e  \, {\rm Au}$) and $Br(\mu  \, {\rm Al} \to e  \, {\rm Al}$) in our model using input values as
 $\omega_{capt}=4.64\times 10^{-19}$, $S^{(p)}=0.0153\times m_{\mu}^{\frac{5}{2}}$, $S^{(n)}=0.0163\times m_{\mu}^{\frac{5}{2}}$  for Al,
 $\omega_{capt}=8.60\times 10^{-18}$, $S^{(p)}=0.0523\times m_{\mu}^{\frac{5}{2}}$, $S^{(n)}=0.0610\times m_{\mu}^{\frac{5}{2}}$ for Au.
Fig. \ref{fig;mueconversion} shows our predictions of the $\mu-e$ conversion processes.
The thick and dashed blue lines corresponds to our predictions in the NO case with $\Lambda_{du}=100$ TeV and $5$ TeV, respectively. $Y^\nu_{12}$ is evaluated assuming the mass hierarchy in the case (i) with $M^{-1}_{\nu1}=0$.
On these lines, $Y^\nu_{33}$ satisfies $Y^\nu_{33}=0.1$. $Y^\nu_{11}$ is a free parameter in the case (i), but it is not relevant to
this LFV process.
$\Lambda^{(ue)}_{uu}$ and $\Lambda^{(ue)}_{du}$ are fixed at $(\Lambda^{(ue)}_{uu}, \, \Lambda^{(ue)}_{du})=(100 \, {\rm TeV}, \, 100 \,(10) \, {\rm TeV})$ on the (dotted) lines.

The green region is excluded by the SINDRUM experiment: $Br(\mu  \, {\rm Au} \to e  \, {\rm Au}) < 3 \times 10^{-13}$ \cite{Bertl:2006up}. As we see the left panel in Fig. \ref{fig;mueconversion}, the dashed line with $\Lambda_{du}=5$ TeV
is already covered by the experiment. This could be more severe than the bound on $\Lambda_{ud}$ from $B_{s} \to \mu \mu$, depending on $Y^\nu_{ij}$. Following Eq. (\ref{eq;Lambdadu}), 
We note that $|\Lambda_{ud}|$ is expected to be the same order as $|\Lambda_{du}|$.

The future prospect for $Br(\mu \, {\rm Al} \to e \, {\rm Al}$) is ${\cal O}(10^{-16})$
at the COMET-II experiment \cite{Kuno:2013mha} and denoted by the dashed line on the right panel in Fig. \ref{fig;mueconversion}, so that  the parameter region with ${\cal O}(0.1)$ Yukawa couplings is expected to be covered by the future experiment.
The dotted lines shows the $\Lambda^{(ue)}_{uu}$ and $\Lambda^{(ue)}_{du}$ dependences.
When $Y^\nu_{33}$ is around ${\cal O}(0.1)$, we also obtain the lower bounds on the scales as
\beq
\label{mueconversion-bound}
\Lambda^{(ue)}_{uu} \, \gtrsim \, 1 \, {\rm TeV}, ~\Lambda^{(ue)}_{du} \, \gtrsim \,  2 \, {\rm TeV}.
\eeq

\subsection{Hadronic $\tau$ decay}
\label{tau decay}
We give a comment on the hadronic $\tau$ decay.
The four-fermi interactions involving $\tau$ lepton and light quarks induce
the hadronic $\tau$ decay; e.g., $\tau \to \mu \eta$ and $\tau \to \mu \pi$.
Following Refs. \cite{Sher:2002ew,Black:2002wh,Celis:2013xja}, we obtain the predictions of our model.
The branching ratio of the hadronic decay of $\tau$ is given by
\beq
Br(\tau \to l M_{qq^\prime}) = \frac{3}{32 \pi} \left | (C^{de}_4)^{l \tau}_{qq^\prime} \right |^2 m_\tau \tau_\tau R^2_{M_{qq^\prime}} F^2_{M_{qq^\prime}} \left ( 1- \frac{m^2_{M_{qq^\prime}}}{m^2_\tau} \right )^2,
\eeq
where $M_{qq^\prime}$ is the light meson: $M_{ss}=\eta$ and $M_{sd}=K$. 
 $R_{M_{dd}}$ is defined as in Eq. (\ref{eq;RM}), replacing $m_{B_q}$ with the meson mass.
 The light quark masses in $R_{M_{qq^\prime}}$ should be $m_d+m_s$ for $K$ and $m_u+m_d+4m_s$ for $\eta$, instead of $m_b+m_q$ in $R_{B_q}$. $Br(\tau \to l \pi)$ is also given in the same manner:  $R_{\pi}=m^2_\pi/(m_u+m_d)$. Note that  $|(C^{ue}_4)^{l \tau}_{uu}|^2$ should be added to $|(C^{de}_4)^{l \tau}_{dd}|^2$ in this case. 
In those processes, the large dimensional parameters, such as $\Lambda_{uu}$ and $\Lambda_{ud}$, sufficiently suppress the branching ratios.

\subsection{Leptonic meson decays in association with the active neutrinos}
\label{LeptonicDecay2}
So far, we concentrate on the flavor violation induced by the neutral scalar exchanging.
In our model, the leptonic meson decays associated with the active neutrinos are
also deviated by the charged Higgs exchanging at the tree level.
The charged currents, generated by the charged Higgs interactions, are written down in the following form;
\begin{eqnarray}
{\cal H}^{C}_{eff}=-( \widetilde C^{de}_{4})^{kl}_{ij} (\overline{d^i_R} u^j_L)  (\overline{\nu^k_L}  e^l_R)- ( \widetilde C^{ue}_4)^{kl}_{ij} \left (\overline{d^{ i}_L}  u^{j}_R \right ) \,  \left ( \overline{\nu^{k}_L} e^{ l}_R \right )+h.c.. \label{eq;deltaF1-C}
\end{eqnarray}
$(\widetilde C^{de}_{4})^{kl}_{ij}$ and $(\widetilde C^{ue}_{4})^{kl}_{ij}$ depend on $(C^{de}_{4})^{kl}_{ij}$ and $( C^{ue}_{4})^{kl}_{ij}$ as
\beq
( \widetilde C^{de}_{4})^{kl}_{ij}= ( C^{de}_{4})^{k^\prime l}_{i j^\prime} \,  V^*_{j j^\prime } \, (V_{PMNS})^*_{ k^\prime k},~( \widetilde C^{ue}_{4})^{kl}_{ij}=- ( C^{ue}_{4})^{k^\prime l}_{i^\prime j} \,  V^*_{i^\prime i} \, (V_{PMNS})^*_{ k^\prime k}.
\eeq
 In addition, there may be couplings involving right-handed neutrinos,
 when the Dirac neutrino scenario is considered. 
 The descriptions of the operators involving right-handed neutrinos are shown in Appendix \ref{appendix1}.

Finally, let us discuss the deviations of the leptonic meson decays in association with active neutrinos 
in the final state. In our scenario, the right-handed neutrinos are very heavy, and then
the leptonic charged $B_q$ decay can be described as 
\beq
\label{eq;Btolnu}
Br(B_q \to e_l \nu)=\sum_{k}Br_{{\rm SM}} (B_q) \left | (V_{PMNS})^*_{lk}+ \frac{R_{B_q}  }{m_{e_l}}  \frac{(\widetilde C^{de}_{4})^{kl}_{bq} -(\widetilde C^{ue}_{4})^{kl}_{bq} }{C^{bq}_{ \pm {\rm SM}}}  \right |^2
\eeq
where $Br_{{\rm SM}} (B_q)$ is the SM prediction of the branching ratio and $C^{bq}_{ \pm {\rm SM}}= 4 G_F V^*_{qb}/\sqrt{2}$ is defined. The interference between the SM prediction and the charged Higgs contribution 
may be large. 
The branching ratio of $B_q \to e_l \nu$ can be approximately evaluated as
\beq
\frac{Br(B_q \to e_l \nu)}{Br_{{\rm SM}} (B_q) }=\left| \delta_{kl}+ \frac{R_{B_q} y^b_d }{m_{e_l} (4 G_F/\sqrt{2})}  (\Lambda^{-2}_{du}+\Lambda^{(ue)-2 }_{du}  )Y^\nu_{kl} +( \Lambda^{-2}_{dd}+\Lambda^{(ue) -2}_{dd}  )  y^e_l\delta_{kl} \right|^2,
\eeq
where $ \widetilde Y^e=V^\dagger_{PMNS} y_{e} V_{PMNS} $ is defined.
The branching ratios of $K \to  e_l \nu$ can be also obtained replacing $R_{B_q} y^b_d$ with $R_{K} y^s_d$.
In these processes, we can obtain the bound on $\Lambda_{dd}$.
\begin{figure}[!t]
\begin{center}
{\epsfig{figure=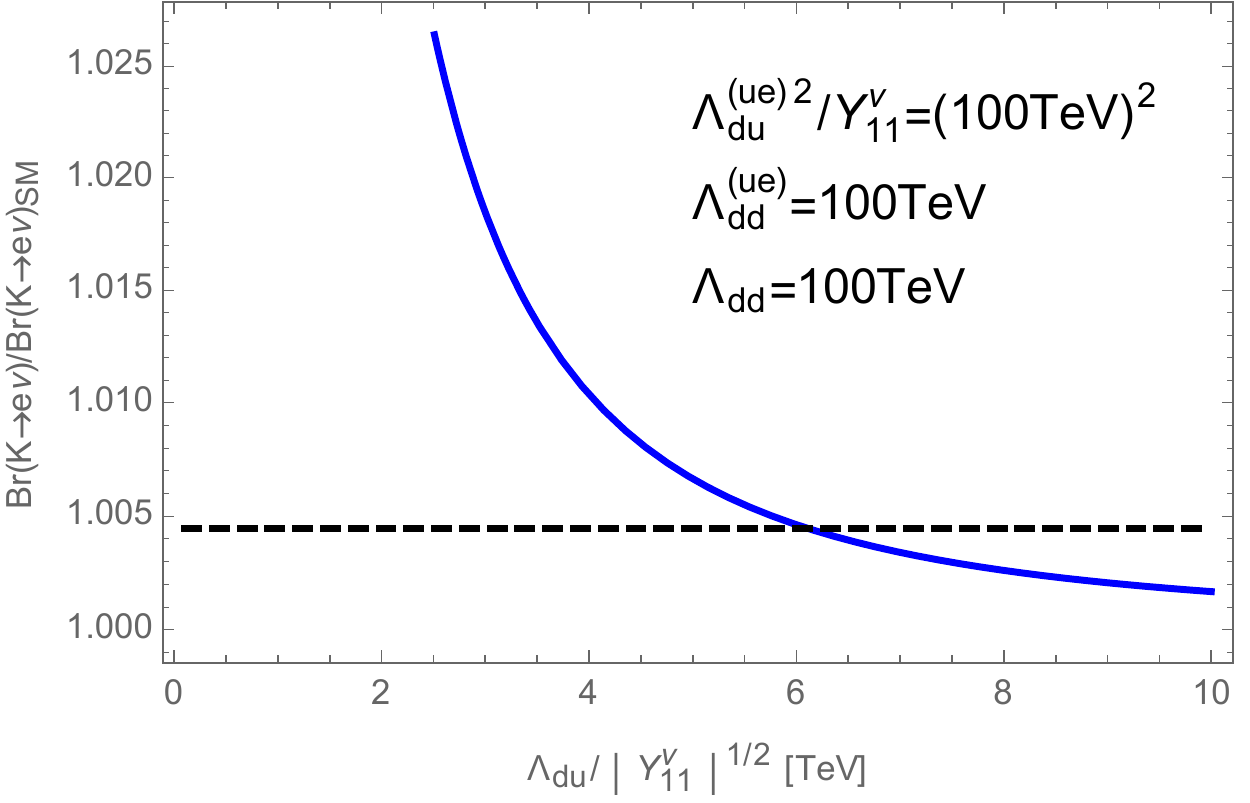,width=0.6\textwidth}} 
\caption{The predictions of $Br(K\rightarrow e \, \nu)$, depending on $\Lambda_{du}/\sqrt{|Y^\nu_{11}|}$.  
The thick blue line
depicts our predictions in the NO case with $\Lambda_{dd}=\Lambda^{(ue)}_{dd}=\Lambda^{(ue)}_{du}/\sqrt{|Y^\nu_{11}|}=100$ TeV.
 }
\label{fig;Ktoenu}
\end{center}
\end{figure}

Especially, $K \to e \nu$ gives the stringent bound on our model. 
The prediction for the process is depicted in Fig. \ref{fig;Ktoenu}. 
The vertical line is the ratio between our prediction and our SM prediction of $Br(K \to e \nu)$.
 The process, $K \to e\nu$, is experimentally measured well: $Br(K \to e \nu)=(1.582\pm0.007)\times10^{-5}$ \cite{PDG}.
 Then, the deviation from the new physics should be less than about 0.4 \%.
 The dashed black line depicts the upper bound from the experimental result.
 The thick blue line corresponds to our predictions in the NO case with $\Lambda_{dd}=\Lambda^{(ue)}_{dd}=\Lambda^{(ue)}_{du}/\sqrt{|Y^\nu_{11}|}=100$ TeV.
As we see, the lower bound on $|\Lambda_{du}|/\sqrt{|Y^\nu_{11}|}$ is about 6 TeV, that is the same as the one derived from the $\mu-e$ conversion.
$\Lambda_{dd}$ and $\Lambda^{ue}_{dd}$ are not so relevant to this process, because of the suppression from $y^e_e$.
Thus, we obtain the lower bound on $\Lambda_{du}$ and $\Lambda^{(ue)}_{du}$:
\beq
\sqrt{ \left |\Lambda_{du} \right |^2+\left | \Lambda^{(ue) }_{du} \right |^2}/\sqrt{|Y^\nu_{11}|} \gtrsim 6 \, {\rm TeV}.
\eeq

\subsection{Semileptonic meson decays}
\label{semileptonicDecay}
We can find the deviations from the SM predictions in the semi-leptonic decays, induced by the scalar exchanging.
The coefficient, $(C^{de}_4)^{kl}_{sb}$, in Eq. (\ref{eq;deltaF1}) contributes to the $b-s$ transition
in association with two  leptons in the final state. Recently, the LHCb collaboration has
reported the excesses in the observables of $B \to K^{(*)} ll$ ($l= e, \, \mu$) processes.
One interesting result is about the lepton flavor universal violation of $B \to K^{(*)} ll$.
The semi-leptonic decay processes, however, encounter the constraints from the leptonic decays discussed in Sec. \ref{LeptonicDecay}. In particular, the contribution in the leptonic decay is enhanced by the lepton mass,
while the semi-leptonic is not. Thus, we can not expect the large deviation from the SM prediction in our model,
as discussed in Ref. \cite{Hurth:2008jc}.

The semi-leptonic $K$ decay also constrains our model.
Following Ref. \cite{Mescia:2006jd}, we can estimate the contribution of $ (C^{de}_4)^{kl}_{sb}$ to $K \to \pi  \mu \mu$.
$\Lambda_{uu}$ and $\Lambda_{du}$ are, however, strongly constrained by the $\epsilon_K$ and the leptonic $B$ decay,
so that the new physics contribution to the branching ratio is at most ${\cal O}(10^{-15})$, even if $Y^\nu_{22}$ is ${\cal O}(1)$.

Next, we investigate $B \to D \, e_l \nu$ processes, where the excesses are reported in the observables concerned with the lepton universality. One interesting possibility to explain the excesses is charged Higgs particle with large flavor changing current with quarks and leptons.  
In fact, $( \widetilde C^{de}_4)^{kl}_{bc}$ and $( \widetilde C^{ue}_4)^{kl}_{bc}$ may be able to 
improve the discrepancy between the theoretical predictions and the experimental results,
since $\Lambda_{dd}$, $\Lambda^{(ue)}_{uu}$ and $\Lambda^{(ue)}_{du}$ in the coefficients can evade
the strong bounds from the flavor physics.
On the other hand, large new physics contribution is required to explain the discrepancy, compared to the SM prediction.
In our model, $( \widetilde C^{de}_4)^{kl}_{bc}$ and $( \widetilde C^{ue}_4)^{kl}_{bc}$ are suppressed by
$V^*_{cb}$ so that it seems that it is difficult to enhance the lepton universality of this decay.
In fact, we can estimate the deviation of the lepton universality in $B \to D \tau \nu$ and it is less than a few percent 
even if  $\Lambda_{dd}$, $\Lambda^{(ue)}_{uu}$ and $\Lambda^{(ue)}_{du}$ are about 500 GeV.

\section{Summary and Discussion}
\label{summary}

Imposing the LR symmetry is one of the attractive extensions. 
The symmetry can resolve the strong CP problem, so that the phenomenology has been widely studied so far. 
The new physics contributions are sufficiently large, if the LR breaking scale is around TeV scale.
Thus, the new particles predicted by the LR symmetry have been surveyed by the LHC experiments.
Based on the current experimental results, the LR symmetry breaking scale seems to be much higher than the EW scale. 
Then, we may conclude that the new physics scale is extremely high compared to the energy scale that 
the LHC can reach. 

In this paper, we assume that the LR symmetry breaking scale is much higher than the EW scale.
The LR symmetry breaking induces the Majorana mass terms for the right-handed neutrinos, and then 
the tiny neutrino masses could be generated by the seesaw mechanism. 
In such a scenario, the neutrino Yukawa couplings are expected to be large, so that
the sizable Yukawa interactions may be crucial to test our scenario. 
In our setup, the field to break the LR symmetry decouples to the Higgs fields, so
we find that the extra $SU(2)_L$ doublets that flavor-dependently couple to the SM fermions 
generally appear around the sfermion scale.
The LR symmetry predicts the explicit correlations between the observed Yukawa couplings 
and the Yukawa couplings involving the extra doublets. Then, we have proposed
that the sizable charged LFV is predicted by the neutrino Yukawa couplings, 
even if the right-handed neutrinos are integrated out at the very high-energy scale. 
The Higgs doublets are expected to reside around ${\cal O}(100)$ TeV to realize the 125 GeV Higgs mass, 
so that the induced flavor violation is expected to be enough large to test the extended SM.

The motivation of this paper is to demonstrate how large the contributions of the extra Higgs doublets to the flavor physics 
can be. We simply assume that the RG corrections and the threshold corrections are at most ${\cal O}(10)$ \%. 
Then, we derive the explicit relations between the observed fermion mass matrices and
the predicted FCNCs. There are actually several free dimensional parameters, but 
we can derive the explicit predictions for the physical observables of the flavor physics. 
$\epsilon_K$ gives the strongest bound to our model, and the lower limit of the scale where the extra Higgs doublets live
is fixed around ${\cal O}(100)$ TeV. The energy scale can be compatible to the sfermion scale in the high-scale SUSY scenario. The parameter relevant to $\epsilon_K$ dominantly contributes to the LFV process as well.
We find that the $\mu \to 3 e$ is crucial to our model and 
our model can be tested if the right-handed neutrinos are heavier than ${\cal O} (10^{13})$ GeV.
The interesting point of this model is that the contributions of the extra Higgs doublets to flavor violation processes
become negligible if the right-handed neutrino masses are light. 
This means that we can obtain the upper bound on the right-handed neutrino mass scale that corresponds to the LR breaking scale. 

We also investigated the other processes, such as the LFV $\tau$ decay and the meson decays in association with
leptons in the final state. One of the stringent constraints comes from $B_q \to ll$ and the $\mu-e$ conversion process in nuclei. The future experiment can cover our prediction, depending on the dimensional parameters in our model.
The search for the contact interaction at the LHC is also important, depending on the parameters.
The obtained results are summarized in Table \ref{table;SummaryTable}.
Note that we discussed the constraints from flavor physics by using the parameters defined in Sec. \ref{sec;4-fermi}, $\Lambda_{\alpha \beta}$ and $\Lambda_{\alpha \beta}^{(ue)}$, that depends on the masses squared for Higgs doublets in Eq. (\ref{HiggsMassMatrix}). 
Therefore, if we consider some explicit model following our setup, constraints on parameters of the model can be easily estimated from our results.

 Many studies have been done in the extended SM with the LR symmetry.
 In most of the cases, the LR symmetry breaking scale is low and the phenomenology has been discussed involving not only the extra scalars and the extra gauge bosons induced by the $SU(2)_R$ symmetry \cite{Huitu:1994zm,Dev:2016dja,Babu:2014vba,Frank:2011jia,Frank:2014kma,Ball:1999mb,Kiers:2002cz,Haba:2017jgf,Zhang:2007da,Guadagnoli:2010sd,Blanke:2011ry,Bertolini:2014sua,Bernard:2015boz,ValeSilva:2016czu,FileviezPerez:2017zwm}.
 In this paper, we propose another possibility that the LR breaking scale is much higher than the TeV-scale,
 motivated by the Type-I seesaw scenario. Moreover, we assume that the sfermion scale is also higher than the scale that the LHC can reach, based on the recent LHC results. In this case, it may be difficult to test our model directly.
 Then, we study the flavor physics predicted by the extra Higgs doublets especially and find some
 parameter region that the future flavor experiments can cover.
As mentioned above, our analysis has not yet explicitly included the LR symmetry breaking effects that are induced by 
the RG correction, the threshold correction and $\Delta_L$. 
We focus on the parameter region where the LR symmetry breaking effect 
is at most ${\cal O}(10)$ \%. If the effects are quantitatively taken into account, we
could survey the parameter region that predicts the larger flavor violating couplings. 
In addition, the flavor violating processes radiatively induced have not been studied in this work,
assuming the mass scale of the scalars are enough high. In order to take into account both
the LR symmetry breaking effects and the radiative corrections to flavor physics, we need to consider 
the mass spectrum and the mixing among the scalars as well. The detailed study will be given near future. 

Before closing our discussion, let us comment on the application of our analysis to the other underlying theories.
In Refs. \cite{Hisano:2015pma,Hisano:2016afc}, the authors
propose that the extra gauge boson can be a good probe to
test the $SO(10)$ GUT in the high-scale SUSY scenario.
We can also consider the case that the extra Higgs doublets are predicted by the $SO(10)$ GUT model.
 In the $SO(10)$ GUT, all matter fields unified into a $16$-representational field in each generation.
 Then, the realization of the realistic Yukawa couplings is, for instance, achieved by introducing several Higgs fields, 
 ${\bf 10}$, $\overline{{\bf 126}}$ and ${\bf 120}$ \cite{Dutta:2004zh,Dutta:2005ni,Senjanovic:2006nc,Dueck:2013gca}:
 \beq
 Y_{10}^{ij} {\bf 16 }_i  {\bf 16}_j {\bf 10} + Y_{126}^{ij} {\bf 16 }_i  {\bf 16}_j \overline{{\bf 126}}+ Y_{120}^{ij} {\bf 16 }_i  {\bf 16}_j {\bf 120} .
 \eeq
In this setup, the several Higgs doublets are predicted after the GUT symmetry breaking. 
we could expect that the Higgs doublets 
originated from ${\bf 10}$, $\overline{{\bf 126}}$ and ${\bf 120}$ have flavor violating Yukawa couplings, so that 
we could discuss flavor physics in the effective multi-Higgs doublet model in the same manner.
This possibility would be also deserved to be discussed more clearly.

\begin{table}[t]
\begin{center}
  \begin{tabular}{|c|c|c|} 
  \hline
 Observables  & bounds& Sec.\\ \hline \hline
 $\epsilon_K$ &$\Lambda_{uu} \gtrsim 180 \, {\rm TeV}$&3.1 \\\hline
 $Br(B_s \to \mu \mu)$&$\left | \Lambda_{ud}  \right |\gtrsim \, 2 \,{\rm TeV}$&3.3 \\\hline
$\sigma(pp \to e^+e^-)|_{\sqrt{s}=13 {\rm TeV}}$ &$\left| \Lambda^{(ue)}_{du} \right | \gtrsim \sqrt{|Y^\nu_{11}|} \times 115\, {\rm GeV},  ~\left| \Lambda^{(ue)}_{uu} \right | \gtrsim \sqrt{|Y^\nu_{11}|} \times 384\, {\rm GeV}$,&3.3  \\ 
 &$\left| \Lambda_{du} \right| \gtrsim \sqrt{|Y^\nu_{11}|} \times 533\, {\rm GeV}$&\\\hline
 $\mu-e$ conversion&$\Lambda^{(ue)}_{uu} \, \gtrsim \, 1 \, {\rm TeV}, ~\Lambda^{(ue)}_{du} \, \gtrsim \,  2 \, {\rm TeV}$&3.4 \\ \hline
  $Br(K\to e\nu)$&$\sqrt{ \left |\Lambda_{du} \right |^2+\left | \Lambda^{(ue) }_{du} \right |^2} \gtrsim \sqrt{|Y^\nu_{11}|} \times 6 \, {\rm TeV}$&3.6\\\hline
   \end{tabular}
 \caption{We summarize bounds from processes considered in this paper.}
  \label{table;SummaryTable}
  \end{center}
\end{table}

\section*{Acknowledgments}
The work of Y. M. is supported by the International Postdoctoral Exchange Fellowship Program (IPEFP) and National Natural Science Foundation of China (NNSFC) under contract Nos. 11435003, 11775092 and 11521064.
The work of Y. O. is supported by Grant-in-Aid for Scientific research from the Ministry of Education, Science, Sports, and Culture (MEXT), Japan, No. 17H05404.
The work of Y. S. is supported in part by the Japan Society for the Promotion of Science (JSPS) Research Fellowships for Young Scientists, No. 16J08299.

\appendix

\section{The alignment of the Dirac neutrino Yukawa couplings }
\label{appendix3}
We can quantitatively estimate the Dirac neutrino Yukawa couplings, assuming that the Majorana mass terms
are also hierarchical. We discuss the Yukawa couplings in the base where the Majorana mass terms 
are diagonal.
Then, we consider the three cases: 
\begin{enumerate}
\renewcommand{\labelenumi}{(\roman{enumi})}
\item $|M_{\nu1}| \gg  |M_{\nu2}|$,  $|M_{\nu3}|$, 
\item $|M_{\nu2}| \gg  |M_{\nu1}|$, $|M_{\nu3}|$,
\item $|M_{\nu3}| \gg  |M_{\nu1}|$, $|M_{\nu2}|$.
\end{enumerate}
In the each case, we can find that some elements of the Dirac neutrino Yukawa couplings are 
irrelevant to the observables concerned with the active neutrinos; e.g.,
\begin{enumerate}
\renewcommand{\labelenumi}{(\roman{enumi})}
\item $\Hat m_{\nu}$ does not depend on $Y^\nu_{11}$ in the case (i), 
\item $\Hat m_{\nu}$ does not depend on $Y^\nu_{22}$ in the case (ii),
\item $\Hat m_{\nu}$ does not depend on $Y^\nu_{33}$ in the case (iii).
\end{enumerate}  

\begin{figure}[H]
\begin{center}
{\epsfig{figure=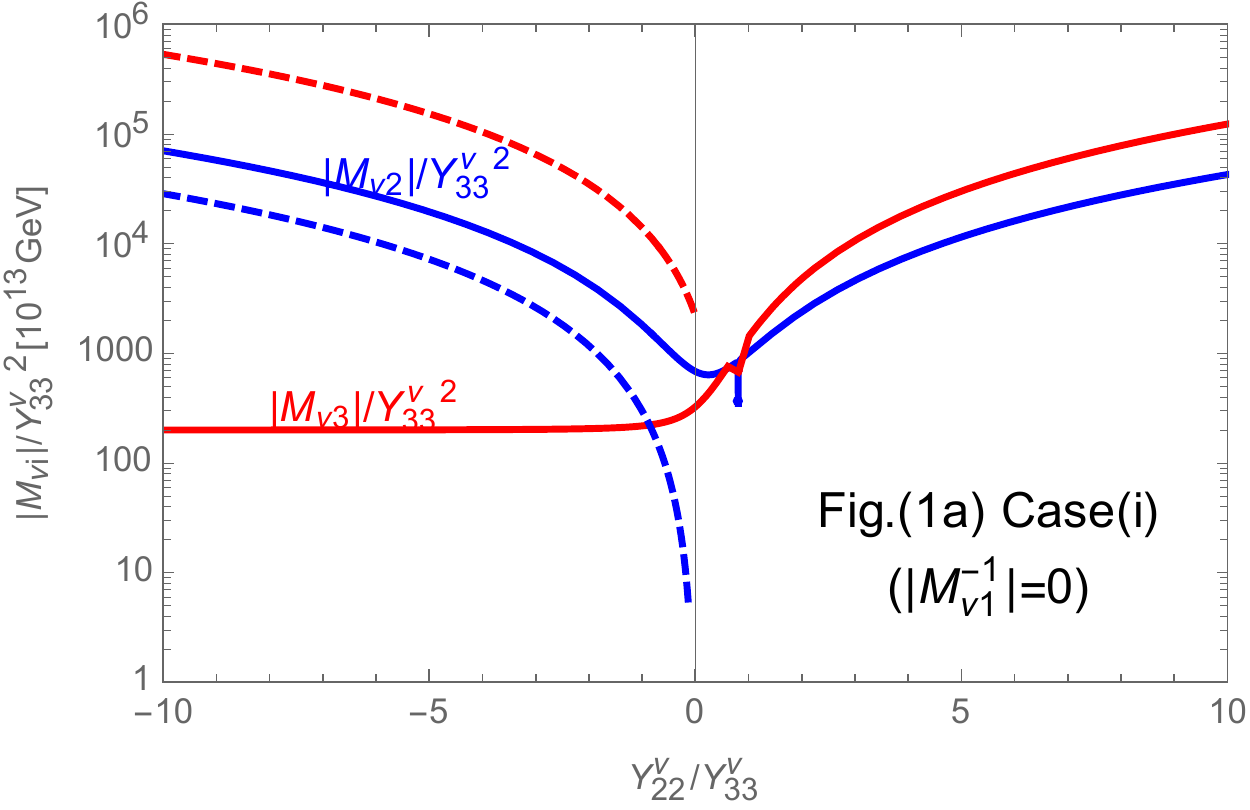,width=0.4\textwidth}}{\epsfig{figure=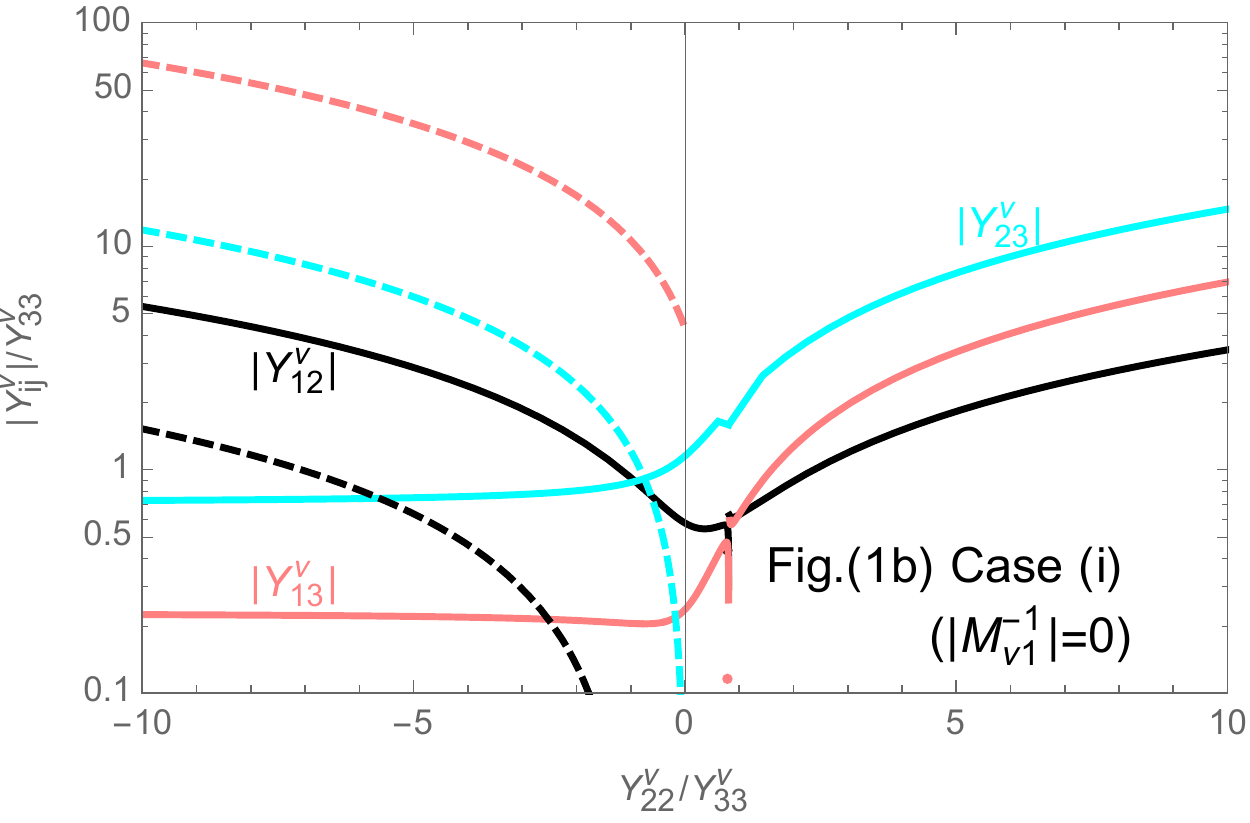,width=0.4\textwidth}}
{\epsfig{figure=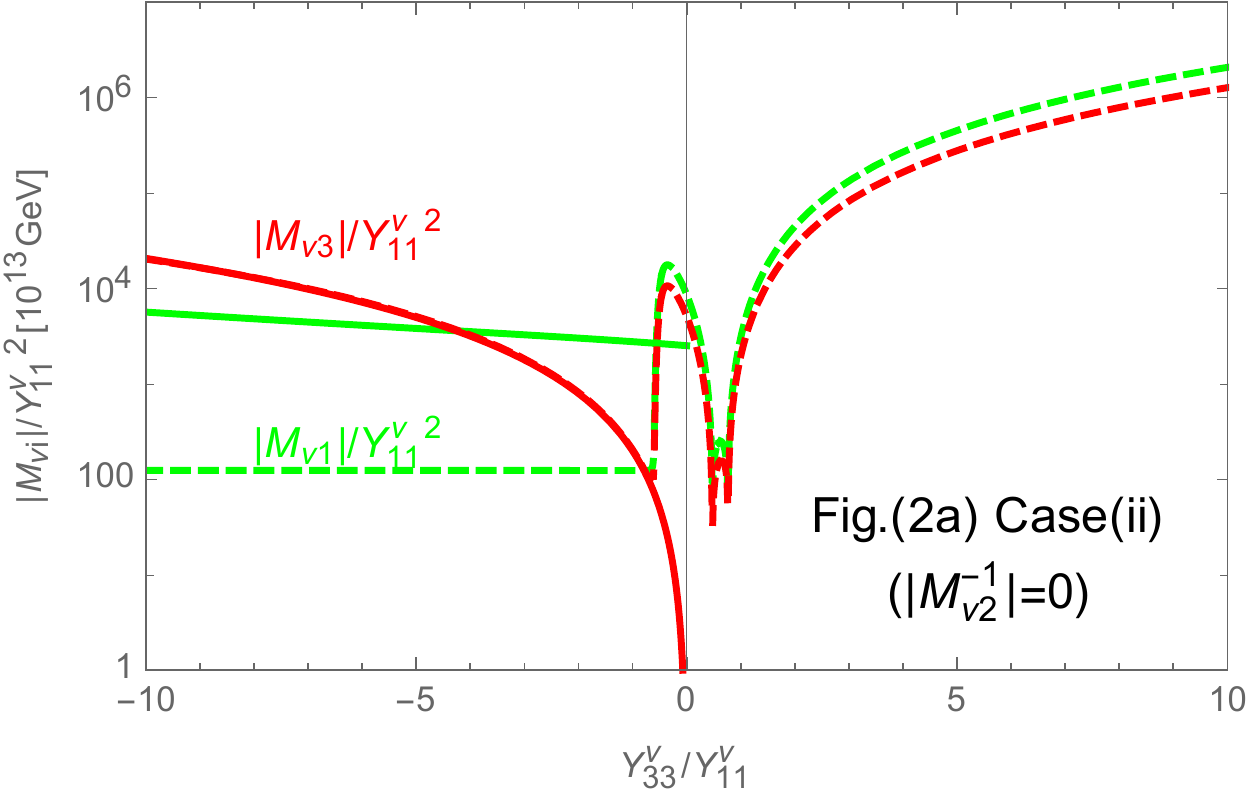,width=0.4\textwidth}}{\epsfig{figure=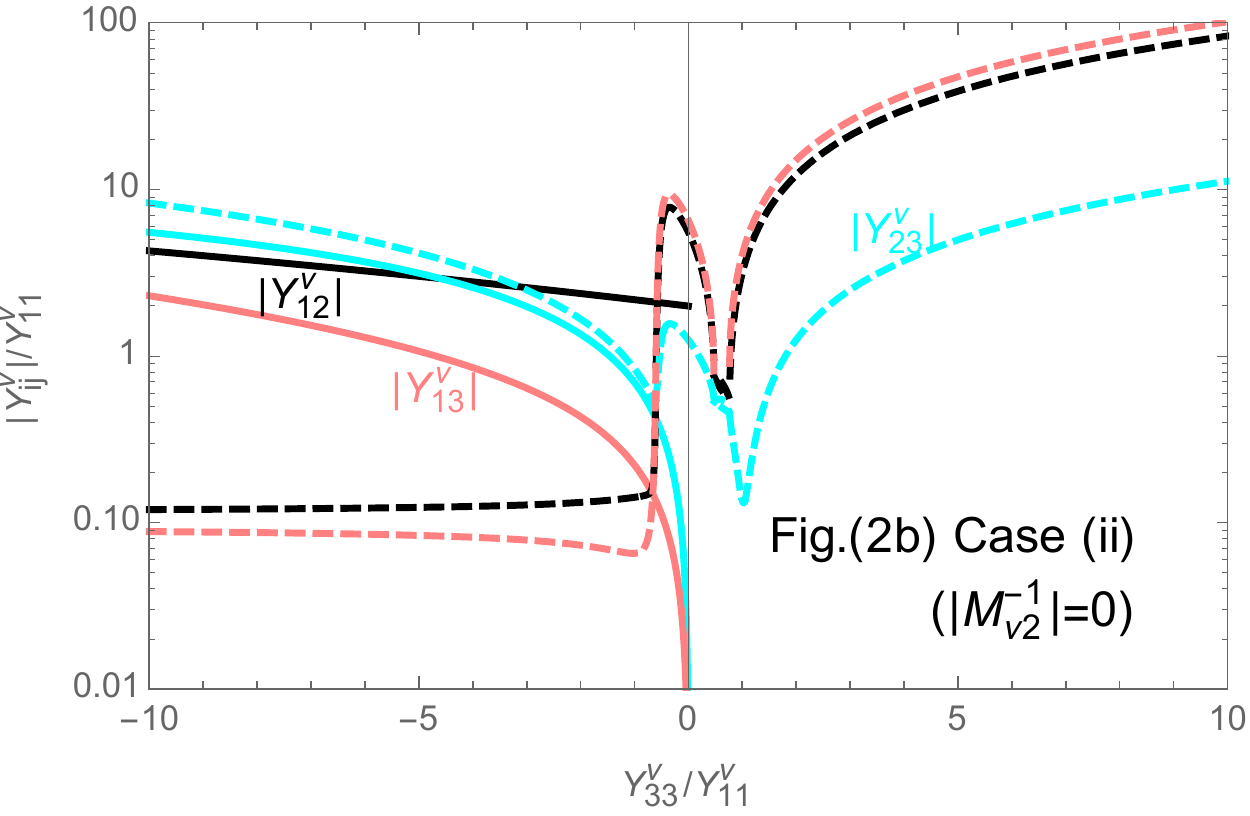,width=0.4\textwidth}}
{\epsfig{figure=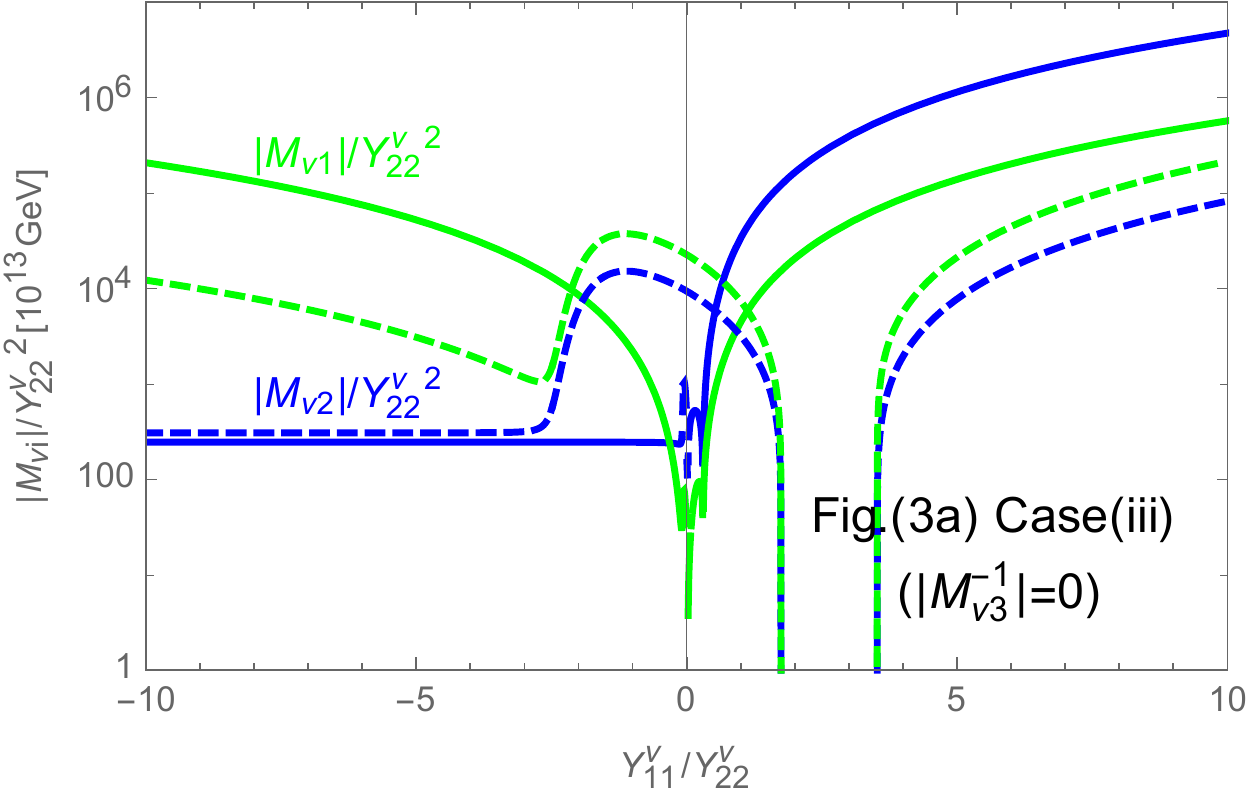,width=0.4\textwidth}}{\epsfig{figure=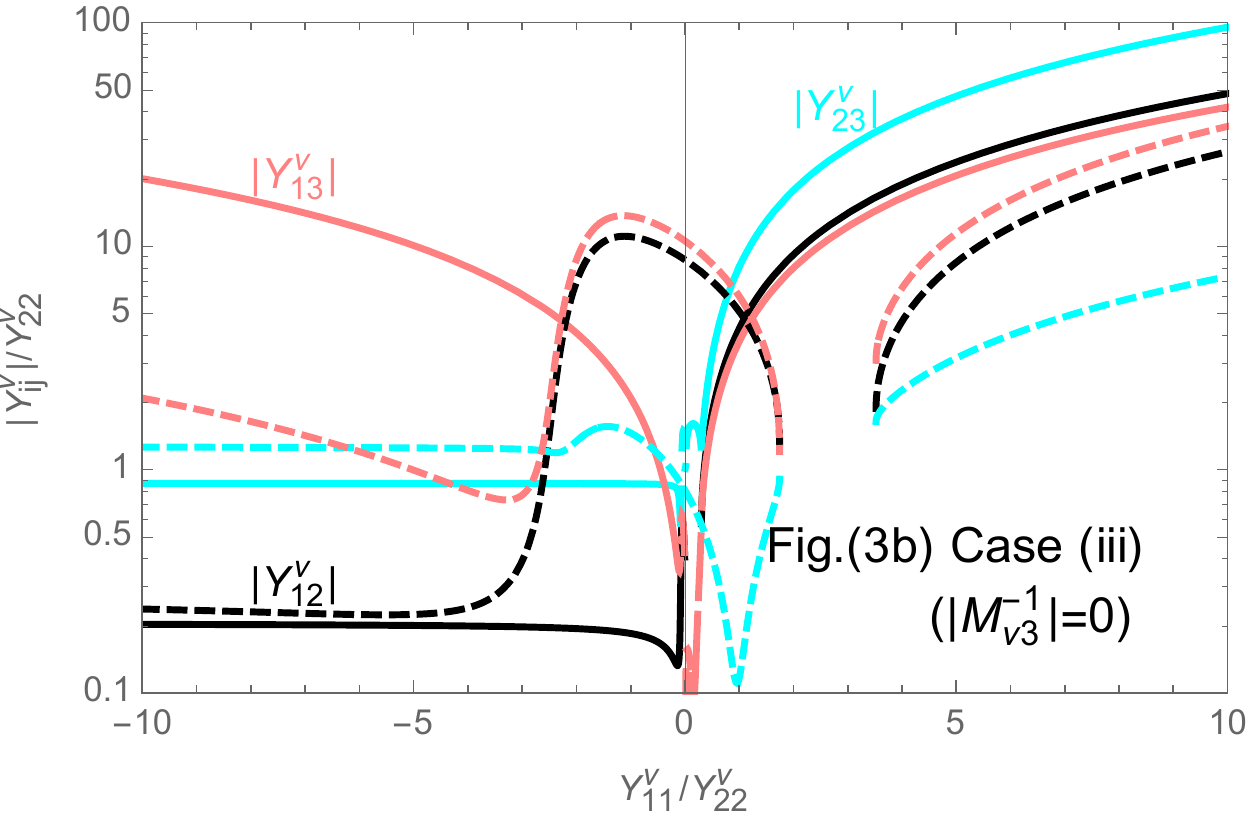,width=0.4\textwidth}}
\caption{ The Majorana mass terms and the Dirac neutrino Yukawa couplings, when
 $Y^\nu_{ij}$ is the hermitian matrix and the Majorana phases in $\Hat m_{\nu}$ are vanishing.
In Fig. (1a), Fig. (2a) and Fig. (3a), the red, blue and green (dashed) line correspond
to the $M_{\nu_3}$, $M_{\nu2}$ and $M_{\nu1}$ in the normal (inverted) ordering. In Fig. (1b), Fig. (2b) and Fig. (3b), the sizes of $Y^\nu_{12}$ (black), $Y^\nu_{13}$ (pink), $Y^\nu_{23}$ (cyan) are shown, in the cases (i), (ii) and (iii), respectively. }
\label{fig;neutrinoYukawa}
\end{center}
\end{figure}

In the limit that the contribution of the heaviest mode is vanishing in $\Hat m_{\nu}$,
we can explicitly estimate the alignment of the Dirac neutrino Yukawa couplings in the each case,
as shown in Fig. \ref{fig;neutrinoYukawa}.
The lightest active neutrinos are assumed to be massless in the each figure.
The input parameters used to plot are summarized in Table \ref{table;input2}.

In Fig. (1a), Fig. (2a) and Fig. (3a), the sizes of the relatively light Majorana mass terms are drawn
in the case (i), the case (ii) and the case (iii), respectively. The red (dashed) line corresponds
to $M_{\nu3}$ normalized by $(Y^\nu_{33})^2$ (Fig. (1a)) or $(Y^\nu_{11})^2$ (Fig. (2a)) in the normal (inverted) ordering case of
the active neutrinos. The blue (dashed) lines correspond
to $M_{\nu2}$ normalized by $(Y^\nu_{33})^2$ (Fig. (1a)) and $(Y^\nu_{22})^2$ (Fig. (3a)) in the normal (inverted) ordering, respectively.
The green (dashed) lines show $M_{\nu1}$ normalized by  $(Y^\nu_{11})^2$ (Fig. (2a)) and $(Y^\nu_{22})^2$ (Fig. (3a)) in the normal (inverted) ordering, respectively.

In Fig. (1b), Fig. (2b) and Fig. (3b), the sizes of the off-diagonal elements of $Y^\nu_{ij}$ normalized by $Y^\nu_{33}$, $Y^\nu_{11}$ and $Y^\nu_{22}$ are shown, assuming that $Y^\nu_{ij}$ is the hermitian matrix. 
As we see, large off-diagonal elements of $Y^\nu$ are predicted, depending on the mass hierarchy of $M_\nu$.
Note that the Majorana phases in $\Hat m_{\nu}$ are vanishing in these plots.
In Fig. (1b), the black, pink and cyan (dashed) lines show the magnitudes of $Y^\nu_{12}/Y^\nu_{33}$, $Y^\nu_{13}/Y^\nu_{33}$ and $Y^\nu_{23}/Y^\nu_{33}$, respectively, when the mass hierarchy of the active neutrinos is normal (inverse). In Fig. (2b) and Fig. (3b), the Yukawa couplings are normalized by $Y^\nu_{11}$ and $Y^\nu_{22}$, instead of
$Y^\nu_{33}$.

\section{$SU(2) _R \times U(1)_{B-L}$ Model}
\label{appendix2}
  The superpotential for the MSSM fields is given by
\beq
W_{vis}= Y^a_{ij}  Q^i_L \tau_2 \Phi_a \tau_2 Q^j_R +Y^a_{ij}  L^i \tau_2 \Phi_a \tau_2 L^j_R + \lambda^{\nu}_{ij}  L_R^i \tau_2 \Delta_R  L^j_R  + \mu^{ab} Tr \left( \tau_2 \Phi^T_a  \tau_2 \Phi_b\right).
\eeq  
The third term generates the Majorana mass term for the right-handed neutrino, and
the last term corresponds to the $\mu$-term of the Higgs superfields.

In addition, we introduce the following term to break $SU(2)_R \times U(1)_{B-L}$:
\beq
W_{SB}= m(S) \, Tr\left( \Delta_R \overline \Delta_R \right ) + w(S).
\eeq  
Note that $S$ is gauge singlet.
There are also $SU(2)_L$ triplet fields, $\Delta_L$ and $\overline{\Delta_L}$, to respect the LR symmetry.
Although the physics involving the $SU(2)_L$ triplets is not discussed in our paper,
the couplings are given by Eq. (\ref{eq:WDeltaL}).

Then, the F-terms and D-terms relevant to the $SU(2)_R$ triplets and $S$ are as follows:  
\begin{eqnarray}
 \pa_{\Delta_R} W&=& m(S)  \,  \overline \Delta_R ,  \\
 \pa_{ \overline \Delta_R} W&=& m(S)  \,  \Delta_R,  \\
  \pa_{ S} W&=&   Tr\left( \Delta_R \overline \Delta_R \right )  \pa_{ S} m(S) + \pa_S w(S),  \\
 D^A_{SU(2)_R}&=&2 Tr \left( \Delta^\dagger_R \tau^A \Delta_R \right ) +2 Tr \left( \overline \Delta^\dagger_R \tau^A \overline \Delta_R \right ), \\
 D_{B-L}&=& \xi - 2 \,  Tr \left( \Delta^\dagger_R  \Delta_R \right ) + 2 \, Tr \left( \overline \Delta^\dagger_R  \overline \Delta_R \right ). 
 \end{eqnarray} 
 Here, we simply assume that the MSSM fields do not develop VEVs.
Now, we require that the vacuum does not break EW symmetry, so that the vacuum alignment should be
\beq
\Delta_R = \begin{pmatrix} 0 & 0 \\ v &0 \end{pmatrix}, \, \overline \Delta_R = \begin{pmatrix} 0 & \overline v \\ 0 &0 \end{pmatrix}.
\eeq
Then, $ \pa_{\Delta_R} W= \pa_{ \overline \Delta_R} W=0$ requires
\beq
m(S)=0,~ v \overline{v}  =- \frac{\pa_S w(S)}{ \pa_{ S} m(S)}.
\eeq
At the point, the D-terms are evaluated as
\beq
 D^1_{SU(2)_R}=D^2_{SU(2)_R}=0, \, D^3_{SU(2)_R}= |v|^2- |\overline v|^2, D_{B-L}=\xi-2( |v|^2- |\overline v|^2).
\eeq 
Thus, the condition, $|v|= |\overline v|$ and $\xi=0$, leads the SUSY conserving vacuum that breaks $SU(2)_R \times U(1)_{B-L}$ to $U(1)_Y$.

Note that we may wonder how break the SUSY and how mediate the $SU(2)_{{\rm R}}$ breaking effect.
See, for instance, Ref. \cite{Kobayashi:2017fgl}.

\section{The summary of the induced four-fermi couplings}
\label{appendix1}
In this section, we summarize the four-fermi couplings, that are not mentioned in Sec. \ref{sec;4-fermi}.
Assuming that the components of $H_A$ ($H$) in the (non-)supersymmetric LR models
are degenerate, the heavy neutral scalar exchanging gives the four-fermi couplings,
\beq
{\cal L}^{q}_{eff}=(C^d_4)_{kl}^{ij} \left (\overline{d^{ i}_R}  d^{j}_L \right ) \,  \left ( \overline{d^{k}_L} d^{ l}_R \right )+(C^u_4)_{kl}^{ij} \left (\overline{u^{ i}_R}  u^{j}_L \right ) \,  \left ( \overline{u^{k}_L} u^{ l}_R \right )+(C^{ud}_4)_{kl}^{ij} \left (\overline{u^{ i}_L}  u^{j}_R \right ) \,  \left ( \overline{d^{k}_L} d^{ l}_R \right )+h.c.,
\eeq
\beq
{\cal L}^{l}_{eff}=(C^e_4)_{kl}^{ij} \left (\overline{e^{ i}_R}  e^{j}_L \right ) \,  \left ( \overline{e^{k}_L} e^{ l}_R \right )+(C^{ue}_4)^{kl}_{ij} \left (\overline{u^{ i}_L}  u^{j}_R \right ) \,  \left ( \overline{e^{k}_L} e^{ l}_R \right )+(C^{de}_4)^{kl}_{ij} \left (\overline{d^{ i}_R}  d^{j}_L \right ) \,  \left ( \overline{e^{k}_L} e^{ l}_R \right )+h.c..
\eeq
The coefficients, $(C^d_4)^{ij}_{kl}$, are given by
\beq
(C^d_4)^{ij}_{kl} =  \begin{pmatrix} Y^{u \, \dagger}_{ij} &  y^{d }_i \,  \delta_{ij} \end{pmatrix}
\begin{pmatrix}  \Lambda^{-2}_{uu} & \Lambda^{-2}_{ud} \\  \Lambda^{-2}_{du} &  \Lambda^{-2}_{dd} \end{pmatrix} \begin{pmatrix} Y^u_{kl} \\ y^{d }_{k}  \,  \delta_{kl} \end{pmatrix},
\eeq
where the dimensional parameters satisfy
\begin{eqnarray}
\label{Lambda-uu}
|\Delta_h|^2\Lambda^{-2}_{uu} &=& (M^{-2}_H)_{33} |U_{4h}|^2 +(M^{-2}_H)_{44} |U_{3h}|^2-(M^{-2}_H)_{34} U^*_{3h} U_{4h}-(M^{-2}_H)_{43} U^*_{4h} U_{3h}, \\
|\Delta_h|^2\Lambda^{-2}_{ud} &=& - (M^{-2}_H)_{33} U^*_{2h} U^*_{4h} -(M^{-2}_H)_{44}U^*_{3h} U^*_{1h}+(M^{-2}_H)_{34} U^*_{3h} U^*_{2h}+(M^{-2}_H)_{43} U^*_{4h} U^*_{1h}, \\
\Lambda^{-2}_{du} &=& \Lambda^{-2 \, *}_{ud},  \label{eq;Lambdadu} \\
|\Delta_h|^2 \Lambda^{-2}_{dd} &=&  (M^{-2}_H)_{33} |U_{2h}|^2 +(M^{-2}_H)_{44} |U_{1h}|^2-(M^{-2}_H)_{34} U^*_{2h} U_{1h}-(M^{-2}_H)_{43} U^*_{1h} U_{2h} +|\Delta_h|^2 \mu_h^{-2}. \nonumber  \\
\end{eqnarray}
The coefficients, $(C^u_4)^{ij}_{kl}$, are given by
\beq
(C^u_4)^{ij}_{kl}=\begin{pmatrix} y^{u }_i \delta_{ij} &  (Y^{d \, \dagger})_{ij} \end{pmatrix}
\begin{pmatrix}  \left (\Lambda^{ D }_{uu}  \right )^{-2}& \left ( \Lambda^{D }_{ud}  \right )^{-2} \\  \left ( \Lambda^{D }_{du}  \right )^{-2} &  \left (  \Lambda^{D }_{dd}  \right)^{-2} \end{pmatrix} \begin{pmatrix} y^{u }_k \delta_{kl} \\ (Y^{d })_{kl} \end{pmatrix},
\eeq
with
\begin{eqnarray}
|\Delta_h|^2 \left (\Lambda^{ D }_{uu}  \right )^{-2}&=& (M^{-2}_H)_{11} |U_{4h}|^2 +(M^{-2}_H)_{22} |U_{3h}|^2-(M^{-2}_H)_{21} U^*_{3h} U_{4h}-(M^{-2}_H)_{12} U^*_{4h} U_{3h}+|\Delta_h|^2 \mu_h^{-2},  \nonumber 
\\
&&  \\
|\Delta_h|^2 \left (\Lambda^{ D }_{ud}  \right )^{-2}&=& - (M^{-2}_H)_{11} U^*_{2h} U^*_{4h} -(M^{-2}_H)_{22}U^*_{3h} U^*_{1h}+(M^{-2}_H)_{21} U^*_{3h} U^*_{2h}+(M^{-2}_H)_{12} U^*_{4h} U^*_{1h}, \\
 \left (\Lambda^{ D }_{du}  \right )^{-2} &=&  \left (\Lambda^{ D }_{ud}  \right )^{-2 \, *},  \\
|\Delta_h|^2 \left (\Lambda^{ D }_{dd}  \right )^{-2}&=&  (M^{-2}_H)_{11} |U_{2h}|^2 +(M^{-2}_H)_{22} |U_{1h}|^2-(M^{-2}_H)_{21} U^*_{2h} U_{1h}-(M^{-2}_H)_{12} U^*_{1h} U_{2h}. \label{Lambda-dd}
\end{eqnarray}
The other coefficients, $(C^{e}_4)_{ij}^{kl}$ and $(C^{ue}_4)_{ij}^{kl}$, are described as
\beq
(C^e_4)^{ij}_{kl}=\begin{pmatrix} (Y^{\nu \, \dagger})_{ij} &  y^e_i \, \delta_{ij} \end{pmatrix}
\begin{pmatrix}  \Lambda^{-2}_{uu} & \Lambda^{-2}_{ud} \\  \Lambda^{-2}_{du} &  \Lambda^{-2}_{dd} \end{pmatrix} \begin{pmatrix} (Y^{\nu })_{kl} \\ y^{e }_{k}  \,  \delta_{kl} \end{pmatrix},
\eeq
\beq
(C^{ue}_4)_{ij}^{kl} = \begin{pmatrix} y^{u }_{i} \delta_{ij} &  ( V y^{d} V^\dagger_R )_{ij} \end{pmatrix}
\begin{pmatrix}  \left  (\Lambda^{(u e) }_{uu} \right )^{-2} & \left (\Lambda^{(u e) }_{ud} \right)^{-2}  \\   \left (\Lambda^{(u e) }_{du} \right )^{-2}  &   \left (\Lambda^{(u e)}_{dd} \right )^{-2}  \end{pmatrix} \begin{pmatrix} (Y^{\nu })_{kl} \\ y^{e }_{k} \delta_{kl} \end{pmatrix}.
\eeq

$\Lambda^{(ue) }_{ab}$ are related to $M^2_H$ as follows:
\begin{eqnarray}
|\Delta_h|^2 \left (\Lambda^{(u e)}_{uu}  \right )^{-2}&=& (M^{-2}_H)_{31} U_{4h}^2 +(M^{-2}_H)_{42} U_{3h}^2-(M^{-2}_H)_{32} U_{3h} U_{4h}-(M^{-2}_H)_{41} U_{4h} U_{3h},  
\\
|\Delta_h|^2 \left (\Lambda^{ (u e) }_{ud}  \right )^{-2}&=& -(M^{-2}_H)_{31} U_{4h} U^*_{2h} -(M^{-2}_H)_{42} U_{3h}U^*_{1h}+(M^{-2}_H)_{32} U_{3h} U^*_{2h}+(M^{-2}_H)_{41} U_{4h} U^*_{1h} \nonumber \\
&& +|\Delta_h|^2 \mu_h^{-2},  \\
|\Delta_h|^2 \left (\Lambda^{ (u e) }_{du}  \right )^{-2} &=& -(M^{-2}_H)_{31} U_{4h} U^*_{2h} -(M^{-2}_H)_{42} U_{3h}U^*_{1h}+(M^{-2}_H)_{32} U_{4h} U^*_{1h} +(M^{-2}_H)_{41} U_{3h} U^*_{2h},   \nonumber \\
 && \\
|\Delta_h|^2 \left (\Lambda^{ (u e) }_{dd}  \right )^{-2}&=&  (M^{-2}_H)_{31} U^{* \, 2}_{2h} +(M^{-2}_H)_{42} U^{* \, 2}_{1h}-(M^{-2}_H)_{32} U^*_{1h} U^*_{2h}-(M^{-2}_H)_{41} U^*_{1h} U^*_{2h}.
\end{eqnarray}
Note that $(C^{de}_4)_{ij}^{kl}$ depend on $ \Lambda^{-2}_{uu}$, $ \Lambda^{-2}_{ud}$, $ \Lambda^{-2}_{du}$ and $ \Lambda^{-2}_{dd}$:
\beq
(C^{de}_4)_{ij}^{kl} =\begin{pmatrix} Y^{u \, \dagger }_{ij} &  y^{d }_i \delta_{ij} \end{pmatrix}
\begin{pmatrix}  \Lambda^{-2}_{uu} & \Lambda^{-2}_{ud} \\  \Lambda^{-2}_{du} &  \Lambda^{-2}_{dd} \end{pmatrix} \begin{pmatrix} (Y^\nu )_{kl} \\ y^{e }_{k} \delta_{kl} \end{pmatrix}.
\eeq

We can also find the four-fermi couplings involving the light neutrinos. They become important in the Dirac neutrino case.
In the Dirac neutrino case, the couplings relevant to the meson decays are 
\beq
{\cal L}^{\nu}_{eff}=(C^{u \nu}_4)_{kl}^{ij} \left (\overline{u^{ i}_R}  u^{j}_L \right ) \,  \left ( \overline{\nu^{k}_L} \nu^{ l}_R \right )+(C^{d \nu}_4)_{kl}^{ij} \left (\overline{d^{ i}_L}  d^{j}_R \right ) \, \left ( \overline{\nu^{k}_L} \nu^{ l}_R \right ) +h.c..
\eeq
$(C^{u \nu}_4)^{ij}_{kl}$ and $(C^{d \nu}_4)^{ij}_{kl}$ are described as
\beq
(C^{d \nu}_4)^{ij}_{kl} = \begin{pmatrix} (V^\dagger_R y^{u } V)_{ij} &  y^{d }_i \delta_{ij} \end{pmatrix}
\begin{pmatrix}  \left  (\Lambda^{(u e) }_{uu} \right )^{-2} & \left (\Lambda^{(u e) }_{ud} \right)^{-2}  \\   \left (\Lambda^{(u e) }_{du} \right )^{-2}  &   \left (\Lambda^{(u e)}_{dd} \right )^{-2}  \end{pmatrix}   \begin{pmatrix} y^{\nu }_k \delta_{kl} \\ (V^\dagger_{PMNS} y^{e } V^\nu_R)_{kl}  \end{pmatrix}
\eeq
and 
\beq
(C^{u \nu}_4)^{ij}_{kl} = \begin{pmatrix} y^{u }_{i} \delta_{ij} &  (Y^d)_{ij}^\dagger \end{pmatrix}
\begin{pmatrix}  \left  (\Lambda^{D}_{uu} \right )^{-2} & \left (\Lambda^{D }_{ud} \right)^{-2}  \\   \left (\Lambda^{D}_{du} \right )^{-2}  &   \left (\Lambda^{D}_{dd} \right )^{-2}  \end{pmatrix} \begin{pmatrix} y^{\nu }_k \delta_{kl} \\ (V^\dagger_{PMNS} y^{e } V^\nu_R)_{kl}  \end{pmatrix}.
\eeq

\section{RG flow of the LR breaking effects}
\label{appendix4}
In this appendix, we summarize the detail of the RG flow of the LR breaking effects. The LR breaking is induced by the $SU(2)_L \times U(1)_Y$ gauge interactions and the leptonic Yukawa interaction.
 In particular, the LR breaking would be enhanced, if $Y^\nu_{ij}$ is large. 

In the supersymmetric model, the running Yukawa couplings for quarks are described as
\beq
Y^{u \, a}_{ij} (\mu)=(Z^{\dagger }_{Q_L})^{ik} \, Y^{ b}_{km} (\Lambda)  \, Z^{mj}_{u_R} \,  Z^{H_u}_{ba}, ~ Y^{d \, a}_{ij} (\mu)=(Z^{\dagger }_{Q_L})^{ik} \, Y^{ b}_{km} (\Lambda)  \, Z^{mj}_{d_R} \,  Z^{H_d}_{ba}.
\eeq
In the same manner, the running Yukawa couplings for leptons are given by
\beq
Y^{l \, a}_{ij} (\mu)=(Z^{\dagger }_{L})^{ik} \, Y^{l \, b}_{km} (\Lambda)  \, Z^{mj}_{R} \,  Z^{H_d}_{ba}.
\eeq
In our study, we do not touch the detail of the setup and discuss the phenomenology.
Then, we approximately parameterize the LR breaking contributions, decomposing the wave function renormalization factors as follows:
\begin{eqnarray}
Z^{ij}_{Q_L} &=& Z^{EW}_{Q_L} \, Z^{ij}_q, ~ Z^{ij}_{u_R} = Z^{EW}_{u_R} \, Z^{ij}_q,~ Z^{ij}_{d_R} = Z^{EW}_{d_R} \, Z^{ij}_q,  \\ 
Z^{ij}_{L} &=&  Z^{EW}_{L} \,  Z^{ij}, ~ Z^{ij}_{R} =  Z^{EW}_{R}  \, Z^{ik} \, Z^{l}_{kj}, \\
Z_{a b}^{H_d}&=& Z^{EW}_{H} \,  Z^H_{ac} Z^d_{cb},~ Z_{a b}^{H_u}= Z^{EW}_{H} \,  Z^H_{ab}.
\end{eqnarray}
Then, the running Yukawa couplings are described as 
\begin{eqnarray}
Y^{l \, a}_{ij} &=& \widetilde Y^{l \, b}_{ik} \,  Z^{l}_{kj} \, Z^d_{ba}  \times Z^{EW}_{L} Z^{EW}_{R}= \Hat Y^{l \, b}_{ik} \,  Z^{l}_{kj} \, Z^d_{ba}  , \\
Y^{d \, a}_{ij} &=&  \widetilde Y^{ b}_{ij} \, Z^d_{ba} \times Z^{EW}_{Q_L} \, Z^{EW}_{d_R}, \\
Y^{u \, a}_{ij} &=&    \widetilde Y^{ a}_{ij} \times Z^{EW}_{Q_L} \, Z^{EW}_{u_R}.
\end{eqnarray}
$\widetilde Y^{l \, b}_{kj}$ and $\widetilde Y^{ a}_{ij}$ are interpreted as the couplings in the LR symmetric limit
and $\Hat Y^{l \, b}_{kj}$ is the hermitian matrix.
$Z^{EW}_{\phi}$ ($\phi=Q_L, \, u_R, \, d_R, \, L, \, e_R$) denotes the contribution from $SU(2)_L \times U(1)_Y$ gauge interactions. Then, the RG equations at the one-loop level are 
\beq
\mu \frac{d}{d  \mu} \ln Z^{EW}_{\phi}= - c^{\phi}_Y \, \frac{\alpha_Y}{4 \pi}- c^{\phi}_2\,  \frac{\alpha_2}{4 \pi},
\eeq
where $ c^{\phi}_Y$ and $c^{\phi}_2$ are the constants given by $U(1)_Y$ and $SU(2)_L$ gauge symmetries.
When we assume that $\Hat Y^{l \, a}_{ij}$ is the only interaction that leads the LR breaking contribution,
the LR breaking factors satisfy the following one-loop RG equations, 
\begin{eqnarray}
\mu \frac{d}{d  \mu} \ln Z^{l}_{ij}&=&  \frac{1}{(4 \pi)^2}   (Y^{l \, a})_{ik}  (Y^{l \, a \, \dagger})_{kj} \approx  \frac{1}{(4 \pi)^2}   (\Hat Y^{l \, a})_{ik}  (\Hat Y^{l \, a \, \dagger})_{kj} ,  \\
\mu \frac{d}{d  \mu} \ln Z^{d}_{ab}&=&  \frac{1}{(4 \pi)^2}  (Y^{l \, a})_{ik}  (Y^{l \, b \, \dagger})_{ki} \approx \frac{1}{(4 \pi)^2}  (\Hat Y^{l \, a})_{ik}  (\Hat Y^{l \, b \, \dagger})_{ki}. 
\end{eqnarray}  

Note that the correction from $Z^{H_{u,d}}_{ab}$ can be interpreted as the mixing effects of the Higgs doublets in the supersymmetric LR model. $Z^{H_{u,d}}_{ab}$ effectively enhance/suppress the Yukawa couplings, as well.
In our analysis, those effects would be characterized by the new physics scale denoted by $\Lambda_{uu}$ etc..
Thus, $Y^\nu_{ij}$ and $Y^e_{ij}$ shown in Eq. (\ref{eq;heavy-Yukawa-LR-lepton-SUSY}) are changed to
  \begin{eqnarray}
 \label{eq;heavy-Yukawa-LR-lepton-SUSY-2}
 \begin{pmatrix} Y^\nu_{A\, ij}  \\ Y^e_{A\, ij}   \end{pmatrix}  &=& \frac{1}{\Delta^r_h}  \begin{pmatrix}   U^*_{1A} &   U^*_{2A} \\   U^r_{3A} & U^r_{4A} \end{pmatrix} \begin{pmatrix}  U^r_{4h} &- U^*_{2h} \\  -U^r_{3h} & U^*_{1h} \end{pmatrix}\begin{pmatrix}  Y^\nu_{ ij}  \\  Y^e_{ ij}\end{pmatrix},
\end{eqnarray}
where $U^r_{3h} \equiv Z^d_{11}  U_{3h} + Z^d_{21}  U_{4h}$ and $U^r_{4h} \equiv Z^d_{22} U_{4h} + Z^d_{12}  U_{3h}$
are defined. 
On the other hand, the quark Yukawa couplings shown in Eq. (\ref{eq;heavy-Yukawa-LR-quark-SUSY}) are modified by the RG corrections as
\begin{equation}
 \label{eq;heavy-Yukawa-LR-quark-SUSY-2}
 \begin{pmatrix} Y^u_{A\, ij}  \\ Y^d_{A\, ij}   \end{pmatrix}  = \frac{1}{\Delta^r_h}  \begin{pmatrix}   U^*_{1A} & U^*_{2A} \\  Z^{-1}_{EW} U^r_{3A} & Z^{-1}_{EW} U^r_{4A} \end{pmatrix}  \begin{pmatrix}  U^r_{4h} & -Z_{EW}  U^*_{2h} \\  -U^r_{3h} & Z_{EW}  U^*_{1h} \end{pmatrix}\begin{pmatrix}  Y^u_{ ij}  \\  Y^d_{ ij}\end{pmatrix}.
\end{equation}
$Z_{EW}$ is the EW contribution described as $Z_{EW}=Z^{EW}_{u_R}/Z^{EW}_{d_R}$. Thus, the new physics scales, discussed in Sec. \ref{sec;4-fermi}, are redefined, taking the RG corrections into account.
In our analysis on the phenomenology, the constraints from flavor physics on the scales and the Yukawa couplings will be discussed. The scales are the ones including the RG corrections, as shown in Eqs. (\ref{eq;heavy-Yukawa-LR-lepton-SUSY-2}) and (\ref{eq;heavy-Yukawa-LR-quark-SUSY-2}). The EW correction will give the difference
between the scales for quark and for lepton. In our study, such a scale difference is ignored and the improved analysis, taking into account the RG corrections more precisely, will be given in the future. 

\section{RG flow to 100 TeV}
\label{app_MZto100TeV}

For predicting the phenomenology, we calculate the Wilson coefficients of the four-fermi interactions at the integrated Higgs mass scale, here we assume it to be 100 TeV.
These Wilson coefficients consist of the Yukawa couplings and dimensional parameters $\Lambda_{\alpha \beta}$.
Thus, we evaluate the Yukawa couplings at 100 TeV.

We evaluate the Yukawa couplings at 100 TeV under following steps.
First, we evaluate running masses at the $M_Z$ scale by using the central values of experimental measurements summarized in Table \ref{table;input2} and \ref{table;input}.
To evaluate the quark running masses we use Mathematica package RunDec \cite{Chetyrkin:2000yt} and to translate the lepton pole masses to the running masses we follow Ref. \cite{Arason:1991ic}. 
We calculate the Yukawa couplings at 1 TeV, using the SM RG running at the two-loop level \cite{Luo:2002ey}.
In our scenario all gaugino has around 1 TeV masses, so that we translate the $\overline{\text{MS}}$ scheme into the $\overline{\text{DR}}$ scheme at 1 TeV by following Ref. \cite{Martin:1993yx}.
Then, the RG correction from 1 TeV to 100 TeV includes the gaugino contributions \cite{Martin:1993zk}.


\begin{thebibliography}{99}


\bibitem{Babu:1989rb} 
  K.~S.~Babu and R.~N.~Mohapatra,
  Phys.\ Rev.\ D {\bf 41}, 1286 (1990).

\bibitem{Babu:2008ep} 
  K.~S.~Babu and R.~N.~Mohapatra,
  Phys.\ Lett.\ B {\bf 668}, 404 (2008)
  [arXiv:0807.0481 [hep-ph]].


\bibitem{Kawamura:2017qey} 
  J.~Kawamura and Y.~Omura,
  JHEP {\bf 1711}, 189 (2017)
  [arXiv:1710.03412 [hep-ph]].


\bibitem{Babu:2014vba} 
  K.~S.~Babu and A.~Patra,
  Phys.\ Rev.\ D {\bf 93}, no. 5, 055030 (2016)
  [arXiv:1412.8714 [hep-ph]].

\bibitem{Dev:2016dja} 
  P.~S.~B.~Dev, R.~N.~Mohapatra and Y.~Zhang,
  JHEP {\bf 1605}, 174 (2016)
  [arXiv:1602.05947 [hep-ph]].


\bibitem{Huitu:1994zm} 
  K.~Huitu and J.~Maalampi,
  Phys.\ Lett.\ B {\bf 344}, 217 (1995)
  [hep-ph/9410342].


 
\bibitem{Frank:2014kma} 
  M.~Frank, D.~K.~Ghosh, K.~Huitu, S.~K.~Rai, I.~Saha and H.~Waltari,
  Phys.\ Rev.\ D {\bf 90}, no. 11, 115021 (2014)
  [arXiv:1408.2423 [hep-ph]].
\bibitem{Frank:2011jia} 
  M.~Frank and B.~Korutlu,
  Phys.\ Rev.\ D {\bf 83}, 073007 (2011)
  [arXiv:1101.3601 [hep-ph]].
 
\bibitem{Glashow:1976nt} 
  S.~L.~Glashow and S.~Weinberg,
  Phys.\ Rev.\ D {\bf 15}, 1958 (1977).
\bibitem{ArkaniHamed:2004fb} 
  N.~Arkani-Hamed and S.~Dimopoulos,
  JHEP {\bf 0506}, 073 (2005)
  [hep-th/0405159].
\bibitem{Giudice:2004tc} 
  G.~F.~Giudice and A.~Romanino,
  Nucl.\ Phys.\ B {\bf 699}, 65 (2004)
  [Erratum-ibid.\ B {\bf 706}, 65 (2005)]
  [hep-ph/0406088].
\bibitem{ArkaniHamed:2004yi} 
  N.~Arkani-Hamed, S.~Dimopoulos, G.~F.~Giudice and A.~Romanino,
  Nucl.\ Phys.\ B {\bf 709}, 3 (2005)
  [hep-ph/0409232].
\bibitem{Wells} 
  J.~D.~Wells,
  Phys.\ Rev.\ D {\bf 71}, 015013 (2005)
  [hep-ph/0411041].
  \bibitem{Giudice:2011cg} 
  G.~F.~Giudice and A.~Strumia,
  Nucl.\ Phys.\ B {\bf 858}, 63 (2012)
  [arXiv:1108.6077 [hep-ph]].
  \bibitem{Hall}
    L.~J.~Hall and Y.~Nomura,
  JHEP {\bf 1201}, 082 (2012)
  [arXiv:1111.4519 [hep-ph]].
  \bibitem{Ibe}
  M.~Ibe and T.~T.~Yanagida,
  Phys.\ Lett.\ B {\bf 709}, 374 (2012)
  [arXiv:1112.2462 [hep-ph]].
  \bibitem{Ibe2}
M.~Ibe, S.~Matsumoto and T.~T.~Yanagida,
  Phys.\ Rev.\ D {\bf 85}, 095011 (2012)
  [arXiv:1202.2253 [hep-ph]].
  \bibitem{ArkaniHamed}
N.~Arkani-Hamed, A.~Gupta, D.~E.~Kaplan, N.~Weiner and T.~Zorawski,
  arXiv:1212.6971 [hep-ph].

 
\bibitem{Omura:2015nja} 
  Y.~Omura, E.~Senaha and K.~Tobe,
  JHEP {\bf 1505}, 028 (2015)
  [arXiv:1502.07824 [hep-ph]].
\bibitem{Omura:2015xcg} 
  Y.~Omura, E.~Senaha and K.~Tobe,
  Phys.\ Rev.\ D {\bf 94}, no. 5, 055019 (2016)
  [arXiv:1511.08880 [hep-ph]].
  
  \bibitem{Sierra:2014nqa} 
  D.~Aristizabal Sierra and A.~Vicente,
  Phys.\ Rev.\ D {\bf 90}, 115004 (2014)
  [arXiv:1409.7690 [hep-ph]].
\bibitem{Crivellin:2015mga} 
  A.~Crivellin, G.~D'Ambrosio and J.~Heeck,
  arXiv:1501.00993 [hep-ph].
\bibitem{deLima:2015pqa} 
  L.~de Lima {\it et al.},
  arXiv:1501.06923 [hep-ph].

  
\bibitem{Ko:2017lzd} 
  P.~Ko, Y.~Omura, Y.~Shigekami and C.~Yu,
  Phys.\ Rev.\ D {\bf 95}, no. 11, 115040 (2017)
  [arXiv:1702.08666 [hep-ph]].
  
\bibitem{Ko:2012sv} 
  P.~Ko, Y.~Omura and C.~Yu,
  JHEP {\bf 1303}, 151 (2013)
  [arXiv:1212.4607 [hep-ph]].
  
\bibitem{Iguro:2017ysu} 
  S.~Iguro and K.~Tobe,
  Nucl.\ Phys.\ B {\bf 925}, 560 (2017)
  [arXiv:1708.06176 [hep-ph]].
  
\bibitem{Iguro:2018qzf} 
  S.~Iguro and Y.~Omura,
  arXiv:1802.01732 [hep-ph].
  
  \bibitem{Crivellin2012ye} 
  A.~Crivellin, C.~Greub and A.~Kokulu,
  Phys.\ Rev.\ D {\bf 86}, 054014 (2012)
  [arXiv:1206.2634 [hep-ph]].

  \bibitem{Celis:2012dk} 
  A.~Celis, M.~Jung, X.~Q.~Li and A.~Pich,
  JHEP {\bf 1301}, 054 (2013)
  [arXiv:1210.8443 [hep-ph]].
  
\bibitem{Tanaka2012nw} 
  M.~Tanaka and R.~Watanabe,
  Phys.\ Rev.\ D {\bf 87}, no. 3, 034028 (2013)
  [arXiv:1212.1878 [hep-ph]].

  
\bibitem{Crivellin:2013wna} 
  A.~Crivellin, A.~Kokulu and C.~Greub,
  Phys.\ Rev.\ D {\bf 87}, no. 9, 094031 (2013)
  [arXiv:1303.5877 [hep-ph]].
  
  
  \bibitem{Crivellin:2015hha} 
  A.~Crivellin, J.~Heeck and P.~Stoffer,
  Phys.\ Rev.\ Lett.\  {\bf 116}, no. 8, 081801 (2016)
  [arXiv:1507.07567 [hep-ph]].
  
\bibitem{Cline:2015lqp}
  J.~M.~Cline,
  Phys.\ Rev.\ D {\bf 93}, no. 7, 075017 (2016)
  [arXiv:1512.02210 [hep-ph]].


 
  


\bibitem{Hu:2016gpe} 
  Q.~Y.~Hu, X.~Q.~Li and Y.~D.~Yang,
  Eur.\ Phys.\ J.\ C {\bf 77}, no. 3, 190 (2017)
  [arXiv:1612.08867 [hep-ph]].
  
\bibitem{Arnan:2017lxi} 
  P.~Arnan, D.~Be\v{c}irevi\'c, F.~Mescia and O.~Sumensari,
  Eur.\ Phys.\ J.\ C {\bf 77}, no. 11, 796 (2017)
  [arXiv:1703.03426 [hep-ph]].

\bibitem{Arhrib:2017yby} 
  A.~Arhrib, R.~Benbrik, C.~H.~Chen, J.~K.~Parry, L.~Rahili, S.~Semlali and Q.~S.~Yan,
  arXiv:1710.05898 [hep-ph].

  
  
  
  
  
  
  
\bibitem{Ko:2011vd} 
  P.~Ko, Y.~Omura and C.~Yu,
  Phys.\ Rev.\ D {\bf 85}, 115010 (2012)
  [arXiv:1108.0350 [hep-ph]].
 
\bibitem{Ko:2011di} 
  P.~Ko, Y.~Omura and C.~Yu,
  JHEP {\bf 1201}, 147 (2012)
  [arXiv:1108.4005 [hep-ph]].
 
 
\bibitem{Ball:1999mb} 
  P.~Ball, J.~M.~Frere and J.~Matias,
  Nucl.\ Phys.\ B {\bf 572}, 3 (2000)
  [hep-ph/9910211].
\bibitem{Kiers:2002cz} 
  K.~Kiers, J.~Kolb, J.~Lee, A.~Soni and G.~H.~Wu,
  Phys.\ Rev.\ D {\bf 66}, 095002 (2002)
  [hep-ph/0205082].
\bibitem{Haba:2017jgf} 
  N.~Haba, H.~Umeeda and T.~Yamada,
  Phys.\ Rev.\ D {\bf 97}, no. 3, 035004 (2018)
  [arXiv:1711.06499 [hep-ph]].
\bibitem{Zhang:2007da} 
  Y.~Zhang, H.~An, X.~Ji and R.~N.~Mohapatra,
  Nucl.\ Phys.\ B {\bf 802}, 247 (2008)
  [arXiv:0712.4218 [hep-ph]].
\bibitem{Guadagnoli:2010sd} 
  D.~Guadagnoli and R.~N.~Mohapatra,
  Phys.\ Lett.\ B {\bf 694}, 386 (2011)
  [arXiv:1008.1074 [hep-ph]].
  
\bibitem{Blanke:2011ry} 
  M.~Blanke, A.~J.~Buras, K.~Gemmler and T.~Heidsieck,
  JHEP {\bf 1203}, 024 (2012)
  [arXiv:1111.5014 [hep-ph]].
\bibitem{Bertolini:2014sua} 
  S.~Bertolini, A.~Maiezza and F.~Nesti,
  Phys.\ Rev.\ D {\bf 89}, no. 9, 095028 (2014)
  [arXiv:1403.7112 [hep-ph]].
\bibitem{Bernard:2015boz} 
  V.~Bernard, S.~Descotes-Genon and L.~Vale Silva,
  JHEP {\bf 1608}, 128 (2016)
  [arXiv:1512.00543 [hep-ph]].
\bibitem{ValeSilva:2016czu} 
  L.~Vale Silva,
  arXiv:1611.08187 [hep-ph].
\bibitem{FileviezPerez:2017zwm} 
  P.~Fileviez Perez and C.~Murgui,
  Phys.\ Rev.\ D {\bf 95}, no. 7, 075010 (2017)
  [arXiv:1701.06801 [hep-ph]].
 
 



\bibitem{PDG}
  C.~Patrignani {\it et al.} [Particle Data Group],
  Chin.\ Phys.\ C {\bf 40}, no. 10, 100001 (2016).




\bibitem{neutrino1} 
  P.~F.~de Salas, D.~V.~Forero, C.~A.~Ternes, M.~Tortola and J.~W.~F.~Valle,
  arXiv:1708.01186 [hep-ph].


\bibitem{CKMfitter}
 CKMfitter global fit results as of Summer 2016 (ICHEP 2016 conference)\\
 \texttt{http://ckmfitter.in2p3.fr/www/html/ckm\_main.html}


  
\bibitem{Aulakh:1997ba} 
  C.~S.~Aulakh, K.~Benakli and G.~Senjanovic,
  Phys.\ Rev.\ Lett.\  {\bf 79}, 2188 (1997)
  [hep-ph/9703434].

\bibitem{Kuchimanchi:1995vk} 
  R.~Kuchimanchi and R.~N.~Mohapatra,
  Phys.\ Rev.\ Lett.\  {\bf 75}, 3989 (1995)
  [hep-ph/9509256].

\bibitem{Mohapatra:1995xd} 
  R.~N.~Mohapatra and A.~Rasin,
  Phys.\ Rev.\ Lett.\  {\bf 76}, 3490 (1996)
  [hep-ph/9511391].



\bibitem{Kobayashi:2017fgl} 
  T.~Kobayashi, Y.~Omura, O.~Seto and K.~Ueda,
  JHEP {\bf 1711}, 073 (2017)
  [arXiv:1705.00809 [hep-ph]].
  


\bibitem{Misiak:2017bgg} 
  M.~Misiak and M.~Steinhauser,
  Eur.\ Phys.\ J.\ C {\bf 77}, no. 3, 201 (2017)
  [arXiv:1702.04571 [hep-ph]].

  
  
  
\bibitem{Buras:2008nn} 
  A.~J.~Buras and D.~Guadagnoli,
  Phys.\ Rev.\ D {\bf 78}, 033005 (2008)
  [arXiv:0805.3887 [hep-ph]].
  
\bibitem{Buras:2010pza} 
  A.~J.~Buras, D.~Guadagnoli and G.~Isidori,
  Phys.\ Lett.\ B {\bf 688}, 309 (2010)
  [arXiv:1002.3612 [hep-ph]].

\bibitem{Inami:1980fz} 
  T.~Inami and C.~S.~Lim,
  Prog.\ Theor.\ Phys.\  {\bf 65}, 297 (1981)

\bibitem{Aoki:2013ldr} 
  S.~Aoki {\it et al.},
  Eur.\ Phys.\ J.\ C {\bf 74}, 2890 (2014)
  [arXiv:1310.8555 [hep-lat]].



\bibitem{Brod:2011ty} 
  J.~Brod and M.~Gorbahn,
  Phys.\ Rev.\ Lett.\  {\bf 108}, 121801 (2012)
  [arXiv:1108.2036 [hep-ph]].

\bibitem{Buras:1990fn} 
  A.~J.~Buras, M.~Jamin and P.~H.~Weisz,
  Nucl.\ Phys.\ B {\bf 347}, 491 (1990).


\bibitem{Brod:2010mj} 
  J.~Brod and M.~Gorbahn,
  Phys.\ Rev.\ D {\bf 82}, 094026 (2010)
  [arXiv:1007.0684 [hep-ph]].

  
\bibitem{Bazavov:2016nty} 
  A.~Bazavov {\it et al.} [Fermilab Lattice and MILC Collaborations],
  Phys.\ Rev.\ D {\bf 93}, no. 11, 113016 (2016)
  [arXiv:1602.03560 [hep-lat]].
  
\bibitem{Belle2}
https://confluence.desy.de/display/BI/B2TiP+ReportStatus


\bibitem{Bellgardt:1987du}
  U.~Bellgardt {\it et al.}  [SINDRUM Collaboration],
  Nucl.\ Phys.\ B {\bf 299}, 1 (1988).

\bibitem{Perrevoort:2016nuv} 
  A.~K.~Perrevoort [Mu3e Collaboration],
  EPJ Web Conf.\  {\bf 118}, 01028 (2016)
  [arXiv:1605.02906 [physics.ins-det]].

\bibitem{Buras:2012ru} 
  A.~J.~Buras, J.~Girrbach, D.~Guadagnoli and G.~Isidori,
  Eur.\ Phys.\ J.\ C {\bf 72}, 2172 (2012)
  [arXiv:1208.0934 [hep-ph]].
  
\bibitem{Bobeth:2013uxa} 
  C.~Bobeth, M.~Gorbahn, T.~Hermann, M.~Misiak, E.~Stamou and M.~Steinhauser,
  Phys.\ Rev.\ Lett.\  {\bf 112}, 101801 (2014)
  [arXiv:1311.0903 [hep-ph]].
  
\bibitem{Buras:2013uqa} 
  A.~J.~Buras, R.~Fleischer, J.~Girrbach and R.~Knegjens,
  JHEP {\bf 1307}, 77 (2013)
  [arXiv:1303.3820 [hep-ph]].
  
\bibitem{Beneke:2017vpq} 
  M.~Beneke, C.~Bobeth and R.~Szafron,
  Phys.\ Rev.\ Lett.\  {\bf 120}, no. 1, 011801 (2018)
  [arXiv:1708.09152 [hep-ph]].
  
\bibitem{Aaij:2017vad} 
  R.~Aaij {\it et al.} [LHCb Collaboration],
  Phys.\ Rev.\ Lett.\  {\bf 118}, no. 19, 191801 (2017)
  [arXiv:1703.05747 [hep-ex]].
\bibitem{Isidori:2003ts} 
  G.~Isidori and R.~Unterdorfer,
  JHEP {\bf 0401}, 009 (2004)
  [hep-ph/0311084].
  
\bibitem{Aaboud:2017buh} 
  M.~Aaboud {\it et al.} [ATLAS Collaboration],
JHEP {\bf 1710}, 182 (2017)
[arXiv:1707.02424 [hep-ex]].

\bibitem{Belyaev2012qa} 
  A.~Belyaev, N.~D.~Christensen and A.~Pukhov,
  Comput.\ Phys.\ Commun.\  {\bf 184}, 1729 (2013)
  [arXiv:1207.6082 [hep-ph]].


\bibitem{Kitano:2002mt}
  R.~Kitano, M.~Koike and Y.~Okada,
  Phys.\ Rev.\ D {\bf 66}, 096002 (2002)
   [Erratum-ibid.\ D {\bf 76}, 059902 (2007)]
  [hep-ph/0203110].

  
\bibitem{Bertl:2006up} 
  W.~H.~Bertl {\it et al.} [SINDRUM II Collaboration],
  Eur.\ Phys.\ J.\ C {\bf 47}, 337 (2006).

  
\bibitem{Kuno:2013mha}
  Y.~Kuno [COMET Collaboration],
  PTEP {\bf 2013}, 022C01 (2013).



\bibitem{Sher:2002ew} 
  M.~Sher,
  Phys.\ Rev.\ D {\bf 66}, 057301 (2002)
  [hep-ph/0207136].
\bibitem{Black:2002wh} 
  D.~Black, T.~Han, H.~J.~He and M.~Sher,
  Phys.\ Rev.\ D {\bf 66}, 053002 (2002)
  [hep-ph/0206056].
\bibitem{Celis:2013xja} 
  A.~Celis, V.~Cirigliano and E.~Passemar,
  Phys.\ Rev.\ D {\bf 89}, 013008 (2014)
  [arXiv:1309.3564 [hep-ph]].


\bibitem{Hurth:2008jc} 
  T.~Hurth, G.~Isidori, J.~F.~Kamenik and F.~Mescia,
  Nucl.\ Phys.\ B {\bf 808}, 326 (2009)
  [arXiv:0807.5039 [hep-ph]].

\bibitem{Mescia:2006jd} 
  F.~Mescia, C.~Smith and S.~Trine,
  JHEP {\bf 0608}, 088 (2006)
  [hep-ph/0606081].


  
  
  

\bibitem{Hisano:2015pma} 
  J.~Hisano, Y.~Muramatsu, Y.~Omura and M.~Yamanaka,
  Phys.\ Lett.\ B {\bf 744}, 395 (2015)
  [arXiv:1503.06156 [hep-ph]].


\bibitem{Hisano:2016afc} 
  J.~Hisano, Y.~Muramatsu, Y.~Omura and Y.~Shigekami,
  JHEP {\bf 1611}, 018 (2016)
  [arXiv:1607.05437 [hep-ph]].

  





\bibitem{Dutta:2004zh} 
  B.~Dutta, Y.~Mimura and R.~N.~Mohapatra,
  Phys.\ Rev.\ Lett.\  {\bf 94}, 091804 (2005)
  [hep-ph/0412105].
\bibitem{Dutta:2005ni} 
  B.~Dutta, Y.~Mimura and R.~N.~Mohapatra,
  Phys.\ Rev.\ D {\bf 72}, 075009 (2005)
  [hep-ph/0507319].
\bibitem{Senjanovic:2006nc} 
  G.~Senjanovic,
  hep-ph/0612312.
\bibitem{Dueck:2013gca} 
  A.~Dueck and W.~Rodejohann,
  JHEP {\bf 1309}, 024 (2013)
  [arXiv:1306.4468 [hep-ph]].


\bibitem{Chetyrkin:2000yt} 
  K.~G.~Chetyrkin, J.~H.~Kuhn and M.~Steinhauser,
  Comput.\ Phys.\ Commun.\  {\bf 133}, 43 (2000).

\bibitem{Arason:1991ic} 
  H.~Arason, D.~J.~Castano, B.~Keszthelyi, S.~Mikaelian, E.~J.~Piard, P.~Ramond and B.~D.~Wright,
  Phys.\ Rev.\ D {\bf 46}, 3945 (1992).
  
\bibitem{Luo:2002ey} 
  M.~x.~Luo and Y.~Xiao,
  Phys.\ Rev.\ Lett.\  {\bf 90}, 011601 (2003).
  
\bibitem{Martin:1993yx} 
  S.~P.~Martin and M.~T.~Vaughn,
  Phys.\ Lett.\ B {\bf 318}, 331 (1993).




 \bibitem{Martin:1993zk} 
  S.~P.~Martin and M.~T.~Vaughn,
  Phys.\ Rev.\ D {\bf 50}, 2282 (1994),
  Erratum: [Phys.\ Rev.\ D {\bf 78}, 039903 (2008)].


\end{thebibliography}
\end{document}